%% file: SUSY-2012-01.tex
\documentclass[final,3p,times,twocolumn,a4paper,12pt]{elsarticle}
\pdfoutput=1
\usepackage{graphicx}
\usepackage{amssymb}
\usepackage{amsthm}
\usepackage{xspace}
\usepackage{atlasphysics}
\usepackage[hyperindex,breaklinks]{hyperref}

\newcommand{\pythia} {{\sc Pythia}\xspace}

\newcommand{\herwig} {{\sc Herwig}\xspace}

\newcommand{\geant} {{\sc Geant4}\xspace}

\newcommand{\spice} {{\sc Spice}\xspace}
\newcommand{\rhad} {\mbox{$R$-hadron}\xspace}
\newcommand{\rhads} {\mbox{$R$-hadrons}\xspace}
\newcommand{\lumi} {4.7~\ifb\xspace} 
\renewcommand{\pt} {\ensuremath{p_{\mathrm{T}}}\xspace}
\newcommand{\ggf} {\ensuremath{gg}\xspace}

\newcommand{\zmumu} {\ensuremath{Z \to \mu\mu}\xspace}
\newcommand{\etmiss} {\ensuremath{E_{\mathrm{T}}^\mathrm{miss}}\xspace}

\renewcommand{\MeV} {\ensuremath{\mathrm{{Me\kern -0.1em V}}}\xspace}
\renewcommand{\GeV} {\ensuremath{\mathrm{{Ge\kern -0.1em V}}}\xspace}
\renewcommand{\TeV} {\ensuremath{\mathrm{{Te\kern -0.1em V}}}\xspace}
\renewcommand{\stau} {\ensuremath{\tilde{\tau}}\xspace}
\newcommand{\MeVgcm} {\ensuremath{\MeV\mathrm{g}^{-1}\mathrm{cm}^{2}}\xspace}
\newcommand{\dedx} {{\ensuremath{\mathrm{d}E/\mathrm{d}x}}\xspace}

\newcommand{\rhadronGluinoFullExpMassExclusion}{1015~\GeV}
\newcommand{\rhadronGluinoFullMassExclusion}{985~\GeV}
\newcommand{\rhadronStopFullExpMassExclusion}{707~\GeV}
\newcommand{\rhadronStopFullMassExclusion}{683~\GeV}
\newcommand{\rhadronSbottomFullExpMassExclusion}{645~\GeV}
\newcommand{\rhadronSbottomFullMassExclusion}{612~\GeV}

\newcommand{\rhadronGluinoAgnosticExpMassExclusion}{988~\GeV}
\newcommand{\rhadronGluinoAgnosticMassExclusion}{989~\GeV}
\newcommand{\rhadronStopAgnosticExpMassExclusion}{647~\GeV}
\newcommand{\rhadronStopAgnosticMassExclusion}{657~\GeV}
\newcommand{\rhadronSbottomAgnosticExpMassExclusion}{615~\GeV}
\newcommand{\rhadronSbottomAgnosticMassExclusion}{618~\GeV}

\newcommand{\rhadronGluinoIDExclusion}{940~\GeV}
\newcommand{\rhadronSbottomIDExclusion}{576~\GeV}
\newcommand{\rhadronStopIDExclusion}{604~\GeV}
\newcommand{\rhadronGluinoIDExpectation}{952~\GeV}
\newcommand{\rhadronSbottomIDExpectation}{571~\GeV}
\newcommand{\rhadronStopIDExpectation}{620~\GeV}

\journal{Physics Letters B}

\usepackage{preprintcover}  
\PreprintCoverPaperTitle{Searches for heavy long-lived sleptons and \rhads with the ATLAS detector in $pp$ collisions at $\sqrt{s} = 7~\TeV$}
\PreprintIdNumber{CERN-PH-EP-2012-236}
\PreprintCoverAbstract{A search for long-lived particles is performed using a data sample of \lumi from proton-proton collisions at a centre-of-mass energy $\sqrt{s} = 7~\TeV$ collected by the ATLAS detector at the LHC. No excess is observed above the estimated background and lower limits, at 95\% confidence level, are set on the mass of the long-lived particles in different scenarios, based on their possible interactions in the inner detector, the calorimeters and the muon spectrometer.
Long-lived staus in gauge-mediated SUSY-breaking models are excluded up to a mass of 300~\GeV for $\tan\beta=5-20$. Directly produced long-lived sleptons are excluded up to a mass of 278~\GeV.
\rhads, composites of gluino (stop, sbottom) and light quarks, are excluded up to a mass of $\rhadronGluinoFullMassExclusion$ ($\rhadronStopFullMassExclusion$, $\rhadronSbottomFullMassExclusion$) when using a generic interaction model.
Additionally two sets of limits on \rhads are obtained that are less sensitive to the interaction model for \rhads. One set of limits is obtained using only the inner detector and calorimeter observables, and a second set of limits is obtained based on the inner detector alone.}
\PreprintJournalName{PLB}

\begin{document}

\begin{frontmatter}
\title{Searches for heavy long-lived sleptons and \rhads with the ATLAS detector in $pp$ collisions at $\sqrt{s} = 7~\TeV$}
\author{ATLAS Collaboration}
\address{}

\begin{abstract}
A search for long-lived particles is performed using a data sample of \lumi from proton-proton collisions at a centre-of-mass energy $\sqrt{s} = 7~\TeV$ collected by the ATLAS detector at the LHC. No excess is observed above the estimated background and lower limits, at 95\% confidence level, are set on the mass of the long-lived particles in different scenarios, based on their possible interactions in the inner detector, the calorimeters and the muon spectrometer.
Long-lived staus in gauge-mediated SUSY-breaking models are excluded up to a mass of 300~\GeV for $\tan\beta=5-20$. Directly produced long-lived sleptons are excluded up to a mass of 278~\GeV.
\rhads, composites of gluino (stop, sbottom) and light quarks, are excluded up to a mass of $\rhadronGluinoFullMassExclusion$ ($\rhadronStopFullMassExclusion$, $\rhadronSbottomFullMassExclusion$) when using a generic interaction model.
Additionally two sets of limits on \rhads are obtained that are less sensitive to the interaction model for \rhads. One set of limits is obtained using only the inner detector and calorimeter observables, and a second set of limits is obtained based on the inner detector alone.
\end{abstract}


\end{frontmatter}

\section{Introduction}
\label{sec:introduction}

Heavy long-lived particles (LLP) are predicted in a range of theories which extend the Standard Model (SM). Supersymmetry (SUSY)~\cite{Miyazawa:1966, Ramond:1971gb, Golfand:1971iw, Neveu:1971rx, Neveu:1971iv, Gervais:1971ji, Volkov:1973ix, Wess:1973kz, Wess:1974tw} models allow long-lived charged sleptons ($\tilde{\ell}$), squarks ($\tilde{q}$) and gluinos ($\tilde{g}$). Heavy LLPs produced at the Large Hadron Collider (LHC) could travel with speed measurably lower than the speed of light. These particles can be identified and their mass, $m$, determined from their speed, $\beta$, and momentum, $p$, using the relation $m=p/\gamma\beta$, with $\gamma$ being the relativistic Lorentz factor. Four different searches are presented in this Letter, using time-of-flight to measure $\beta$ and specific ionisation energy loss, \dedx, to measure $\beta\gamma$. The searches are optimised for the different experimental signatures of sleptons and composite colourless states of a squark or gluino together with SM quarks and gluons, called \rhads.

Long-lived charged sleptons would interact like muons, releasing energy by ionisation as they pass through the ATLAS detector. A search for long-lived sleptons identified in both the inner detector (ID) and in the muon spectrometer (MS) is therefore performed (``slepton search''). The results are interpreted in the framework of gauge-mediated SUSY breaking (GMSB)~\cite{Dine:1981gu,AlvarezGaume:1981wy,Nappi:1982hm,Dine:1993yw, Dine:1994vc,Dine:1995ag,Kolda:1997wt} with the light stau ($\tilde{\tau}$) as the LLP. In these models a substantial fraction of the events would contain two LLP candidates, and this feature is also utilised in discriminating signal from background. Direct pair production of sleptons is also used to interpret the data independently of the mass spectrum of the other SUSY particles.

Coloured LLPs ($\tilde{q}$ and $\tilde{g}$) would hadronise forming \rhads, bound states composed of the LLP and light SM quarks or gluons. They may emerge as charged or neutral states from the $pp$ collision and be converted to a state with a different charge by interactions with the detector material, and thus arrive as neutral, charged or doubly charged particles in the muon spectrometer.

In ATLAS, LLPs can be identified via the timing information in the muon spectrometer or calorimeters and via the measurement of the energy loss in the silicon pixel detector. All of these techniques are combined in this analysis to achieve optimal sensitivity for the ``full-detector \rhad search''. In addition, searches based on only the calorimeter and the inner detector information (``MS-agnostic \rhad search''), and based solely on the inner detector (``ID-only \rhad search'') are performed. The latter two cases are motivated by the limited understanding of \rhad interactions in matter, in particular the possibility that \rhads are electrically neutral in the MS. Furthermore, these searches are sensitive to scenarios in which the \rhads decay before reaching the MS. In all searches the signal particles are assumed to be stable within the ATLAS detector, at least to the point it hits the last relevant component of the subdetector used for detecting it.

Previous collider searches for LLPs have been performed at LEP~\cite{Barate:1997dr,Abreu:2000tn,Achard:2001qw,Abbiendi:2003yd}, HERA~\cite{Aktas:2004pq}, the Tevatron~\cite{Abe:1989es,Abe:1992vr,Acosta:2002ju,PhysRevLett.99.131801,Abazov:2008qu,Aaltonen:2009kea,PhysRevLett.108.121802}, and the LHC~\cite{Khachatryan:2010uf,Khachatryan:2011ts,Aad:2011mb,Aad:2011yf,Aad:2011hz,Aad:2012zn,Chatrchyan:2012sp}.

\section{Data and simulated samples}

The work presented in this Letter is based on \lumi of $pp$ collision data collected at a centre-of-mass energy $\sqrt{s} = 7~\TeV$ in 2011. The events are selected online by muon triggers for the slepton search and by missing transverse momentum and muon triggers for the \rhad searches. Data and Monte Carlo $Z\to\mu\mu$ samples are used for timing resolution studies. Monte Carlo signal samples are used to study the expected signal behaviour and to set limits.

The GMSB samples are generated with the following model parameters: number of super-multiplets in the messenger sector, $N_5=3$, messenger mass scale, $m_{\rm{messenger}}=250~\TeV$, sign of the Higgsino mass parameter, $\rm{sign}(\mu)=1$,  and $C_{\rm{grav}}$, the scale factor for the gravitino mass which determines the \stau lifetime was set to 5000 to ensure that the \stau does not decay in the detector. The ratio of the vacuum expectation values of the two Higgs doublets, $\tan{\beta}$, is varied between 5 and 40 and the SUSY-breaking mass scale $\Lambda$ is varied from 50 to 150~\TeV, corresponding to light \stau masses varying from 122.2 to 465~\GeV. The mass spectra of the GMSB models are obtained from the \spice program~\cite{Engelhard:2009br} and the events are generated using \herwig~\cite{Corcella:2000bw}.

The \rhad samples are generated with gluino (squark) masses from 300--1500~\GeV (200--1000~\GeV). The pair production of gluinos and squarks is simulated in \pythia\cite{Sjostrand:2006za}, incorporating specialised hadronisation routines \cite{hadroutines,Kraan:2004tz,Farrar:2010ps} to produce final states containing \rhads. A 10\% gluino-ball fraction is assumed in the gluino sample production. The simulation of \rhad interactions with matter is handled by dedicated \geant~\cite{Agostinelli:2002hh,rhadgeant} routines based on a generic model~\cite{Mackeprang:2009ad}. All Monte Carlo events pass the full ATLAS detector simulation~\cite{Agostinelli:2002hh,Alexa:2006ee} and are reconstructed with the same programs as the data. All signal Monte Carlo samples are normalised to the integrated luminosity of the data.

\section{The ATLAS detector}

The ATLAS detector~\cite{Aad:2008zzm} is a multipurpose particle physics detector with a forward-backward symmetric cylindrical geometry and near $4\pi$ coverage in solid angle\footnote{ATLAS uses a right-handed coordinate system with its origin at the nominal interaction point in the centre of the detector and the $z$-axis coinciding with the axis of the beam pipe. The $x$-axis points from the interaction point to the centre of the LHC ring, and the $y$-axis points upward. Cylindrical coordinates ($r$, $\phi$) are used in the transverse plane, $\phi$ being the azimuthal angle around the beam pipe. The pseudorapidity is defined in terms of the polar angle $\theta$ as $\eta = - \ln \tan(\theta/2)$.}. The ID consists of a silicon pixel detector, a silicon micro-strip detector, and a transition radiation tracker. The ID is surrounded by a thin superconducting solenoid providing a 2~T magnetic field, and by high-granularity liquid-argon sampling electromagnetic calorimeters (LAr). An iron/scintillator-tile calorimeter provides coverage for hadrons in the central rapidity range. The end-cap and forward regions are instrumented with liquid-argon calorimeters for both electromagnetic and hadronic measurements. The MS surrounds the calorimeters and consists of three large superconducting air-core toroids each with eight coils, a system of precision tracking chambers, and detectors for triggering.

The ATLAS trigger system is designed to select the events of most interest with a data-taking rate of about 400~Hz from a beam bunch crossing rate as high as 40~MHz. The first-level trigger (level-1) selection is carried out by custom hardware and identifies detector regions and the bunch crossing for which a trigger element is found. The high-level trigger is performed by dedicated software, seeded by data acquired from the bunch crossing and regions found at level-1. The components of particular importance to this analysis are described in more detail below.

\subsection{The pixel detector}
\label{sec:pixdet}

As the innermost detector system in ATLAS, the silicon pixel detector provides at least three precision measurements for each track in the region $|\eta|<2.5$ at radial distances from the LHC beam line $r<15$~cm. The sensors in the pixel barrel (covering the central $|\eta|$-region) are placed on three concentric cylinders around the beam-line, whereas sensors in the end-cap (covering the high-$|\eta|$ region) are located on three disks perpendicular to the beam axis on each side of the barrel. In the barrel (end-cap) the intrinsic accuracy is $10~\mu$m in the $r\phi$-plane and $115~\mu$m in the $z$($r$)-direction. The data are only read out if the signal is larger than a set threshold. The time for which the signal exceeds that threshold, ToT, is recorded. The larger the initial signal is the longer this time.

\subsubsection{Pixel detector specific ionisation (\dedx) measurement}

The relation between the ToT and the charge deposition in each pixel is measured in dedicated calibration scans and shows a good linearity. Therefore, the ToT measurement is well correlated with the energy loss of a charged particle in the pixel detector. The maximum ToT value corresponds to 8.5 times the average charge released by a minimum ionising particle (MIP) for a track perpendicular to the silicon detectors and leaving all its ionisation charge on a single pixel. If this value is exceeded, the ToT (and therefore the charge) is not correctly measured. In LHC collisions the charge generated by one track crossing the pixel detector is rarely contained in just one pixel. Neighbouring pixels are joined together to form clusters and the charge of a cluster is calculated by summing up the charges of all pixels after calibration correction. The specific energy loss \dedx~is defined as the average of all individual cluster charge measurements for the clusters associated with the track. To reduce the Landau tails, the average is evaluated after having removed the cluster with the highest charge (the two clusters with the highest charge are removed for tracks having five or more clusters).

\subsubsection{Mass measurement with the pixel detector}

The masses of slow charged particles can be measured using solely the ID information by fitting each \dedx and momentum measurement to an empirical Bethe-Bloch function and deducing their $\beta\gamma$ value. The measurable $\beta\gamma$ range lies between 0.2 and 1.5, the lower bound being defined by the overflow in the ToT spectrum, and the upper bound by the overlapping distributions in the relativistic rise branch of the curve. This particle identification method~\cite{dedxnote} uses a five-parameter function to describe how the most probable value of the specific energy loss (${\cal{M}}_{\mathrm{d}E\over \mathrm{d}x}$) depends on $\beta\gamma$:
\begin{equation}
{\cal{M}}_{\mathrm{d}E\over \mathrm{d}x}(\beta\gamma) = {p_1\over \beta^{p_3}}\ln(1+(p_2\beta\gamma)^{p_5})-p_4
  \label{bbfun}
\end{equation}

\begin{figure*}[tb]
  \centering
  \includegraphics[width=0.49\linewidth]{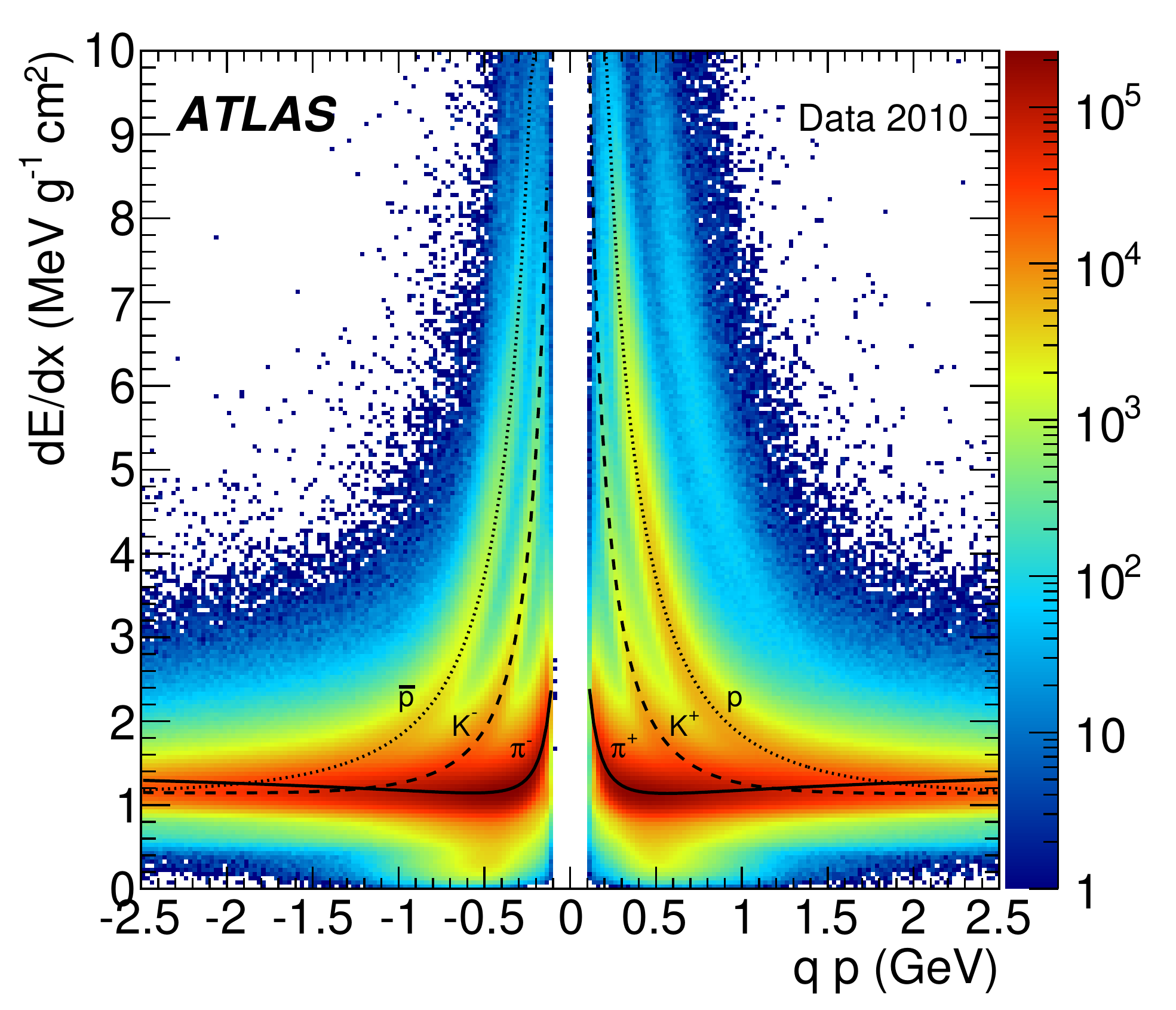}
  \includegraphics[width=0.48\linewidth]{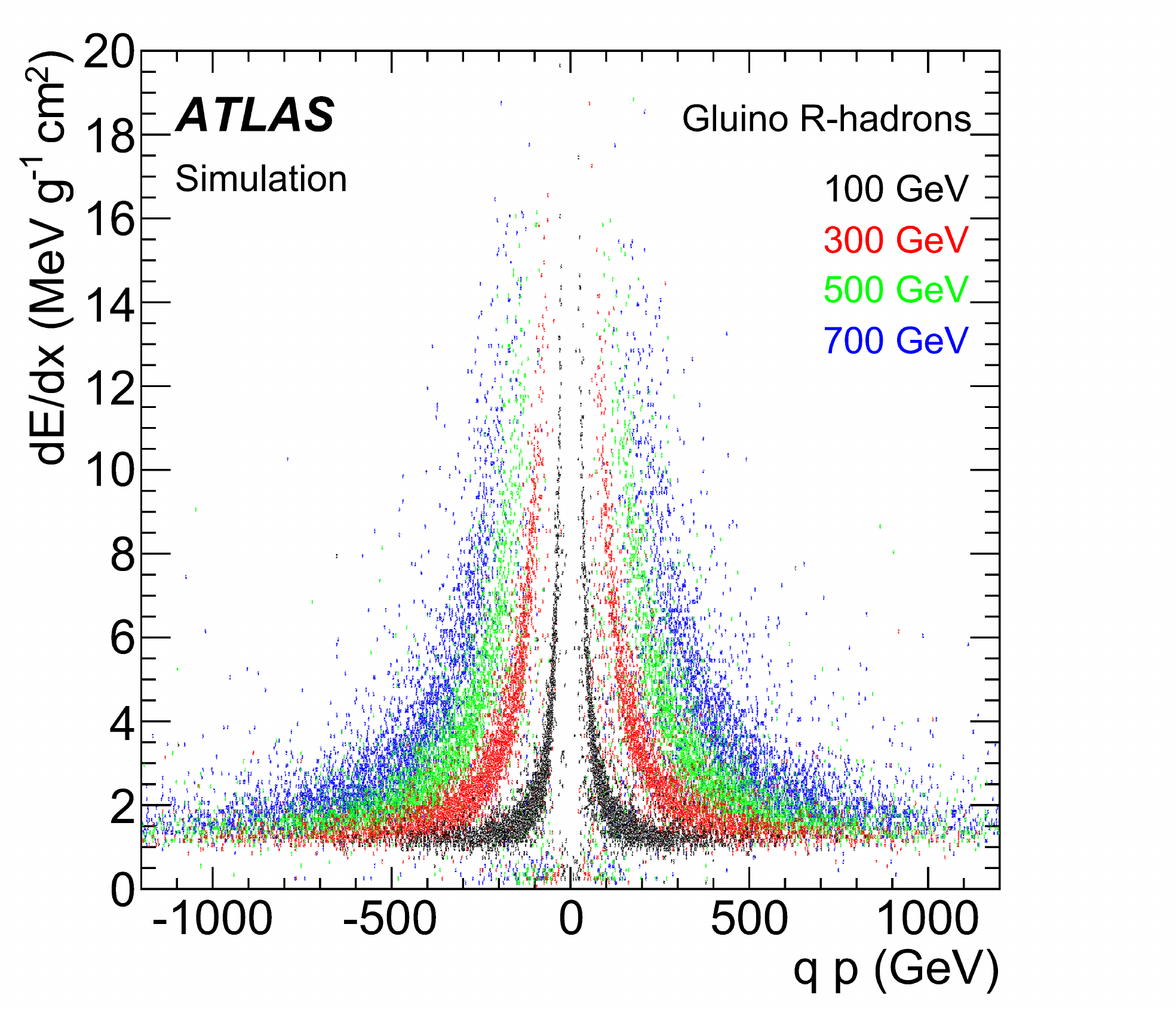}
  \caption{Left: Distribution of \dedx versus charge times momentum for minimum bias collisions in a data sample from 2010. With tracks reconstructed for $\pt > 100~\MeV$ this sample is more suitable for the calibration of the Bethe-Bloch function than 2011 data in which tracks had a cutoff of $\pt=400~\MeV$. The distribution of the most probable value for pions (solid), kaons (dashed) and protons (dotted) are superimposed. The band due to deuterons is clearly visible. Right: Simulated distribution of specific energy loss versus momentum for singly-charged hypothetical \rhads of various masses.}
  \label{fig:dEdx}
\end{figure*}

Fig.~\ref{fig:dEdx}(left) shows how this function describes data for low momentum tracks. Fig.~\ref{fig:dEdx}(right) shows the simulated pixel \dedx spectra for singly-charged hypothetical \rhads of masses 100, 300, 500 and 700~\GeV. As expected, these distributions extend into the high pixel \dedx region even for high momentum tracks. The most probable value of \dedx for MIPs is about 1.2~\MeVgcm with a spread of about 0.2~\MeVgcm and a slight $\eta$ dependence, increasing by about 10\% from low-$\eta$ to high-$\eta$ regions.

For all tracks having a reconstructed momentum $p$ and a measured specific energy loss \dedx above the value for MIPs, a mass estimate $m_{\beta\gamma} = p/\beta\gamma$ is obtained by inverting Eq.~(\ref{bbfun}). The procedure is continuously monitored through precise ($<1\%$) measurements of the mass of known particles (kaons and protons). For LLPs the expected \dedx values are much larger than those of SM particles, allowing for identifying them based on this information.

\subsection{Calorimeters}
\label{sec:calorimeters}

Liquid argon is used as the active detector medium in the electromagnetic (EM) barrel and end-cap calorimeters, as well as in the hadronic end-cap (HEC) calorimeter. All are sampling calorimeters, using lead plates for the EM calorimeters and copper plates for the HEC calorimeter. The barrel EM calorimeter covers the region $|\eta|<1.475$ and consists of three layers and a pre-sampler. The EM end-cap calorimeter consists of three layers in the region $1.375<|\eta|<2.5$ (two for $2.5<|\eta|<3.2$) and a pre-sampler for $1.5<|\eta|<1.8$. The four layers of the HEC calorimeter cover the range $1.5<|\eta|<3.2$.

The ATLAS tile calorimeter is a cylindrical hadronic sampling calorimeter. It uses steel as the absorber material and plastic scintillators as the active layers. It covers radii from 2280 to 4230~mm and the $\eta$ coverage extends to $|\eta| \lesssim 1.7$. The calorimeter is subdivided into a central barrel covering $|\eta| \lesssim 1.0$ and an extended barrel covering $0.8 \lesssim |\eta| \lesssim 1.7$. Both barrel parts are divided into 64 modules and the cells in each module are divided into three layers.

\subsubsection{Calorimeter $\boldmath{\beta}$ measurement}

The ATLAS tile and LAr calorimeters have sufficiently good timing resolutions to distinguish highly relativistic SM particles from the slower moving LLPs. The time resolution depends on the energy deposited in the cell and also the layer type and thickness, but typical resolutions are 2~ns for an energy deposit of 1~\GeV and generally better for the tile calorimeter.

To ensure the highest possible timing accuracy, it is necessary to calibrate the data using particles with known speed. This calibration applies a common shift for each run, and is then performed as a function of calorimeter layer and cell energy. The reliability of such a calibration for this analysis depends on the assumption that the particles used for calibration have similar characteristics to the LLPs in question when depositing energy in the calorimeters. The analysis uses muons for this purpose and it is cross-checked that jets give a consistent result. The effect of a possible bias in the measured time for late-arriving signals was tested by applying an arrival-time-dependent resolution function and found negligible. The $\beta$ measurements from all cells assigned to the extrapolated track are combined in a cell-energy-weighted average, typically using timing measurements from three or four calorimeter cells. The resolution of the resulting $\beta$ measurement can be seen in Fig.~\ref{fig:betagamma_and_beta}(left). For data (MC) the mean $\beta$-value is 0.983 (0.986) and the resolution is 0.090 (0.092).

\subsection{The muon detectors}
\label{sec:muondetectors}

The MS forms the outer part of the ATLAS detector and detects charged particles exiting the calorimeters and measures their momenta in the pseudorapidity range $|\eta| < 2.7$. It is also designed to trigger on these particles in the region $|\eta| < 2.4$. In the barrel the chambers are arranged in three concentric cylindrical shells around the beam axis, while in the two end-caps the muon chambers are arranged in three wheels that are perpendicular to the beam axis.

The precision momentum measurement is performed by monitored drift tube (MDT) chambers, using the $\eta$ coordinate. These chambers consist of three to eight layers of drift tubes. In the forward region ($2.0 < |\eta| < 2.7$), cathode strip chambers are used in the innermost tracking layer. Resistive plate chambers (RPC) in the barrel region ($|\eta|<1.05$) and thin gap chambers in the end-cap ($1.05 < |\eta| < 2.4$) provide a fast level-1 trigger and measure both the $\eta$ and $\phi$ coordinates of the track.

\subsubsection{ $\boldmath{\beta}$ measurement in the MS}

The default reconstruction of particles in the MDT chambers~\cite{Aad:2009wy} relies on the assumption that they travel with the speed of light ($\beta=1$). To improve the track quality for slow LLPs, the individual track segments can be reconstructed with different values for $\beta$. The actual $\beta$ of the particle is estimated from the set of segments with the lowest $\chi^2$. In a successive combined track re-fit, including ID and MS hits, the particle trajectory is estimated more accurately. The time-of-flight to each tube is then obtained using the difference between the time-of-flight corresponding to the re-fitted track position in each tube and the time actually measured. By averaging the $\beta$ values estimated from the time-of-flight in the different tubes an improved MDT $\beta$~estimation can be achieved.

The RPCs have an intrinsic time resolution of $\sim$1~ns while the digitised signal is sampled with a 3.12~ns granularity, allowing a measurement of the time-of-flight. In the RPCs, $\beta$ is first calculated separately for each hit from the independent position and time measurements. A single $\beta$~estimate is obtained by averaging the $\beta$ values from all the hits.

By definition, in a perfectly calibrated detector, any energetic muon coming from a collision at the interaction point will pass the detector at $t_{0}=0$. The $t_0$ distributions in the different detector systems are measured and their means used to correct the calibration. The observed width of these distributions after correction is used as the error on the time measurement in the $\beta$ fit and to smear times in the simulated samples. The time resolution in the MS (about 3~ns) does not depend on the arrival time.

\subsection{Combining $\boldmath{\beta}$ measurements}

There are three possible $\beta$ measurements per track, from the RPCs, the MDTs and the calorimeters. The $\beta$ measurements from the different detectors are only used if $\beta > 0.2$ (the limit of the sensitivity) and if they are consistent internally, i.e.~the $\chi^2$ probability of the average between hits is reasonable (calorimeter) or the RMS of the measurement is consistent with the expected errors (MS). Measurements that are accepted are combined in a weighted average. The weights are obtained from the calculated error of each measurement multiplied by the pulls of the $\beta$ distributions for muons from $Z$ boson decays.

\begin{figure*}[tbp]
  \centering
  \includegraphics[width=0.48\linewidth]{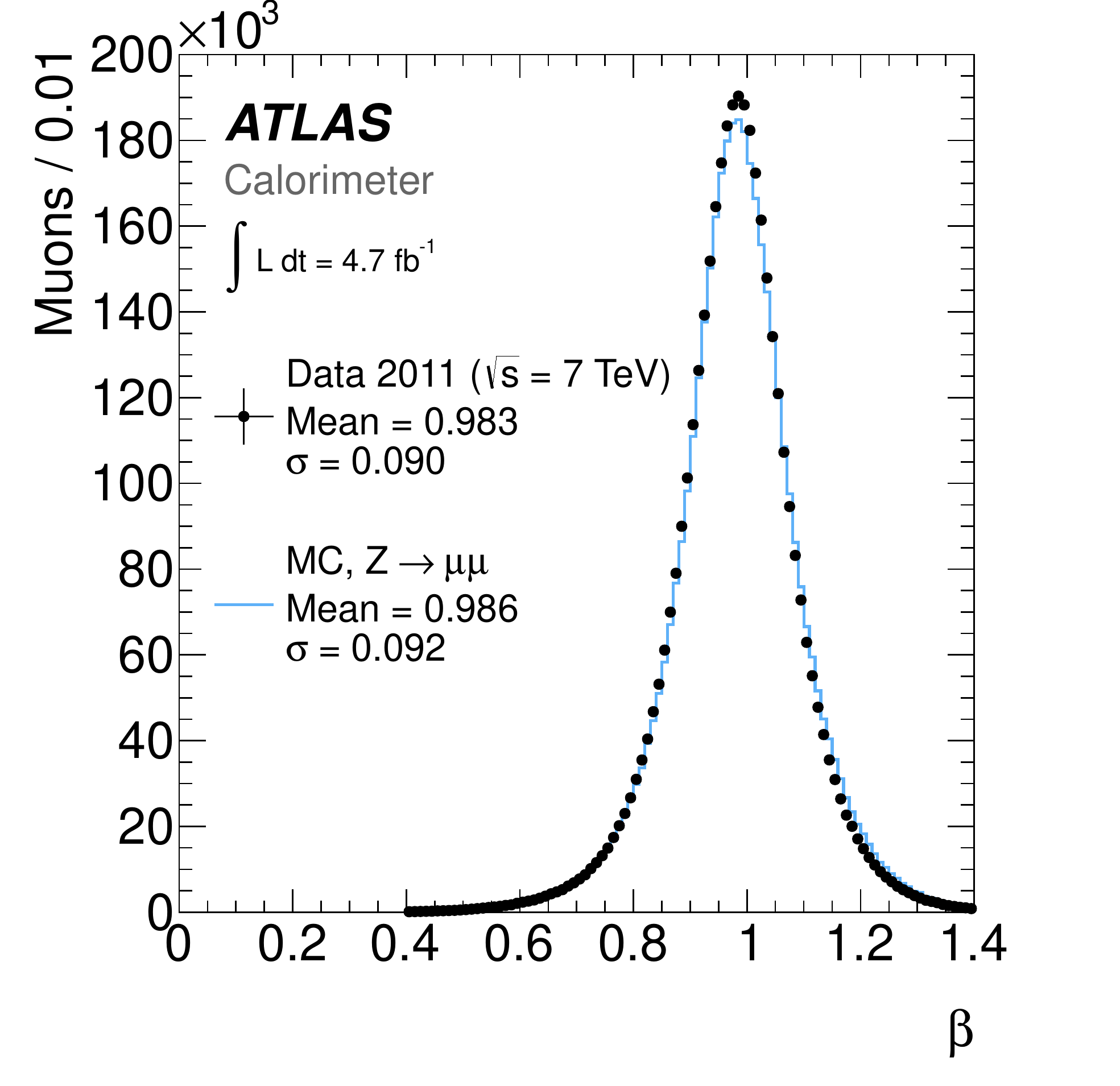}
  \includegraphics[width=0.48\linewidth]{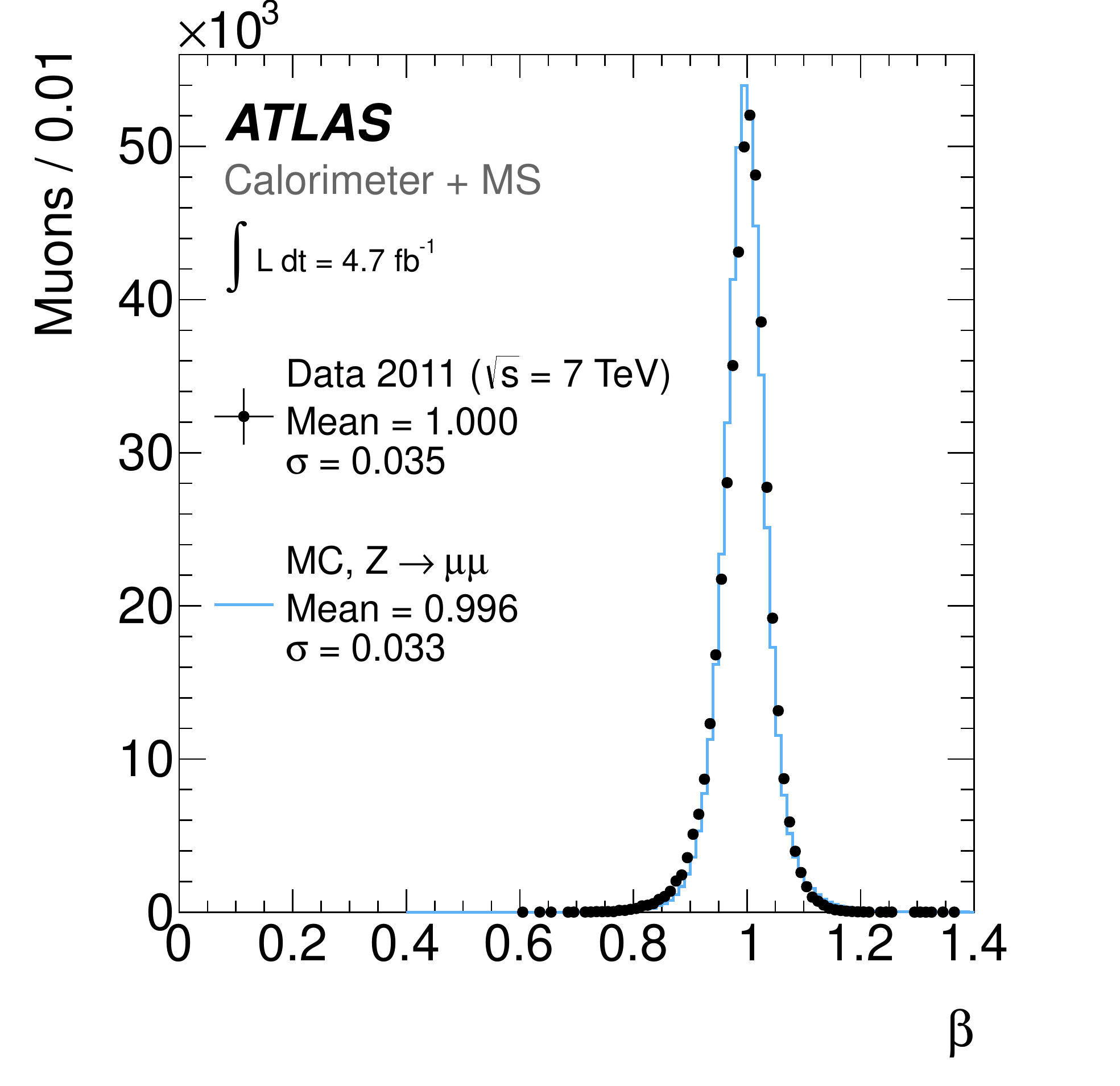}
  \caption{Distribution of $\beta$ for the calorimeter (left) and combined calorimeter + MS measurements (right) obtained for selected $Z\rightarrow\mu\mu$ decays in data and MC simulation. The typical resolutions are quoted in the figures.}
  \label{fig:betagamma_and_beta}
\end{figure*}

Since $\beta$ is estimated from the measured time-of-flight, for a given resolution on the time measurement, a slower particle has a better $\beta$ resolution. Prior to the $\beta$ estimation, the timing values of the hits in the MC samples are smeared to reproduce the resolution measured in the data. Fig.~\ref{fig:betagamma_and_beta}(right) shows the $\beta$ distribution for selected $Z\rightarrow\mu\mu$ candidate events in data compared to simulation after having smeared the hit times in simulation as described above. The data are well reproduced by the smeared MC distribution: the mean values are 1.000 and 0.996 and the resolution values are 0.035 and 0.033 for data and MC, respectively. The smearing mechanism reproduces the measured muon $\beta$ distribution. The same time-smearing mechanism is applied to the signal Monte Carlo samples.

\section{LLP candidate and event selection}
\label{candidate_selection}

\subsection{Trigger selection}

This analysis is based on events collected by two main trigger types: single-muon and missing transverse momentum triggers.

\subsubsection{Single-muon trigger}\label{sec:trigger_muon}

The muon trigger and its performance in 2011 data are described in detail in References~\cite{TriggerSF, Aad:2012xs}. This analysis uses un-prescaled muon triggers with a \pt threshold of 18~\GeV. Offline muons are selected with $\pt > 50~\GeV$, well above the trigger threshold.

Level-1 muon triggers are accepted and passed to the high-level trigger only if assigned to the collision bunch crossing. Late triggers due to late arrival of the particles are thus lost. The trigger efficiency for particles arriving late at the muon spectrometer is difficult to assess from data, where the majority of candidates are in-time muons. This efficiency is obtained from simulated \rhad and GMSB events passing the level-1 trigger simulation. The muon triggers are found to be efficient for GMSB signatures, which contain two typically high-$\beta$ LLPs that reach the MS, and additional muons stemming from neutralino decays. The trigger efficiency increases roughly linearly from zero at $\beta = 0.62$ to its full potential of about 90\% at $\beta = 0.82$ for sleptons and \rhads that reach the MS. The estimated trigger efficiency for GMSB slepton events is between 70\% and 85\%. Muon triggers are less efficient for \rhad events, where one or both of the \rhads may be uncharged as they enter the MS and $\beta$ is typically low.

\subsubsection{ Missing transverse momentum trigger}\label{sec:trigger_met}

Since gluinos and squarks are produced via the strong interaction, \rhad events often contain high-\pt jets from QCD radiation. The modest energy depositions of the \rhads themselves combined with these jets naturally give rise to missing transverse momentum (magnitude denoted as \etmiss).

The \rhad analyses use missing transverse momentum triggers utilising only calorimeter information~\cite{Casadei:1331180,Aad:2012xs} with thresholds between 60 and 70~\GeV (a full
description of the ATLAS trigger system is given in~\cite{Aad:2008zzm}). For the full-detector search the single-muon trigger described in the previous section is used in addition to the \etmiss trigger. Unlike the muon trigger, there is no loss of efficiency for the \etmiss triggers when \rhads arrive late. The efficiency of the \etmiss triggers decreases with increasing \rhad mass as the relative contribution of \ggf-initiated processes decreases, though it remains between 15\% and 20\% for heavy \rhads.

\subsection{Offline selection}
\label{secOfflineSelection}

Two different signal types are studied: sleptons and \rhads. Given their different expected interactions with the ATLAS detector, two dedicated LLP candidate and event selections are used as described in the following sections.

\subsubsection{ Common event selection}\label{sec:event_selection}

Collision events are selected by requiring a good primary vertex, with at least three ID tracks, and with requirements on the position of the reconstructed primary vertex. The primary vertex is defined as the reconstructed vertex with the  highest $\sum p^2_T$ of associated tracks. Events recorded during a time where a problem was present in one of the relevant subdetectors are rejected. Cosmic-ray background is rejected by removing tracks that do not pass close to the primary vertex in $z$. Candidates with an ID track with $|z_{0}^{\mathrm{trk}} - z_{0}^{\mathrm{vtx}}|>10$~mm or $|d_{0}^{\mathrm{trk}}|>2$~mm are removed, where $z_{0}^{\mathrm{trk}}$ is the  coordinate along the beam direction and $d_{0}^{\mathrm{trk}}$ is the transverse impact parameter at the distance of closest approach of the track to the primary vertex. Both requirements are tightened to 1.5~mm for the ID-only analysis. For the analyses involving the muon spectrometer, events with cosmic rays are also rejected by a topological cut on any two candidates with opposite $\eta$ and $\phi$ ($|\eta_1+\eta_2|<0.005$ and $||\phi_1-\phi_2|-\pi|<0.005$).

\subsubsection{ Slepton candidate selection}\label{sec:slepton_selection}

The analysis searching for sleptons requires two muon candidates in each event, because two sleptons are produced, and both have a high probability of being observed in the MS. Two sets of selection criteria are applied. A loose selection with high efficiency is used to select candidates in events where there are two LLP candidates. In events where only one candidate passes the loose selection, that candidate is required to pass a tight selection.

Candidates in the loose slepton selection are required to have $\pt > 50~\GeV$. The \pt~measurements in the ID and MS are required to be consistent, so that the difference between the ID \pt and the combined fit \pt does not exceed a half of their average. Each candidate is required to have  $|\eta| < 2.5$. Any two candidates that combine to give an invariant mass close to the $Z$ boson mass ($\pm10~\GeV$) are both rejected. Candidates are also required to have associated hits in at least two of the three super-layers of precision measurement chambers in the MS.

The number of calorimeter cells plus MS hits contributing to the $\beta$ measurement must exceed the number of detector (sub)systems used by three. For signal LLPs, the $\beta$ values are expected to be consistent between the individual measurements, whereas for muons a low $\beta$-value is typically due to a poor measurement in one detector component and thus not consistent between different components. Therefore the estimated $\beta$ is required to be consistent for measurements in the same detector system, based on the hit time resolutions, and the $\beta$ measurements from the different detector systems are required to be consistent with each other.
The different detector system measurements of $\beta$ are required to be pair-wise consistent at the  3$\sigma$~level, and the combined $\beta$~to be consistent with the $\beta\gamma$~estimated in the pixel detector within 3$\sigma$. Finally, in order to reduce the muon background, the combined $\beta$ measurement is required to be less than 0.95.

To pass the tight selection, a candidate is required in addition to have $\pt > 70~\GeV$, at least two separate detector systems measuring $\beta$, the number of hits minus the number of detector systems participating in the measurement be at least twelve, and the consistency between $\beta$ estimates in different detector systems be within 2$\sigma$. These cuts are optimised to give better background rejection.

Finally, a mass cut is applied on the candidate mass, $m_{\beta}=p/\gamma\beta$, calculated from the candidate's measured momentum and $\beta$. This cut depends on the hypothetical $\tilde\tau$ mass and is different for different points in the GMSB model parameters space, determined by the expected significance of the signal. For the two-candidate sample, both masses are required to be above the cut.

The number of background and expected signal events above the mass cut in the two-candidate and one-candidate signal regions is used to search for the presence of sleptons and \rhads.

Typical efficiencies for signal events to pass all cuts including the mass cut are 20\% for each of the one and two candidate event categories, giving 40\% efficiency in total.

\subsubsection{ \rhad candidate selection}\label{sec:rhad_selection}

Since the \rhad contains light quarks and gluons in addition to the squark or gluino, the charge of the \rhad can change following nuclear interactions with the detector material. This possibility makes it difficult to rely on a single detection mechanism without some loss of detection efficiency, as a neutral state would go undetected until the next nuclear interaction occurs. Some of the main hadronic states regarded in the models considered are neutral, and it is thus natural to take an inside-out approach, starting from the ID track and adding discriminators from outer detector systems if a signal is seen along the extrapolated track. This is reflected in the three different \rhad searches; ``full-detector'', ``MS-agnostic'' (not considering the MS) and ``ID-only'' (relying solely on the ID); making successively fewer assumptions about the \rhad scattering model and lifetime. While the first two differ only in their estimate of $\beta$ (the ``MS-agnostic'' uses only the calorimeters) and the utilised triggers, the ID-only selection is generally more restrictive in order to reject the larger backgrounds.

In the full-detector and MS-agnostic analyses, \rhad candidates are required to have a good quality ID track with $p > 140~\GeV$ and $|\eta| < 2.5$. The track must not be within an $\eta-\phi$ distance $\Delta R = \sqrt{\left( \Delta \eta \right)^2 + \left(\Delta \phi \right)^2} = 0.3$ of any jet with $\pt > 40~\GeV$, reconstructed using the anti-$k_t$ jet algorithm~\cite{jetalgo} with distance parameter set to 0.4. Furthermore, the track must not have any nearby ($\Delta R < 0.25$) tracks with $\pt > 10~\GeV$ nor pixel hits shared with another track. Tracks with momenta greater than 3.5~\TeV are rejected as unphysical. The candidate must have a good \dedx measurement and a good estimate of $\beta$. The uncertainty on $\beta$ is required to be less than 10\% when measured with the calorimeter only, and less than 4\% when the result of a combination.

In the ID-only analysis, selection requirements are further tightened. Vertices must have more than four associated tracks and the \etmiss trigger must be confirmed off-line ($\etmiss>85~\GeV$, including MS contributions) to ensure better background rejection. Candidate \rhad tracks must have more than two (six) pixel (silicon
micro-strip) hits, impact parameters compatible with the primary vertex,  $\pt> 50~\GeV$ and  $p>~100~\GeV$. The isolation cut is also more severe, rejecting events that have a track of $\pt>1~\GeV$ within $\Delta R = 0.25$ of the \rhad candidate. Additionally, tracks are discarded if their momentum resolution exceeds 50\%, or if they are identified as an electron.

In events with multiple \rhad candidates, only one -- randomly chosen -- candidate is used.

The final signal selection in the full-detector and MS-agnostic analyses, optimised for each mass hypothesis, is based on $\beta\gamma$ and $\beta$ requiring $\beta\gamma < 1.5-2.0$ and $\beta < 0.8-0.9$, with lower values for higher masses. A signal region is defined in the $m_{\beta\gamma}$-$m_{\beta}$ plane for each mass point. The lower value of the mass cut is set such that it corresponds to being $2\sigma$ of the mass resolution below the nominal mass value. The upper value is set to 2.5~\TeV. For the ID-only analysis the final selection requires the measured \dedx value to exceed a certain threshold, which has been set as function of $\eta$ such that the rejection of MIPs is independent of $\eta$.
The selection efficiency for gluino \rhads of 900~\GeV mass is about 11\% and 7\% in the full-detector and MS-agnostic analyses, respectively. It is about 6\% in the ID-only analysis.

An alternative \rhad model, which is an extension of the triple-Regge model used to describe squark \rhads \cite{Mackeprang:2009ad} has also been considered. In this model, the signal efficiency is 40\% lower at 300~\GeV and increasing to the same level at 900~\GeV above which it is higher, compared to the model used throughout this analysis.

\section{Background estimation}

The background for both the slepton and the \rhad searches is mostly composed of high-\pt muons with mis-measured $\beta$ or large ionisation. The background estimation is derived from data in all cases. The background mass spectrum is estimated by calculating a mass from the \pt spectrum of candidates and the measured $\beta$ distribution of the background obtained from control samples.

The estimation of the background mass distributions relies on two assumptions: that the signal-to-background ratio before applying cuts on $\beta$ is small, and that the $\beta$ distribution for background candidates is due to the finite resolution of the measurement and is therefore independent of the source of the candidate and its momentum. Checks of the validity of these assumptions are discussed in Section~\ref{subsec:bkgsyst}.

The detector is divided into $\eta$ regions so that the $\beta$ resolution within each region is similar. The muon $\beta$ probability density function (pdf) in each $\eta$ region is the distribution of the measured $\beta$ of all muons in the region normalised to one. The sample used in producing the $\beta$ pdf is enlarged with respect to the main selection of the analysis by lowering the \pt cut to 30 GeV and removing the $Z$ veto, in order to increase the acceptance and reduce possible signal contamination. Similarly, a $\beta\gamma$ pdf is constructed from tracks in a background dominated region.

The reconstructed mass distribution of muons in different regions of the detector depends on both $\beta$ and momentum distributions through $m=p/\gamma\beta$. The regions also differ in the muon momentum distribution; therefore the combination of momentum with random $\beta$ (see below) is done separately in each region and the resulting mass distributions are added together.

\subsection{Slepton search}

The background is determined by convoluting the $\beta$-distribution of muons with the expected muon momentum spectrum. The momentum spectrum is determined by selecting candidates that pass all selection requirements listed in Section~\ref{sec:slepton_selection} apart from the requirements on $\beta$ and $m$. Each candidate then gets assigned a $\beta$-value by drawing it randomly from the muon $\beta$ distribution. If the assigned $\beta$-value is below the cut value it is used (together with the measured momentum) to assign a mass to the candidate. The statistical uncertainty is reduced by using each muon candidate multiple times. The distribution of mass values obtained this way gives the background estimate.

\subsection{\rhad searches}
\label{subsec:RHbackground}

For the \rhad full-detector and MS-agnostic analyses, the momentum, $\beta\gamma$ and $\beta$ are obtained by taking a \pt-spectrum of charged particle tracks from the data and assigning randomly $\beta\gamma$ and $\beta$ values to these tracks. The mass is then calculated from the measured $p$ value and the assigned value of $\beta\gamma$ or $\beta$.
The mass distributions are then normalised to data by scaling to a sideband outside the signal region.

For the ID-only analysis the choice of the control sample takes into account the non-negligible correlations between $p$, \dedx and $\eta$. The ionisation dependence on the path length in the sensor is not linear~\cite{Bichsel88}, so the pixel \dedx depends on $\eta$; the ionisation also depends on the particle $\beta\gamma$ via the Bethe-Bloch formula, and therefore on its momentum, until the Fermi plateau is reached; finally $p$ and $\eta$ are not uncorrelated. The distributions used for the random-sampling are derived from two categories of background events. They are obtained by applying modified selections which ensure that signal contamination is minimised.

\begin{itemize}
  \item A first sample (``low-ionisation'') is used to generate the $\eta$ and $p$ distributions. This is selected in the same way as the event candidates, but without the requirement on high ionisation.  Instead, an upper bound is placed on the \dedx (at (1.8~\MeVgcm)), ensuring orthogonality with the signal selection.
  \item A second background sample (``low-momentum'') is used to generate the \dedx templates. A background sample free of signal but with no upper bound on the \dedx is obtained by considering tracks that have a maximum momentum of 100 GeV. Specifically, tracks in the ``low-momentum'' background sample satisfy all the event candidate requirements except that the transverse momentum cut is looser, $\pt > 10$~GeV, and the momentum $p$ is required to be between 40 and 100~\GeV (where the Fermi plateau has already been reached).
\end{itemize}

A large background sample consisting of two million $p$, $\eta$, \dedx triplets is randomly generated. The momentum is first generated according to a binned function based on ``low-ionisation'' events. Then the pseudorapidity is generated according to the $\eta(p)$ binned functions based on ``low-ionisation'' events. Finally the ionisation is generated according to $\dedx(\eta)$ binned functions based on ``low-momentum'' events.
The normalisation of the generated background to the selected data is obtained by scaling the background to the data before the high \dedx cut and in the region of the mass distribution where no signal is expected (mass below $140$~\GeV).

\section{Systematic uncertainties}

Several possible sources of systematic effects are studied. The resulting systematic uncertainties are summarised in Table~\ref{tab:systematics}. The errors given are those on the expected yields in the signal region.

\subsection{Theoretical cross-sections}

Signal cross-sections are calculated to next-to-leading order in the strong coupling constant, adding the resummation of soft gluon emission at next-to-leading-logarithmic accuracy (NLO+NLL)~\cite{Beenakker:1996ch,Beenakker:1997ut,Kulesza:2008jb,Kulesza:2009kq,Beenakker:2009ha,Beenakker:2010nq,Beenakker:2011fu}. The nominal cross-section and the uncertainty are taken from an envelope of cross-section predictions using different PDF sets and factorisation and renormalisation scales, as described in Reference~\cite{Kramer:2012bx}.
This prescription leads to a 5\% relative uncertainty on the expected signal normalisation in the slepton search, and a 15\% to 30\% uncertainty for the \rhad search, increasing with \rhad mass.

\subsection{Expected signal}

The muon trigger efficiency is calculated using the tag-and-probe technique on $Z\to\mu\mu$ events as described in References~\cite{TriggerSF, Aad:2012xs}. The uncertainty on the single muon trigger efficiency is estimated to be 1\%. The reduction in the muon trigger efficiency due to late arrival of particles is estimated from simulation. The effect of the difference between data and MC simulation in time alignment  of hits in the muon trigger system relative to the LHC clock is estimated by shifting the hit time of the highest $\beta$ candidate in each simulated event by 4~ns. The difference in trigger efficiency when this change is applied is between 0.5\% and 1.5\% for the different GMSB samples, and a systematic uncertainty of 1.5\% is assigned. For \rhads the systematic uncertainty is estimated in the same way and found to be 2\%.

The \etmiss trigger used for the \rhads relies on the emission of jets. Therefore, the trigger efficiency depends on the amount of initial and final state radiation (ISR and FSR). To evaluate the associated uncertainty, 1~\TeV gluino pair-production samples are simulated in \pythia 6.4.26 using the Perugia 2011 tune~\cite{Skands:2010ak}, setting the radiation level low and high. A simple threshold curve modelling of the trigger is applied to all three samples. The largest variation from the central sample is found to be 8.5\%.

The \etmiss triggers use calorimeter energy deposits to calculate the transverse energy, and are thus blind to muons, which therefore can be used for calibration and to study systematic errors. To evaluate the trigger efficiency, the trigger turn-on curve is obtained by fitting the measured efficiency vs. \etmiss in \zmumu events where the Z has a high \pt, both in data and simulation. Simulated events are re-weighted so the distribution of collisions per bunch-crossing match the running conditions of the 2011 data. These efficiency turn-on curves are then applied to the expected \etmiss spectrum from simulated \rhad events. The total uncertainty is estimated from three contributions: the relative difference between the efficiencies obtained using the fitted threshold curves from \zmumu data and simulation and the differences in efficiency obtained from independent $\pm 1\sigma$ variations in fit parameters relative to the unchanged turn-on curve fit for both \zmumu data and MC simulation. The total estimated \etmiss trigger uncertainty, including effects of a 10\% variation of the \etmiss scale, is a 4.5\% relative error on the efficiency for the signal.

In the ID-only analysis, the effect of the offline $\MET$ measurement uncertainty on the total efficiency is evaluated by applying a scale factor of $\pm~10\%$ and a smearing of up to 20\% to the missing transverse momentum~\cite{Aad:2011he}.
The resulting uncertainty depends on the $\rhad$ mass and is between  7.3\% (200~\GeV) and 4.5\% (1500~\GeV) of the central value.

Differences in the selection efficiency between data and MC simulation for the \rhad full-detector and MS-agnostic searches (excluding the final selection on $\beta\gamma$, $\beta$, momentum and mass selection) are evaluated using $Z\to\mu\mu$ events. The overall relative uncertainty is found to be below 1.5\%. The effect of the jet energy scale (JES) uncertainty on the requirement for signal candidates in the \rhad full-detector and MS-agnostic searches to be separated by $\Delta R = 0.3$ from jets with $\pt> 40~\GeV$ is found to be negligible.

To verify that the signal efficiency is insensitive to the pile-up re-weighting, a 900~\GeV gluino sample is divided into samples of high ($N_{\mathrm{vtx}} \ge 8$) and low ($N_{\mathrm{vtx}} < 8$) number of reconstructed primary vertices. The relative difference in reconstruction efficiency is found to be negligible.

The systematic uncertainty due to the track reconstruction efficiency and momentum resolution differences between data and simulation is estimated~\cite{ATLAS-CONF-2011-063} to be 0.5\% on the acceptance for GMSB events. For analyses relying solely on the ID for tracking, the resulting relative uncertainty on the signal efficiency is 1.3\%.

Differences in the pixel \dedx and the subsequent measurement of $\beta\gamma$ between data and simulation are considered by comparing the pixel \dedx distribution for high-\pt muons from $Z\to\mu\mu$ events in data and MC simulation. This uncertainty is estimated to be 5\%. In the ID-only analysis, this effect is evaluated by comparing the ionisation in data and simulation for MIPs and for particles with low $\beta\gamma$. Scale factors are measured in the two cases and found to be consistent. The uncertainty on the signal efficiency derived by applying these scale factors varies between 0.2\% and 5.8\% in the studied mass range, decreasing with the particle's mass.

The signal $\beta$ resolution is estimated by smearing the measured time of hits in the MS and calorimeter according to the spread observed in the time calibration. The systematic uncertainty due to the smearing process is estimated by scaling the smearing factor up and down, so as to bracket the distribution obtained in data. A 4\% (6\%) systematic uncertainty is found in the one-candidate (two-candidate) GMSB signal region. The corresponding uncertainty for \rhads is 3.6\%. The uncertainty due to the timing calibration of the calorimeter hits is further tested by comparing a calibration obtained from jet events to that obtained from muons. This results in a 1\% relative systematic uncertainty on the signal yield.

An uncertainty of 3.9\%~\cite{ATLAS-CONF-2011-116,Lumi2011Paper} is assigned to the integrated luminosity corresponding to this data set.

\subsection{Background estimation}
\label{subsec:bkgsyst}

The assumption that the $\beta$-pdf is independent of $p$ is tested by constructing the $\beta$-pdf separately in each $\eta$-region using a low and a high-momentum sample. This results in differences of up to 10\% on the background estimate.
To quantify the variability of the $\beta$ and momentum distributions within a region and its effect on the background estimation for sleptons, the detector is sub-divided into 25 $\eta$ regions instead of the eight used in the analysis and the background estimated with this division. The resulting systematic uncertainty is 6.5\% for the two-candidate events and 3.1\% for one-candidate events. The uncertainty on the background distribution due to the limited numbers of data events used to estimate the momentum distribution was calculated by dividing the candidate sample randomly into two samples and estimating the background from each sub-sample separately. The resulting error in the slepton search is negligible. A comparison of the muon $\beta$ distribution in inclusive muon events to that from $Z\to\mu\mu$ decays also exhibited negligible differences. The total uncertainty on the background estimate for the slepton search is 13\% for two-candidate events and 11\% for one-candidate events.

The total uncertainty in the background estimation for the full-detector \rhad analysis is estimated as follows. The $\beta\gamma$ and $\beta$ pdfs are obtained by considering sidebands (lower momenta, for \rhads 40--100~\GeV). Similarly, the momentum pdf is obtained from sidebands in $\beta\gamma$ and/or $\beta$. Several variations of the size of the sidebands (and also inclusion of the signal region) result in 8--10\% uncertainty estimates for each of the two variables. To test the momentum dependence of the $\beta$ pdf, the number of $\eta$ bins is varied. The resulting systematic uncertainty is 2\%. To estimate the uncertainty on the background distribution coming from the limited statistics of the data samples used to estimate the momentum distribution, the candidate sample is divided randomly into two samples and the background estimate is derived from each sub-sample separately. The resulting uncertainty on the full-detector \rhad search is 2\% for low mass and 5\% for high mass. The total uncertainty on the background estimate from the above tests is 15\%.

For the ID-only analysis, the statistical uncertainty in the background estimate dominates for most of the mass range, and is up to 20\%. Other effects contribute at most a few percent. These include variations from different binning choices for $p$, $\eta$ and \dedx  in the two background samples, and a contribution from pile-up effects.

\begin{table*}[tbp]
 \begin{center}
   \begin{tabular}{lrrrr}
     \hline\hline
     &
     \multicolumn{2}{c}{GMSB sleptons} &
     \multicolumn{2}{c}{~~~~~~~\rhads} \\
     Source &
     one-cand. &
     two-cand. &
     ID-only &
     other \\
     \hline
     \hline
     Theoretical uncertainty on signal acceptance &
     5 &
     5 &
     \multicolumn{2}{c}{~~~~~~~$15-30$}\\
     \hline
     Uncertainty on signal efficiency &
     \\
     ~~~Signal trigger efficiency &
     1.8 &
     1.8 &
     4.5 &
     4.5 \\
     ~~~QCD uncertainties (ISR, FSR) &
     --&
     --&
     8.5 &
     8.5 \\
     ~~~Signal pre-selection efficiency &
     --&
     --&
     --&
     1.5 \\
     ~~~Momentum resolution &
     0.5 &
     0.5 &
     1.3 &
     1.3 \\
     ~~~Pixel \dedx calibration &
     --&
     --&
     $5.8-0.2$ &
     5 \\
     ~~~Combined $\beta$ timing calibration &
     4 &
     6 &
     --&
     --\\
     ~~~~~~Calorimeter $\beta$ timing calibration &
     --&
     --&
     --&
     1.0\\
     ~~~~~~MS $\beta$ timing calibration &
     --&
     --&
     --&
     3.6 \\
     ~~~Offline \etmiss scale &
     --&
     --&
     $7.3-4.5$ &
     --\\
     Total  uncertainty on signal efficiency &
     4.4 &
     6.3 &
     $13.4-10.6$ &
     11.6 \\
     \hline
     Luminosity &
     3.9 &
     3.9 &
     3.9 &
     3.9 \\
     \hline
     Experimental uncertainty on background estimate & 11 & 13 & $3-20$ & 15 \\
     \hline\hline
   \end{tabular}
   \caption{Summary of systematic uncertainties (given in percent). Ranges indicate a mass dependence for the given uncertainty (low mass--high mass).}
   \label{tab:systematics}
 \end{center}
\end{table*}

\section{Results}

\begin{figure*}[tbp]
  \hspace{0.5cm}%
  \includegraphics[width=0.48\linewidth]{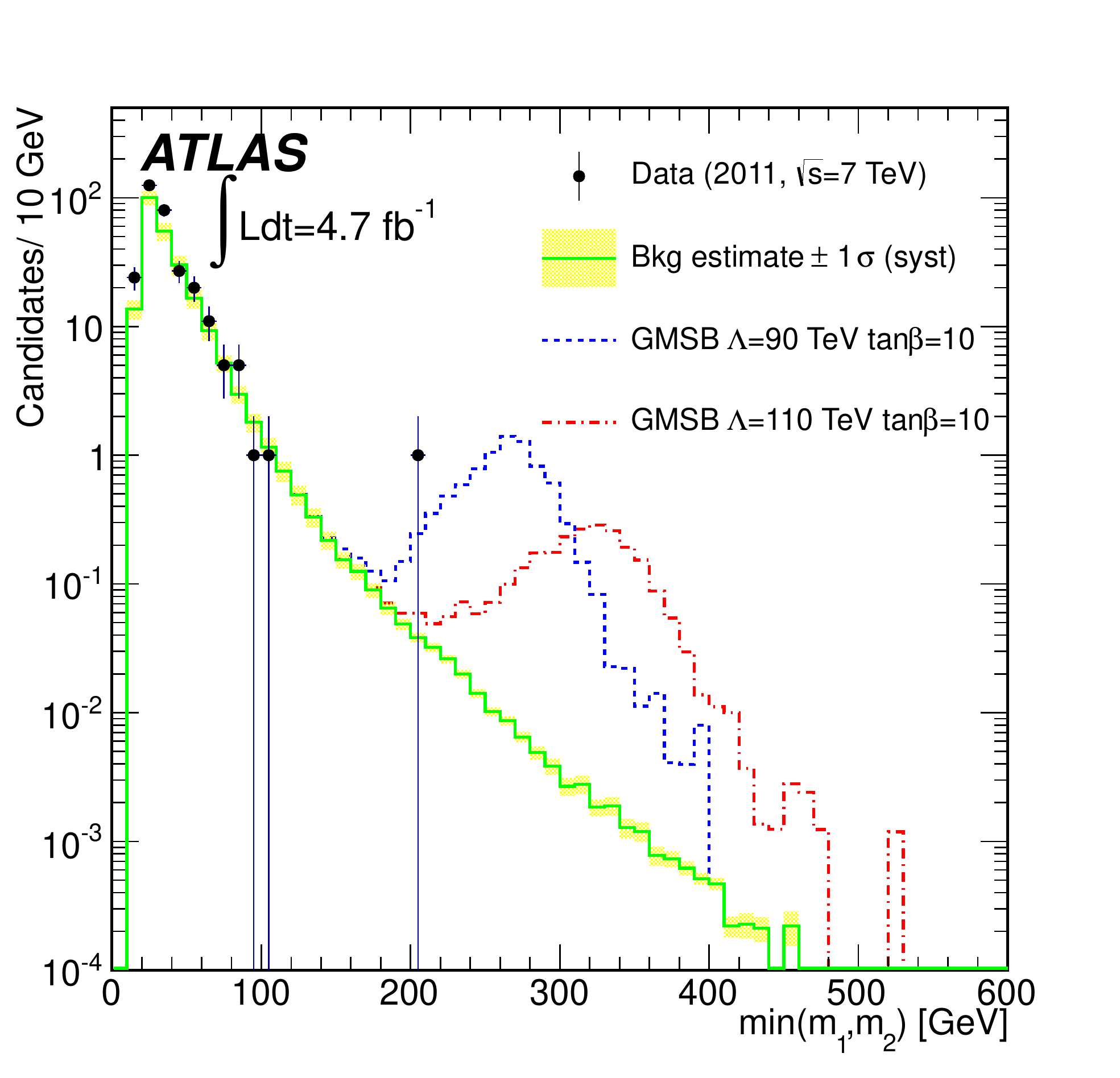}%
  \includegraphics[width=0.49\linewidth]{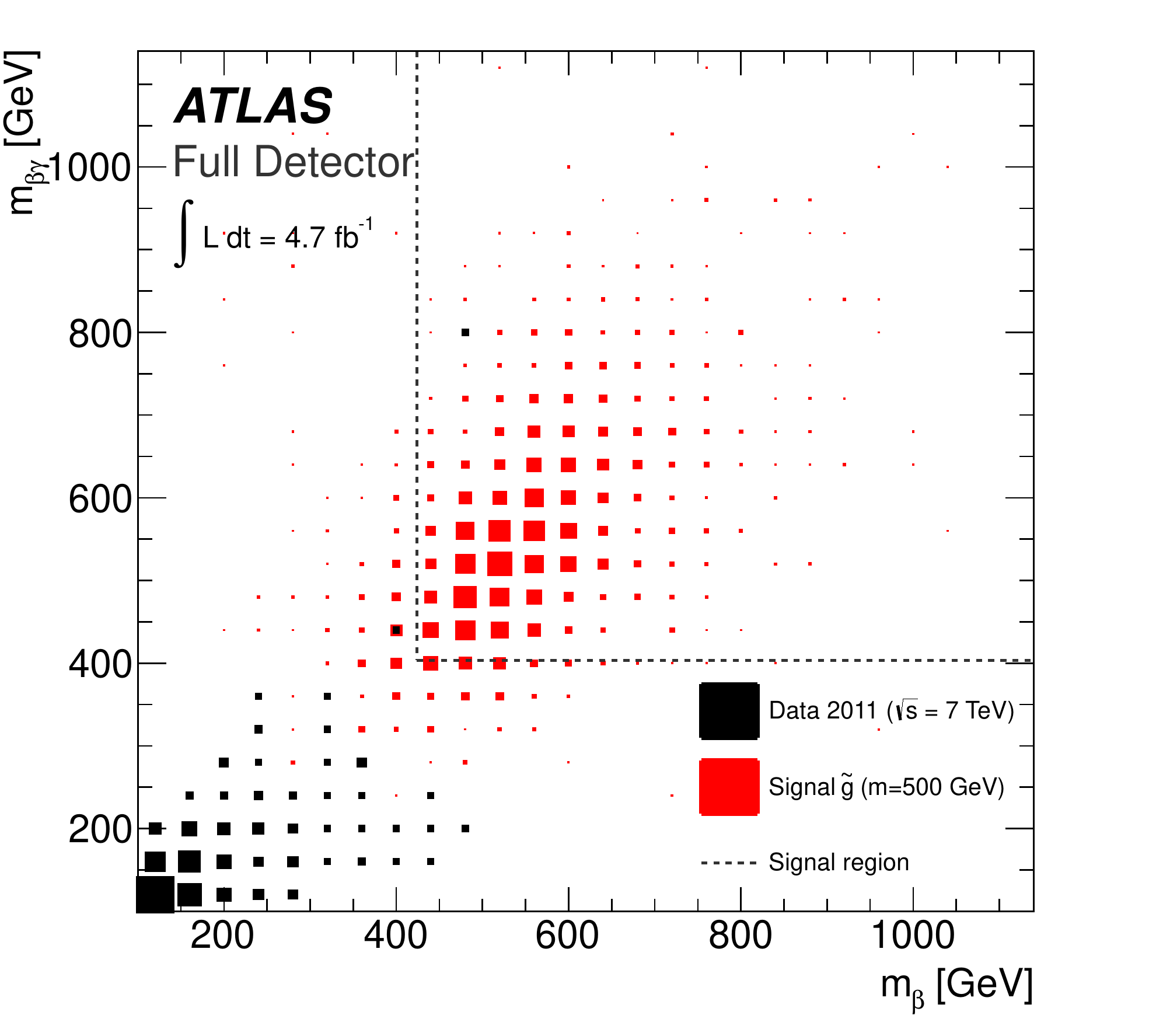}\\%
  \includegraphics[width=0.49\linewidth]{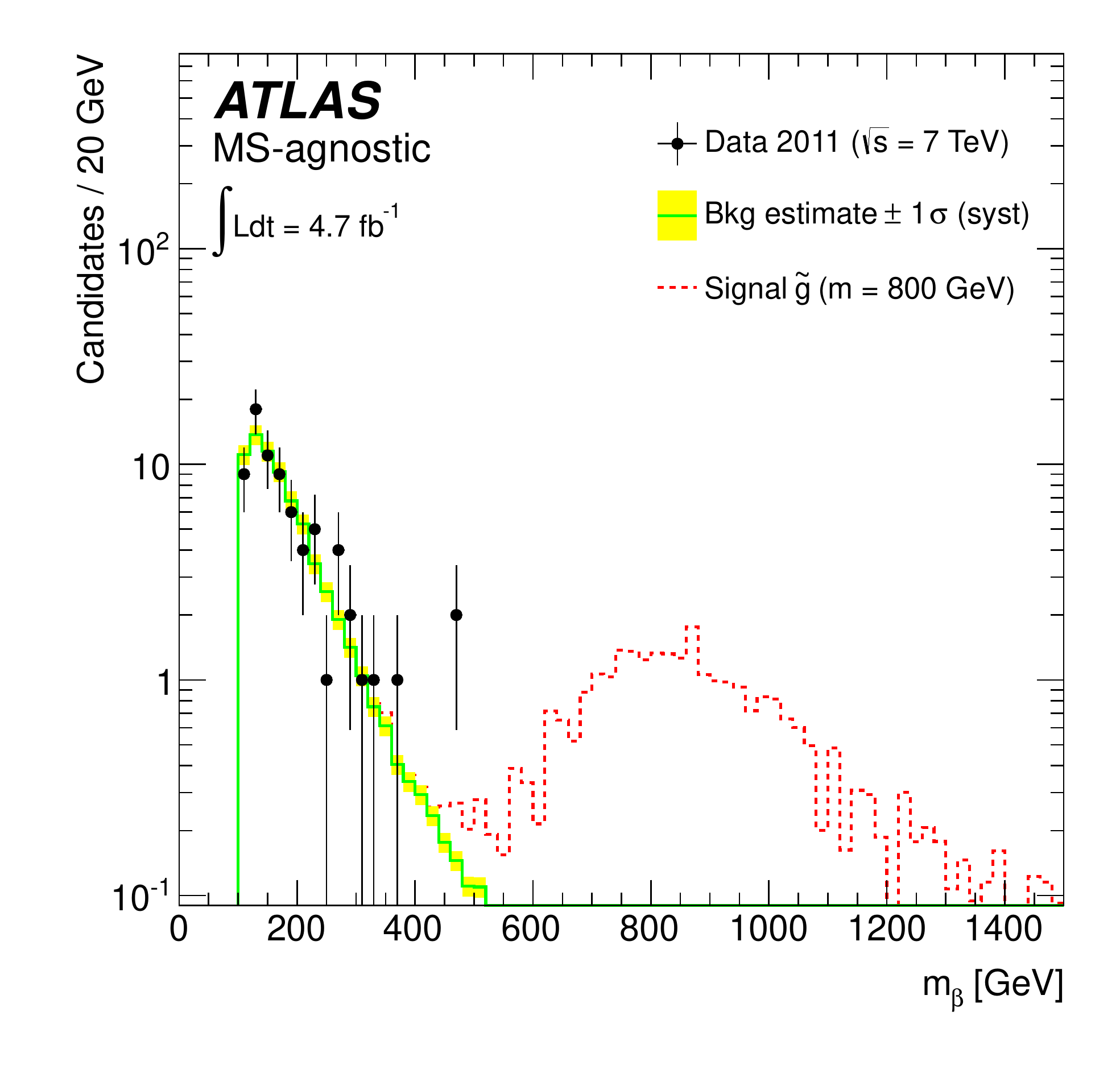}%
  \includegraphics[width=0.49\linewidth]{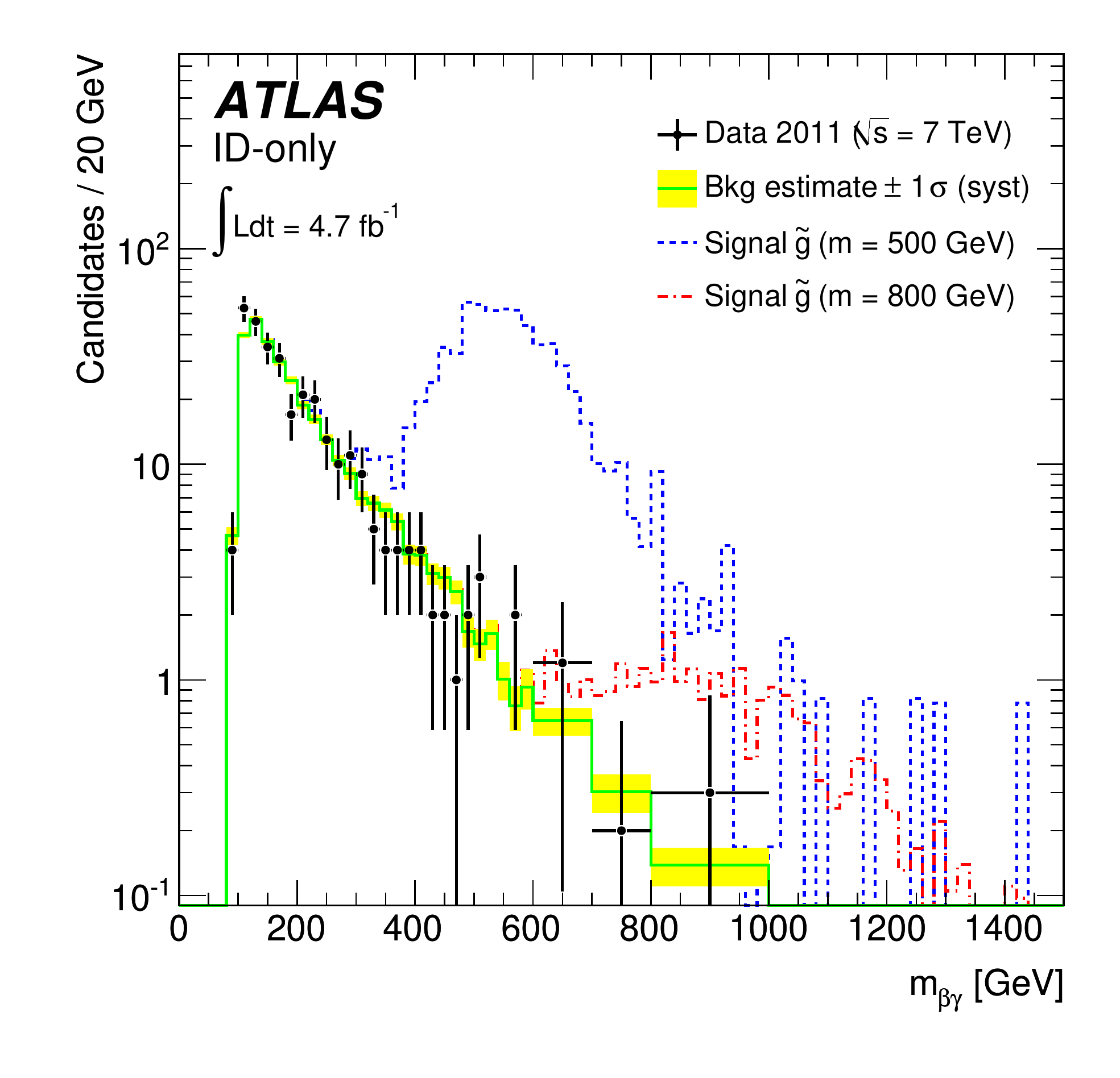}%
  \caption{Observed data, background estimate and expected signal in the two-candidate signal region in the slepton search (top-left), full-detector \rhad search (top-right), MS-agnostic \rhad search (bottom-left) and in the ID-only \rhad search (bottom-right). For the slepton search only the lower of the two masses is plotted and for the full-detector \rhad search there is one candidate in the $m_{\beta\gamma}$ overflow bin not shown in the figure. The dashed lines in the upper-right plot mark the lower edges of the signal region chosen for this \rhad mass point.}
  \label{fig:DSB}
\end{figure*}

The mass distributions observed in data together with the background estimate, its systematic error and examples of signal are shown in Fig.~\ref{fig:DSB}, for the two-candidate signal region in the slepton search (top-left), the full-detector \rhad search (top-right), the MS-agnostic \rhad search (bottom-left) and the ID-only \rhad search (bottom-right).

No indication of signal above the expected background is observed. Upper cross-section limits on new particles are set by counting the number of events passing a set of mass cuts optimised for a given mass point and model. For the ID-only analysis the full mass spectrum of the background and the hypothetical signal is considered.
Cross-section limits are obtained using the $CL_s$ prescription~\cite{Read:451614}. Mass limits are derived by comparing the obtained cross-section limits to the lower edge of the $1\sigma$ band around the theoretically predicted cross-section for each process. A Poisson likelihood function is used for the \rhad limit-setting while a likelihood combining two Poisson functions is used for the slepton search to benefit from the separation into one-candidate and two-candidate signal regions.

\begin{figure*}[htbp!]
  \centering
  \includegraphics[width=0.48\linewidth]{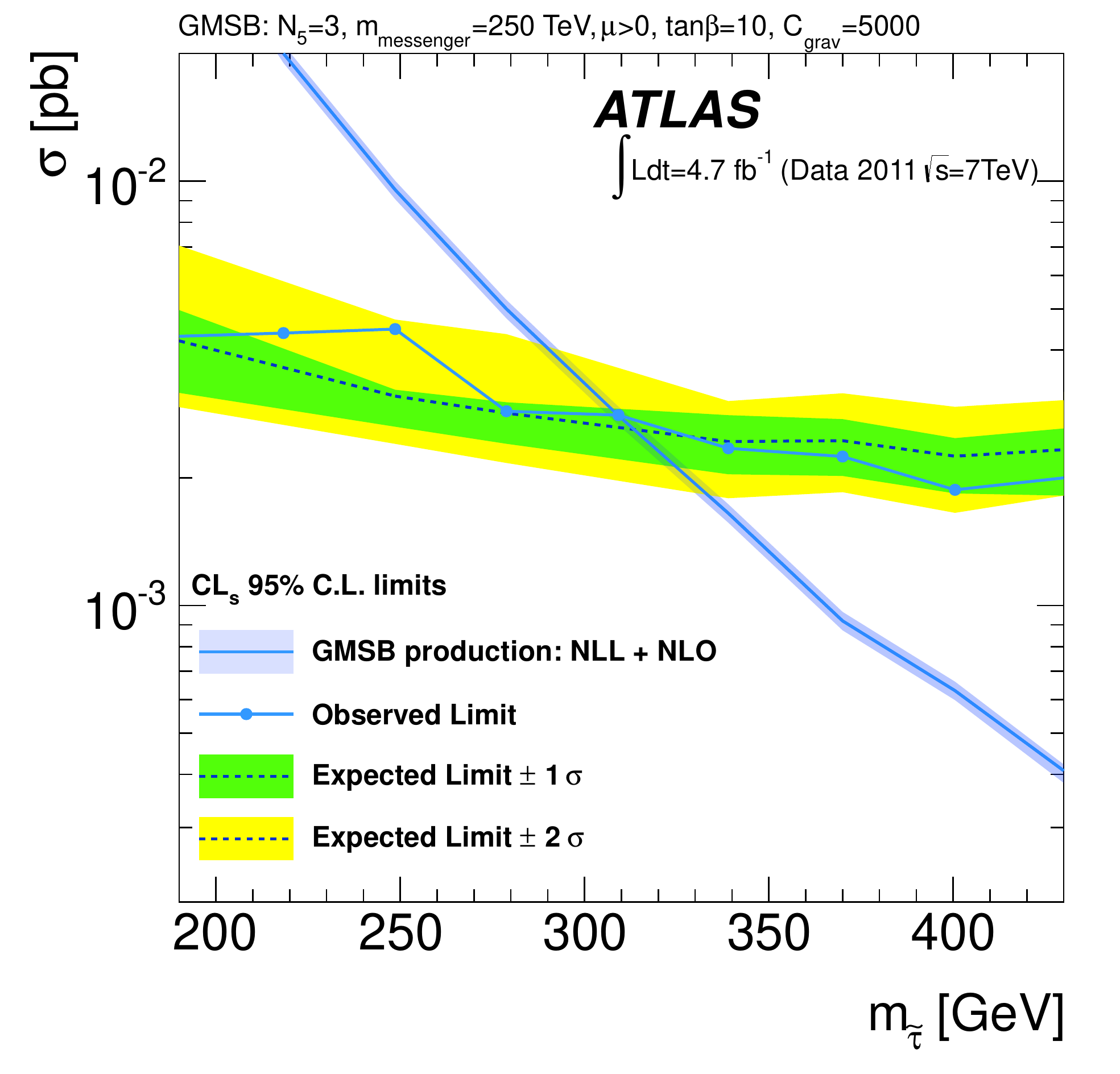}
  \includegraphics[width=0.50\linewidth]{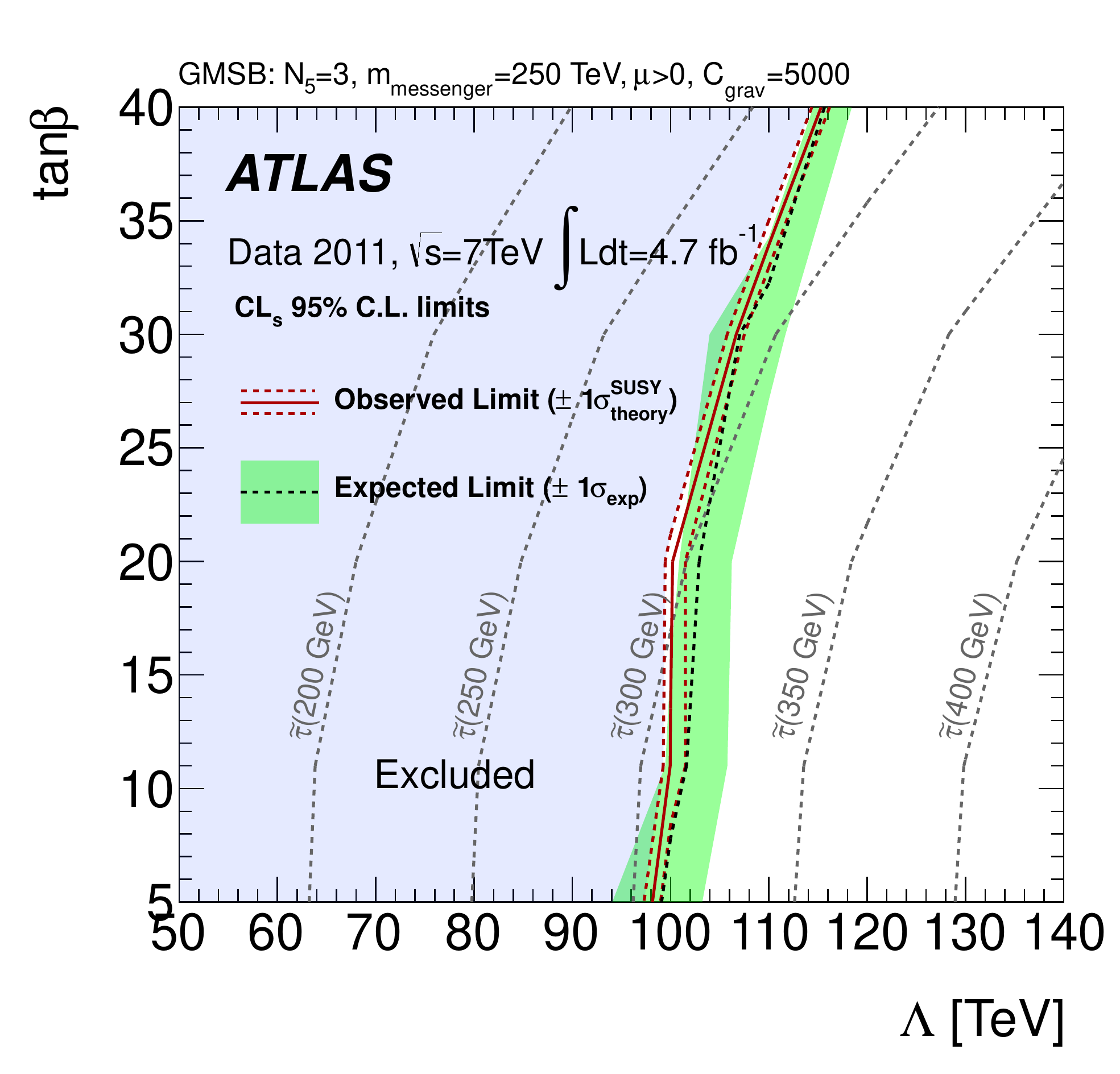}
  \includegraphics[width=0.48\linewidth]{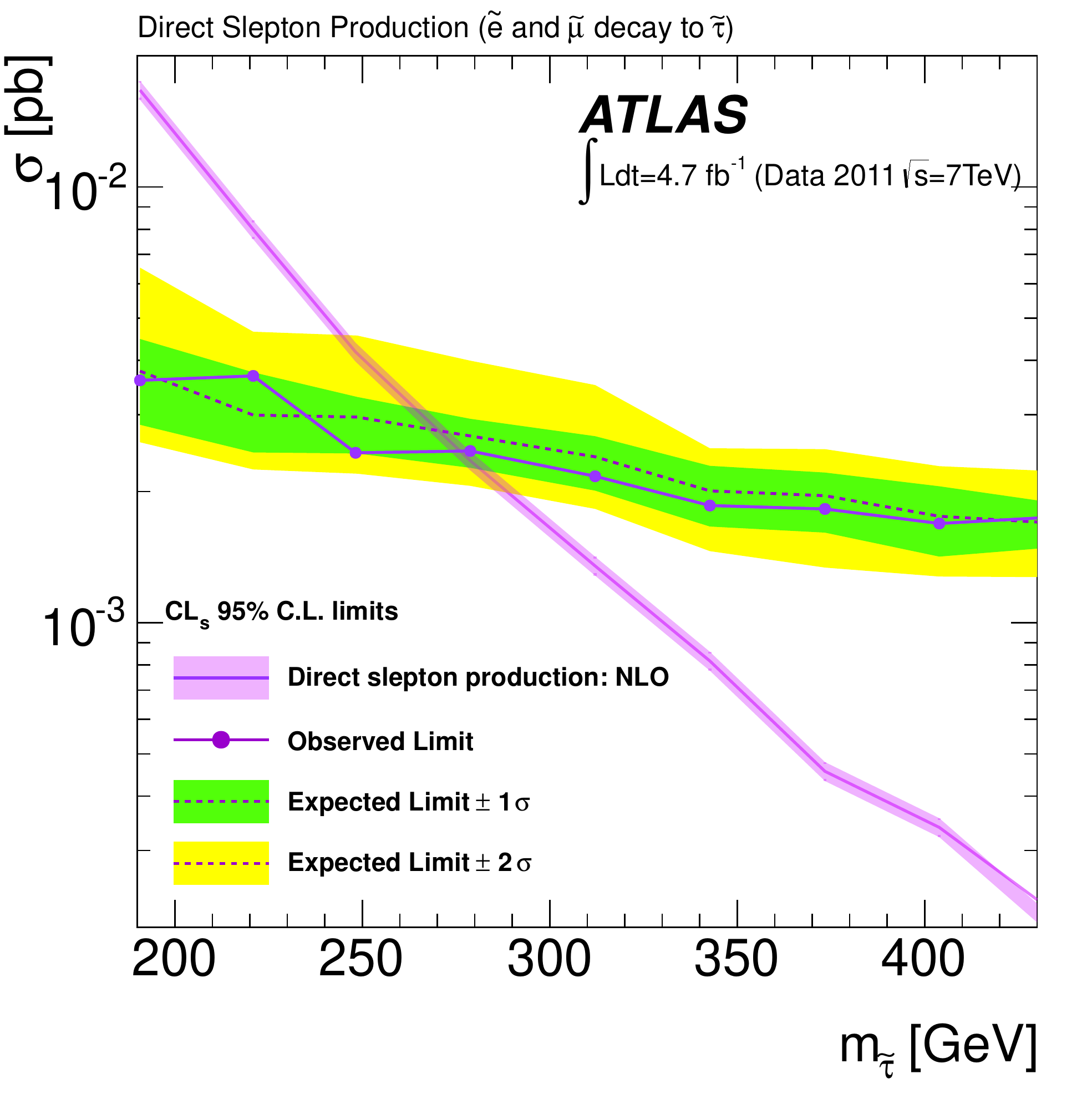}
  \caption{Cross-section limits as a function of the $\tilde{\tau}$ mass in GMSB models (top-left), limits on $\Lambda$ and $\tan\beta$ in GMSB models (top-right) and the cross-section limits as a function of mass on directly produced sleptons (bottom). The top-left and bottom plots show observed and expected limits with $\pm 1\sigma$ and $\pm 2\sigma$ uncertainty bands, as well as the theoretical prediction for the cross-section is shown with a $1\sigma$ band. In the top-right plot, the observed and expected limits are shown, with the $1\sigma$ theoretical uncertainty and the $1\sigma$ band, respectively.}
  \label{fig:limitsSTAU}
\end{figure*}

The resulting production cross-section limits in the GMSB scenario as a function of the stau mass are presented in Fig.~\ref{fig:limitsSTAU} and compared to theoretical predictions. Long-lived staus in GMSB models with $N_5=3$, $m_{\rm{messenger}}=250~\TeV$ and $\mathrm{sign}(\mu)=1$ are excluded at 95\% confidence level (CL) up to a stau mass of 300~\GeV for $\tan\beta=5-20$, and up to stau masses of 285~\GeV and 268~\GeV for $\tan\beta=30$ and $\tan\beta=40$, with expected limits of 305~\GeV, 287~\GeV and 270~\GeV respectively. The lower limit on $\Lambda$ is 99--110~\TeV for values of $\tan \beta$ from 5 to 40. For $\Lambda$ values around this limit, strong production of squarks and gluinos is suppressed due to their large masses. Directly produced sleptons comprise 38--64\% of the GMSB cross-section, and depend only on the slepton mass. Using the same analysis, directly produced long-lived sleptons, or sleptons decaying to long-lived ones, are excluded at 95\%~CL up to a \stau mass of 278~\GeV (273~\GeV expected) for models with slepton mass splittings smaller than 50~\GeV.

\begin{figure*}[tbp]
    \centering
    \includegraphics[width=0.48\linewidth]{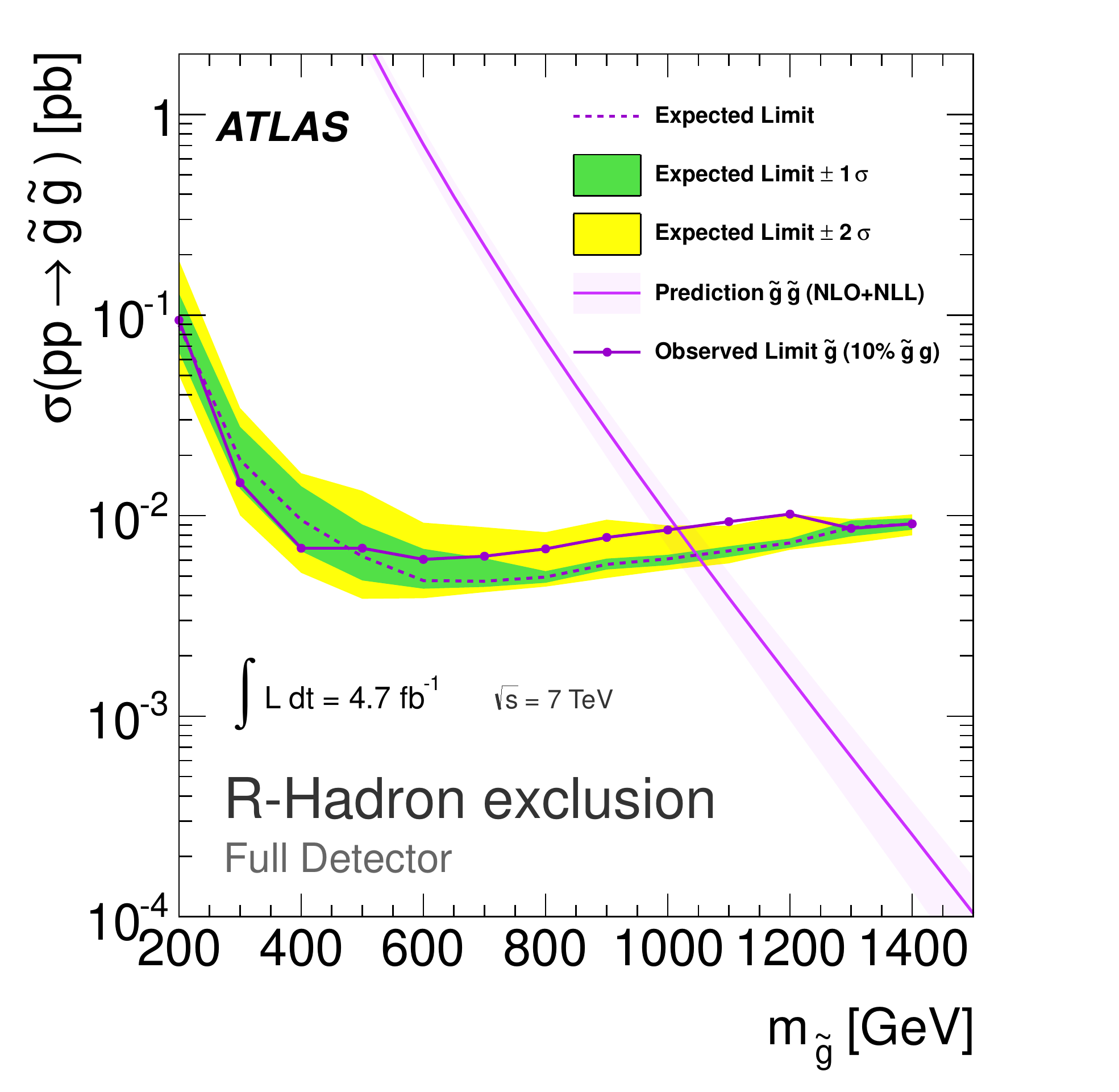}
    \includegraphics[width=0.48\linewidth]{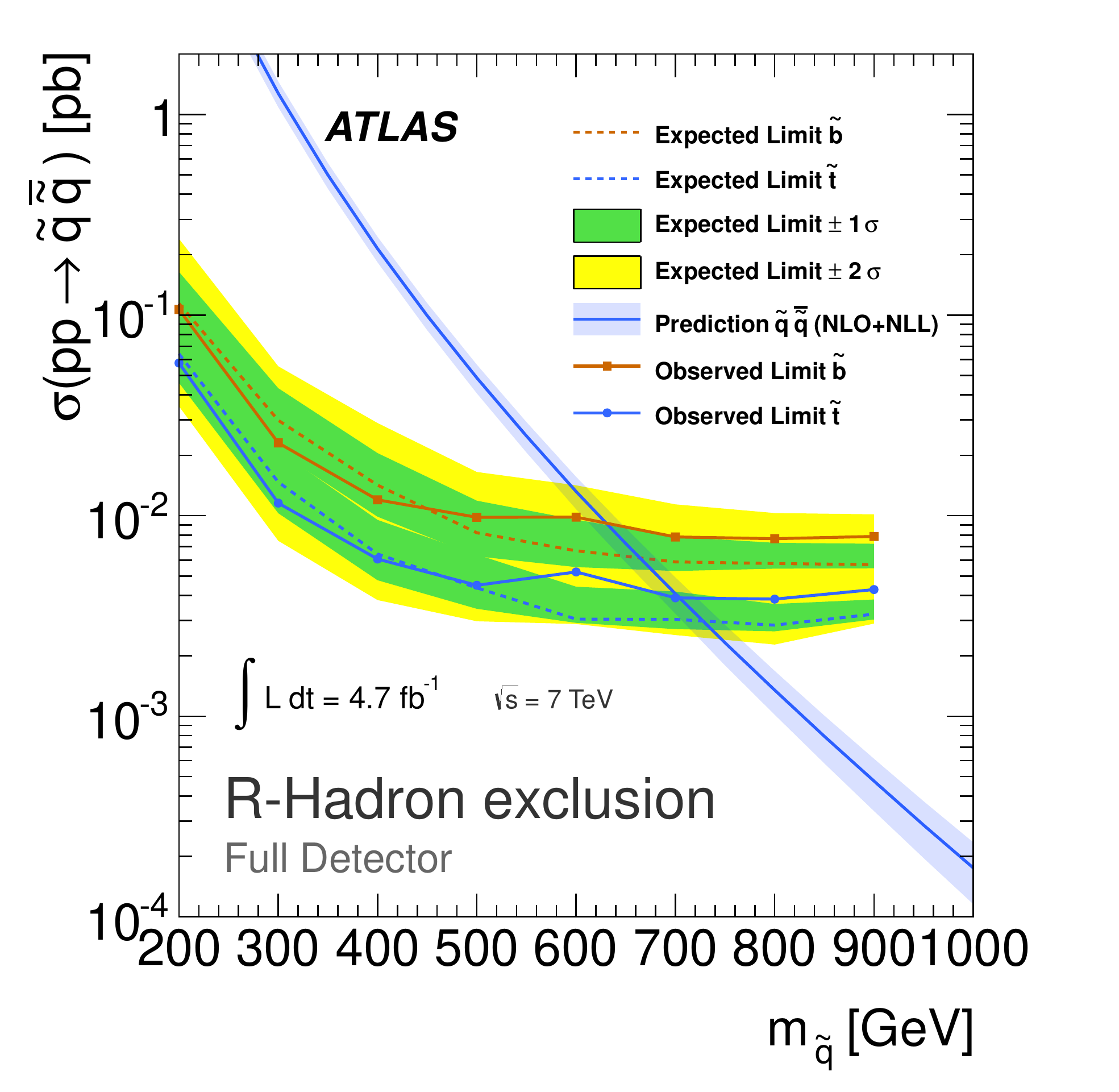}
    \caption{Cross-section upper limits at 95\%~CL for gluino (left) and squark (right) \rhads in the full-detector search are shown. Both the expected limit with $\pm 1\sigma$ and $ \pm 2 \sigma$ uncertainty bands and the observed upper limit are given. In addition the theoretical prediction for the production cross-section calculated at NLO+NLL and its uncertainty are drawn.}
    \label{fig:limitsRH}
\end{figure*}

\begin{figure*}[tbp]
    \centering
    \includegraphics[width=0.48\linewidth]{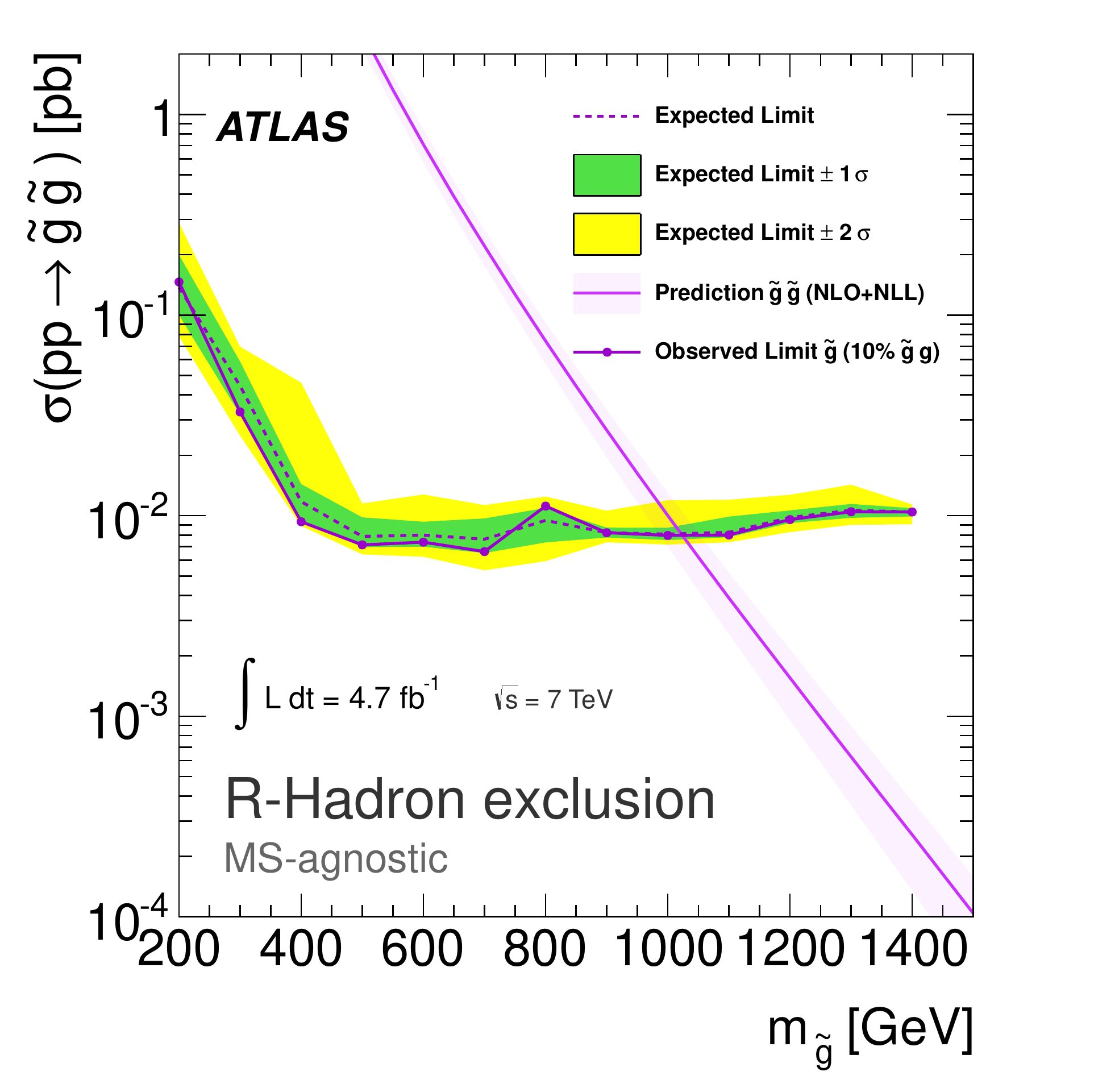}
    \includegraphics[width=0.48\linewidth]{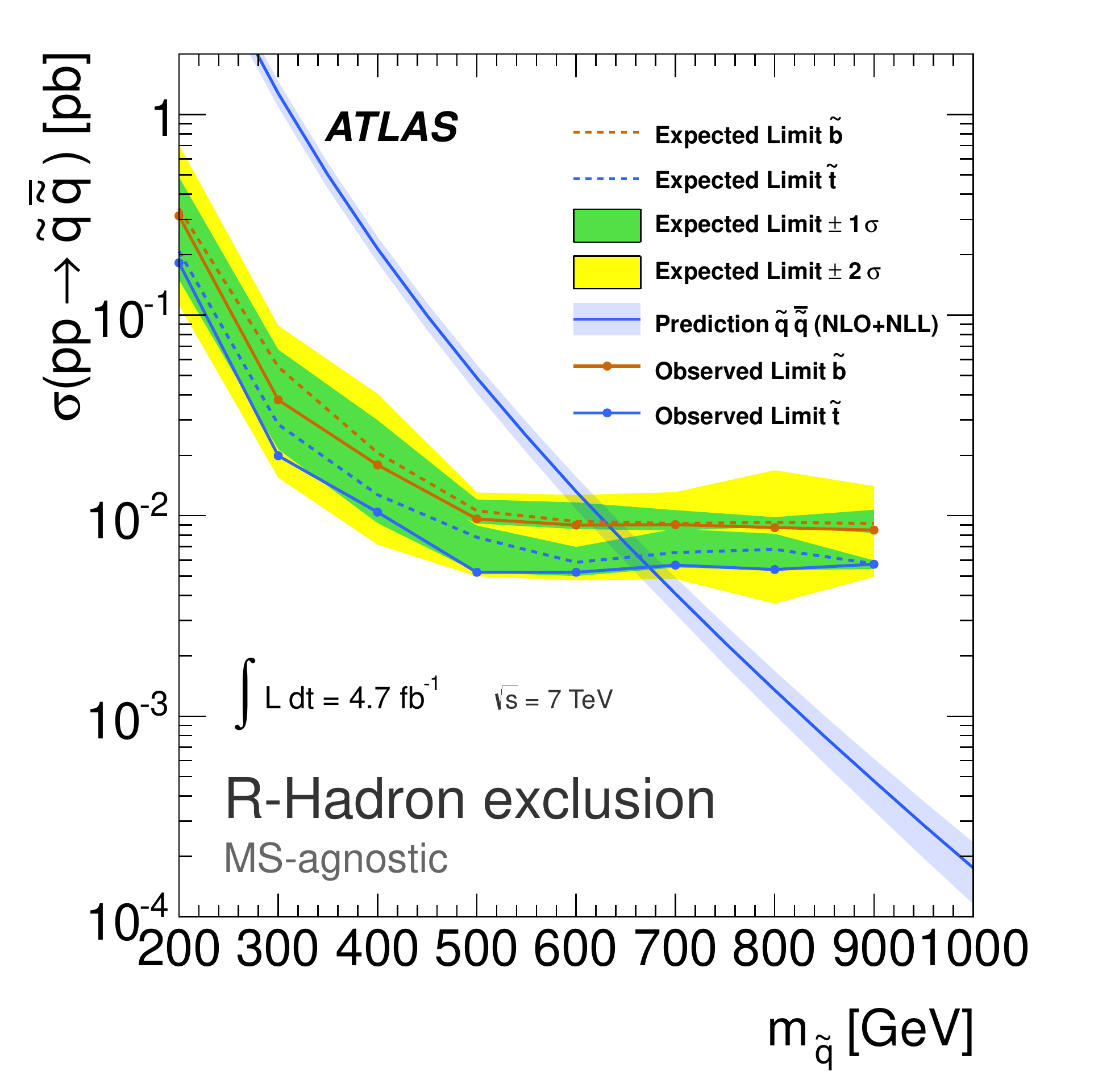}
    \caption{Cross-section upper limits at 95\%~CL for gluino (left) and squark (right) \rhads in the MS-agnostic search, which is independent of the fraction of \rhads that are charged in the MS. Both the expected limit with $\pm 1\sigma$ and $ \pm 2 \sigma$ uncertainty bands and the observed upper limit are given. In addition the theoretical prediction for the production cross-section calculated at NLO+NLL and its uncertainty are drawn.}
    \label{fig:limitsRHagnostic}
\end{figure*}

The resulting limits on \rhads are shown in Figs.~\ref{fig:limitsRH} and~\ref{fig:limitsRHagnostic} compared to the theoretical cross-sections. Gluino \rhads in a generic interaction model are excluded up to a mass of $\rhadronGluinoFullMassExclusion$. Stop \rhads are excluded up to a mass of $\rhadronStopFullMassExclusion$, and sbottom \rhads are excluded up to a mass of $\rhadronSbottomFullMassExclusion$. The according expected limits are \rhadronGluinoFullExpMassExclusion, \rhadronStopFullExpMassExclusion and \rhadronSbottomFullExpMassExclusion respectively. The MS-agnostic search yields a lower mass limit of  $\rhadronGluinoAgnosticMassExclusion$ for gluinos, $\rhadronStopAgnosticMassExclusion$ for stops and $\rhadronSbottomAgnosticMassExclusion$ for sbottoms, with expected limits of \rhadronGluinoAgnosticExpMassExclusion, \rhadronStopAgnosticExpMassExclusion and \rhadronSbottomAgnosticExpMassExclusion respectively. The MS-agnostic limits are independent of the fraction of \rhads that arrive charged in the MS.

\begin{figure*}[tbp]
  \centering
  \includegraphics[width=0.48\linewidth]{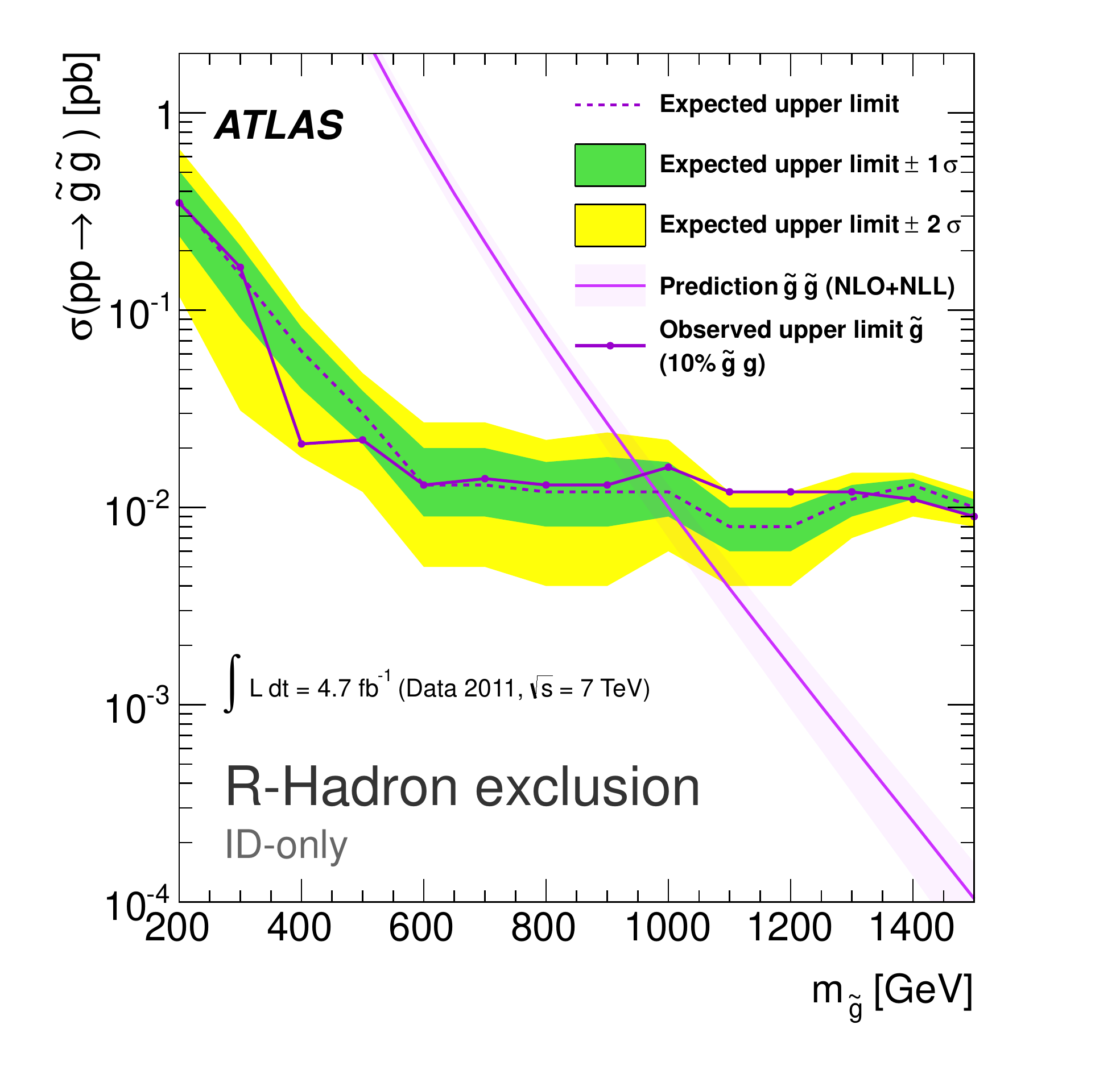}
  \includegraphics[width=0.48\linewidth]{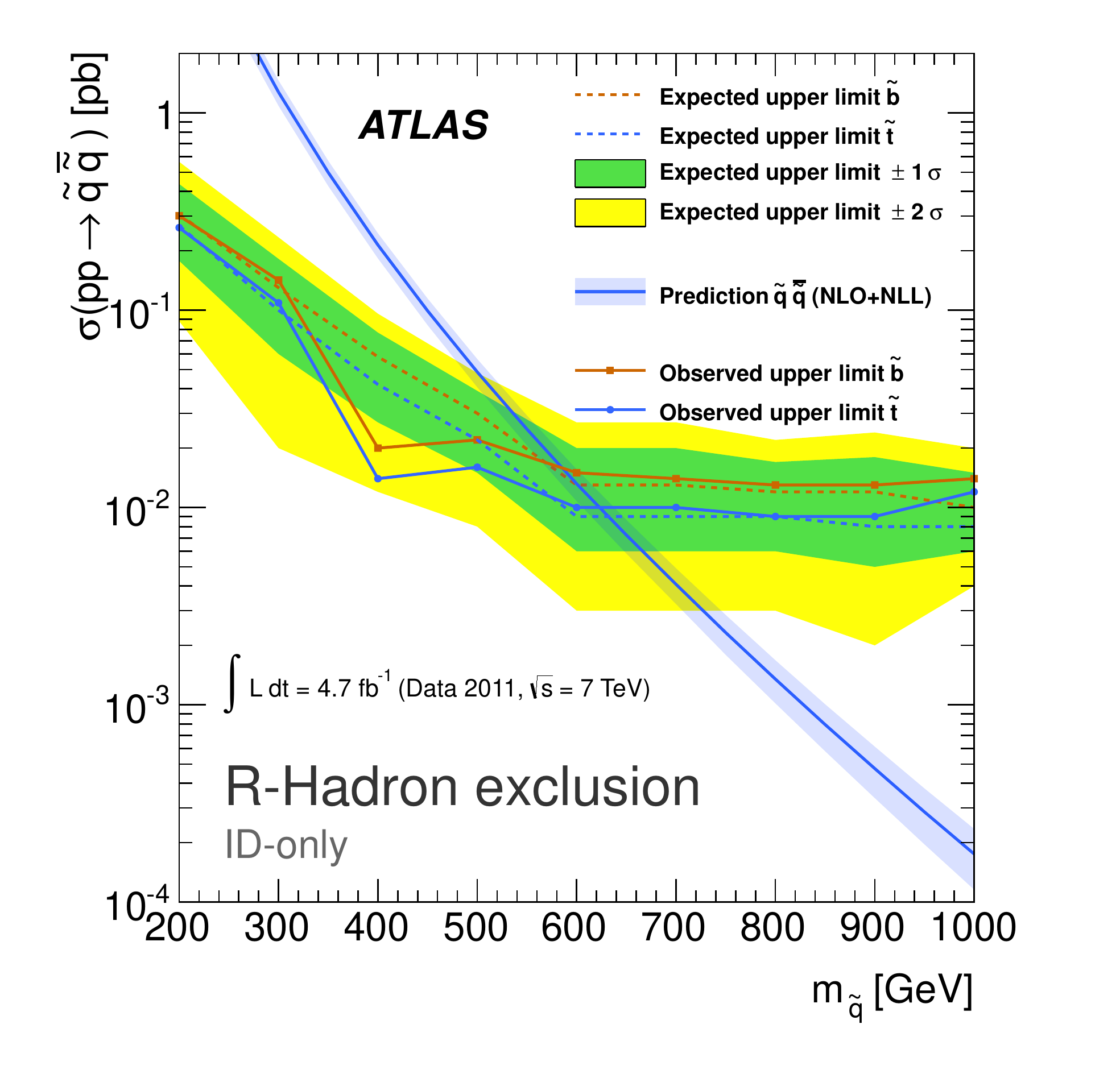}
  \caption{Cross-section upper limits at 95\%~CL for gluino (left) and squark (right) \rhads in the ID-only search. Both the expected limit with $\pm 1\sigma$ and $ \pm 2 \sigma$ uncertainty bands and the observed upper limit are given. In addition the theoretical prediction for the production cross-section calculated at NLO+NLL and its uncertainty are drawn.}
  \label{fig:limitsRHID}
\end{figure*}

The limits from the ID-only search are shown in Fig.~\ref{fig:limitsRHID} compared to the theoretical cross-sections. Gluino \rhads with $m<\rhadronGluinoIDExclusion$, stop \rhads with $m<\rhadronStopIDExclusion$ and sbottom \rhads with $m<\rhadronSbottomIDExclusion$ are excluded at 95\%~CL, with expected limits of \rhadronGluinoIDExpectation, \rhadronStopIDExpectation and \rhadronSbottomIDExpectation respectively. The ID-only search has similar sensitivity also to \rhads with lifetimes as short as a few ns.


\section{Conclusion}

Searches for long-lived massive particles were performed using data from $pp$ collisions at $\sqrt{s} = 7~\TeV$ collected by the ATLAS detector in 2011, corresponding to an integrated luminosity of \lumi. Several different signatures are considered. The data are found to match the Standard Model background expectation in all signal regions. The exclusion limits placed for various models impose new constraints on non-SM cross-sections. Long-lived \stau's in the GMSB model considered, for $\tan\beta = 5-20$, are excluded at 95\%~CL for masses up to 300~\GeV, while directly produced  long-lived sleptons, or sleptons decaying to long-lived ones, are excluded at 95\%~CL up to a \stau mass of 278~\GeV for models with slepton mass splittings smaller than 50~\GeV.

Long-lived \rhads containing a gluino (stop, sbottom) are excluded for masses up to~\rhadronGluinoFullMassExclusion (\rhadronStopFullMassExclusion, \rhadronSbottomFullMassExclusion) at 95\%~CL, for a generic interaction model. Limits obtained independently of the fraction of \rhads that arrive charged in the MS exclude masses up to \rhadronGluinoAgnosticMassExclusion (\rhadronStopAgnosticMassExclusion, \rhadronSbottomAgnosticMassExclusion). Furthermore, using only the inner detector, \rhads are excluded up to \rhadronGluinoIDExclusion (\rhadronStopIDExclusion, \rhadronSbottomIDExclusion).

These results substantially extend previous ATLAS limits~\cite{Aad:2011yf, Aad:2011hz}, and are complementary to the searches for SUSY particles which decay promptly.

\input{Acknowledgements.tex}
\appendix

\bibliographystyle{elsarticle-num-names}
\biboptions{sort&compress}
\bibliography{SUSY-2012-01}

\clearpage

\section{Auxiliary material}
\begin{figure*}[tb]
  \centering
  \includegraphics[width=0.49\linewidth]{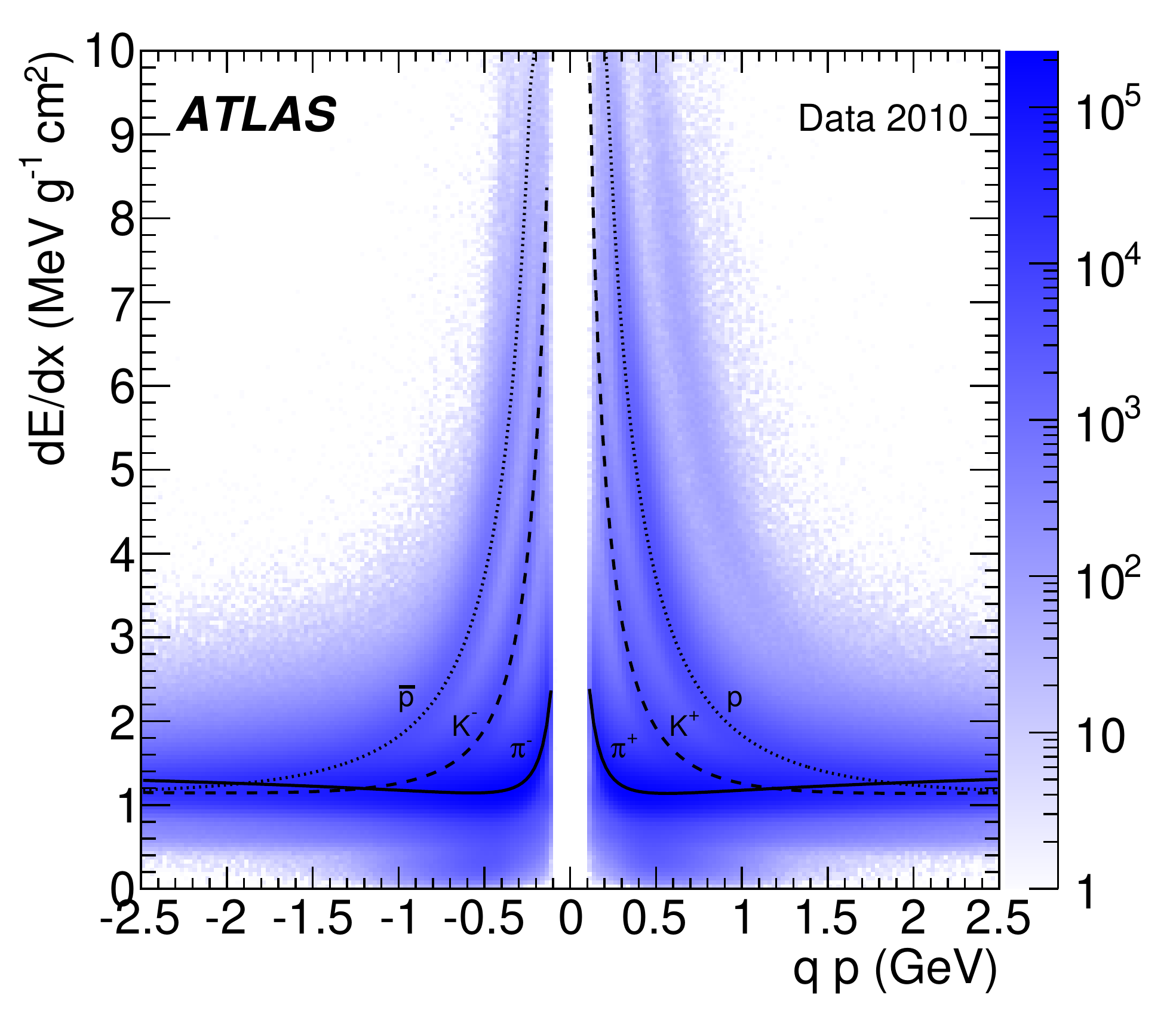}
  \caption{Greyscale-printing version of Fig.~\ref{fig:dEdx}(left): Distribution of \dedx versus charge times momentum for minimum bias collisions in a data sample from 2010. With tracks reconstructed for $\pt > 100~\MeV$ this sample is more suitable for the calibration of the Bethe-Bloch function than 2011 data in which tracks had a cutoff of $\pt=400~\MeV$. The distribution of the most probable value for pions (solid), kaons (dashed) and protons (dotted) are superimposed. The band due to deuterons is clearly visible.}
  \label{fig:dEdx_colour}
\end{figure*}

\begin{figure*}
  \centering
  \includegraphics[width=0.49\linewidth]{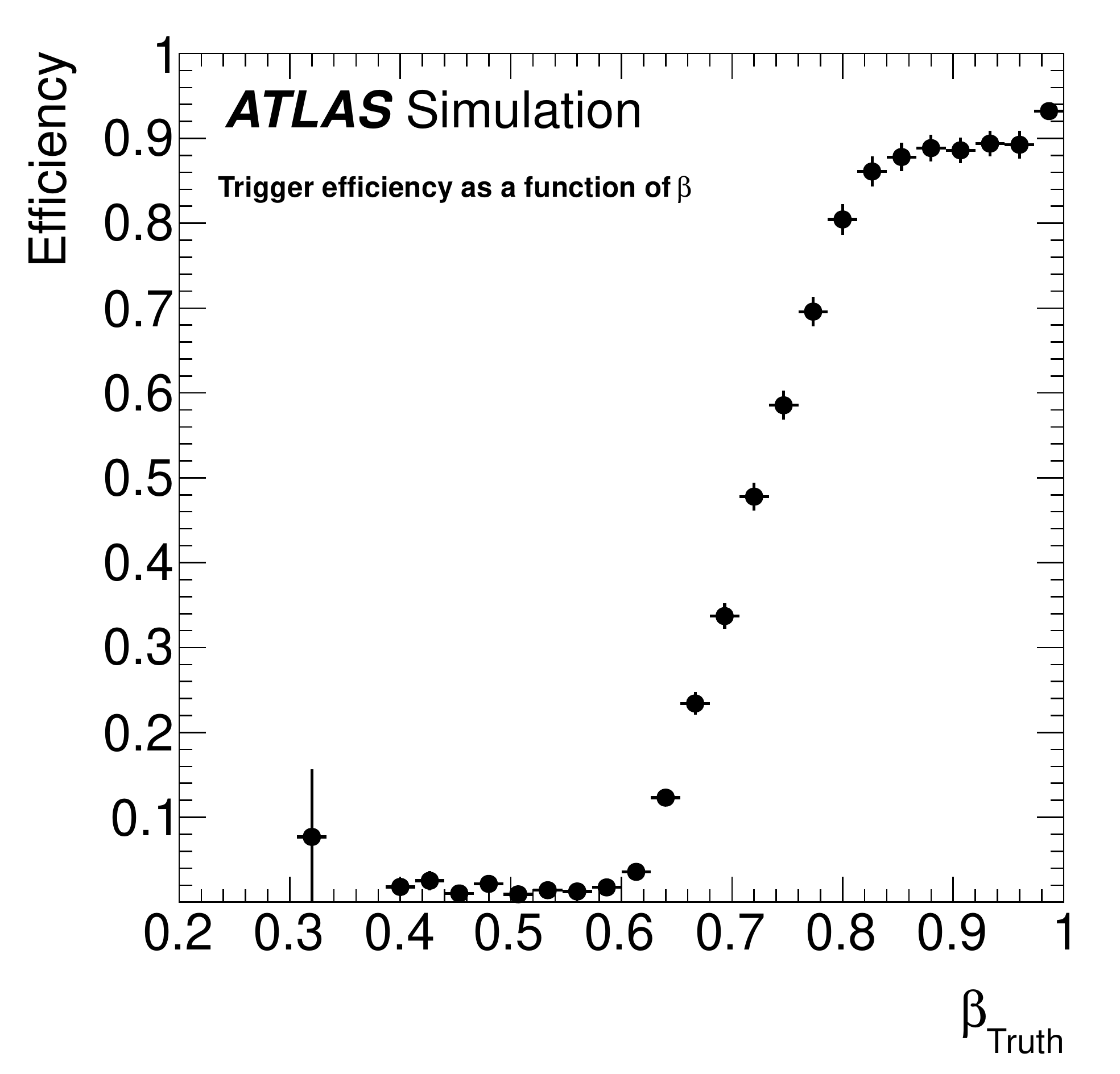}
  \includegraphics[width=0.49\linewidth]{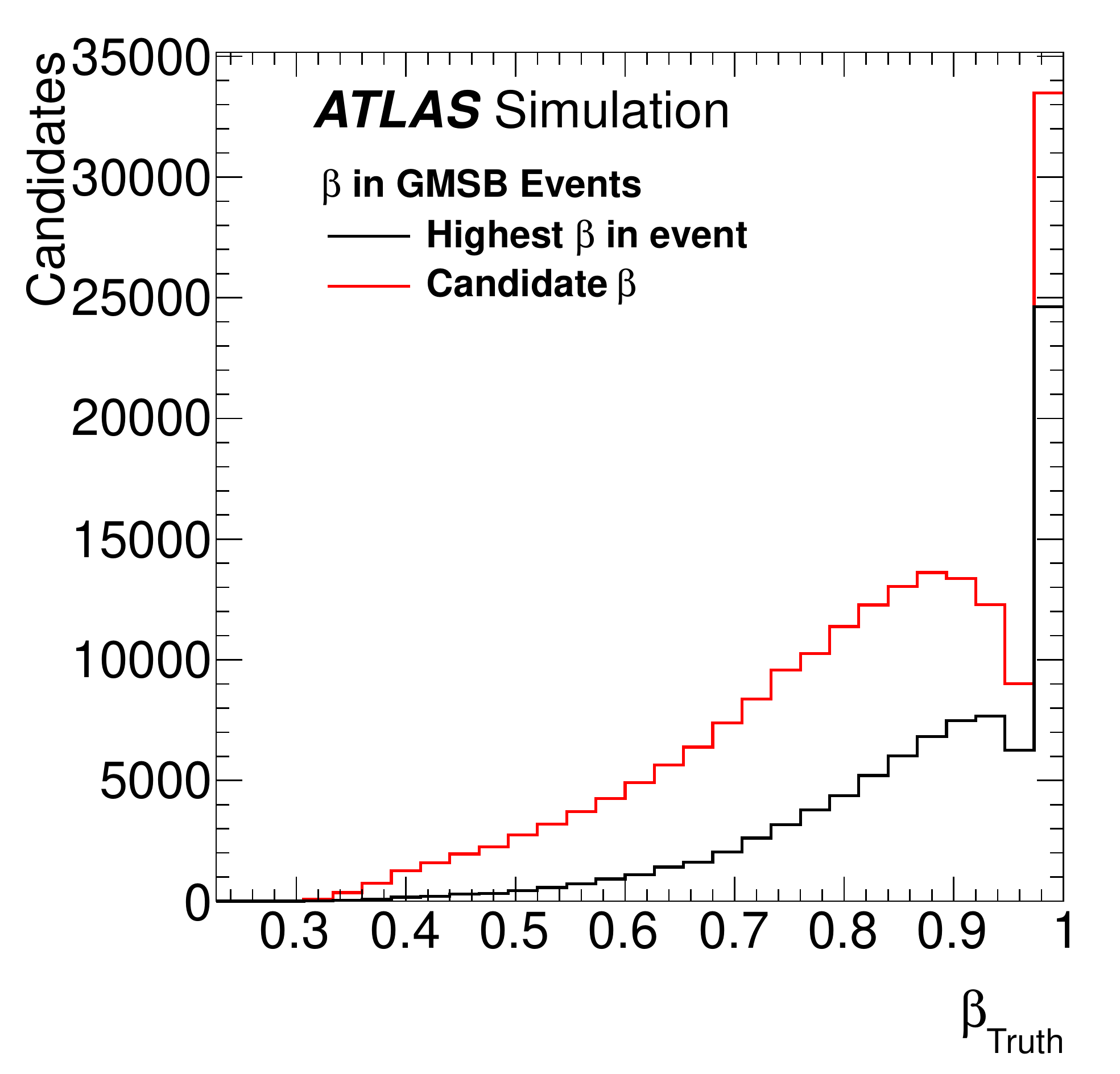}
  \caption{Left: Muon trigger efficiency in simulated GMSB events as a function of the highest $\beta$ candidate in the event. Right: Distribution of $\beta$ for candidates with the highest $\beta$ (black) and for all candidates in the events (red). Low-$\beta$ candidates are found in triggered events because there is a high-$\beta$ slepton or muon in the event that could trigger. (For interpretation of the references to colour in this figure legend, the reader is referred to the web version of this Letter.)}
  \label{fig:MuTrig}
\end{figure*}

\begin{figure*}
  \centering
  \includegraphics[width=0.49\linewidth]{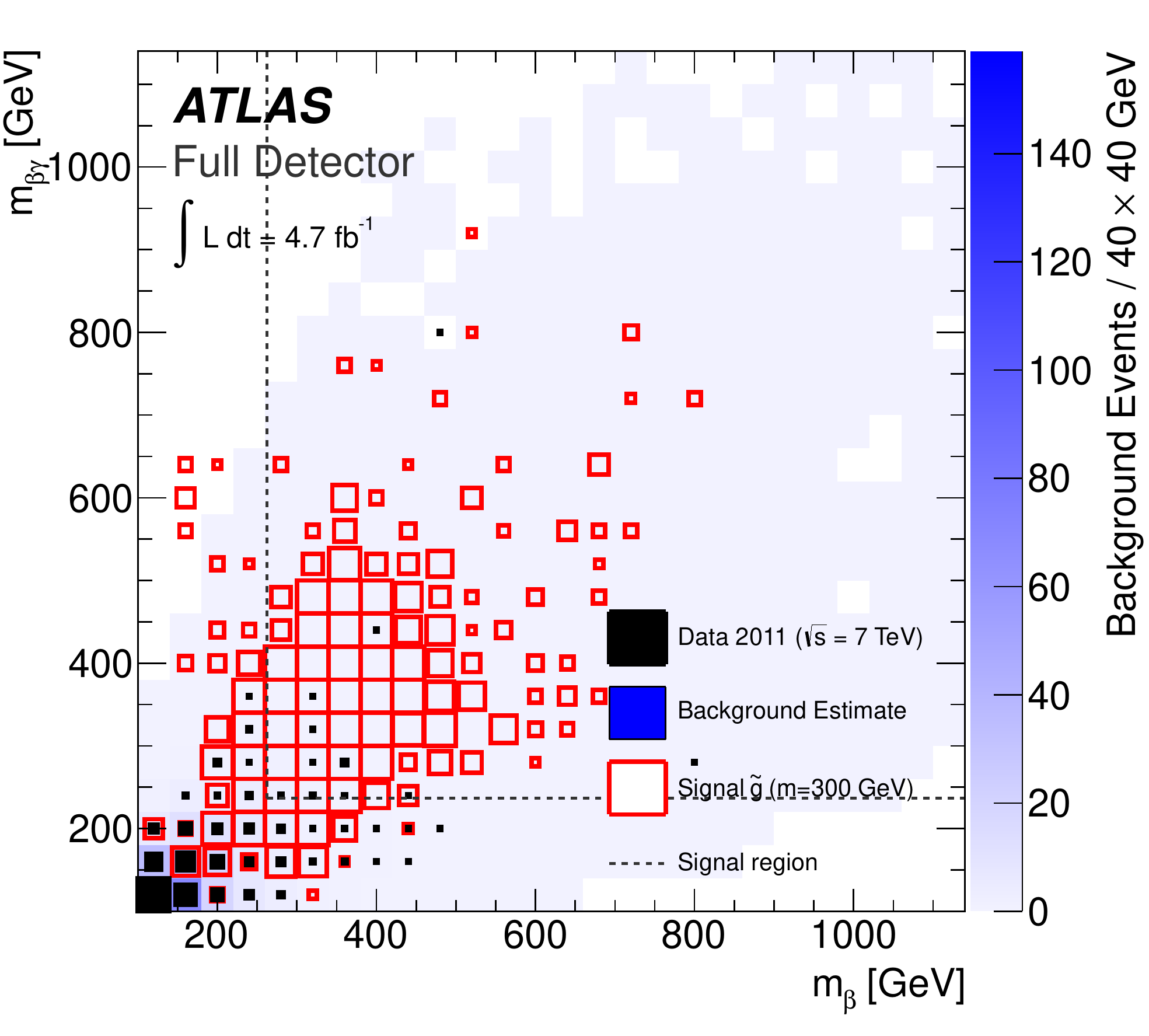}
  \includegraphics[width=0.49\linewidth]{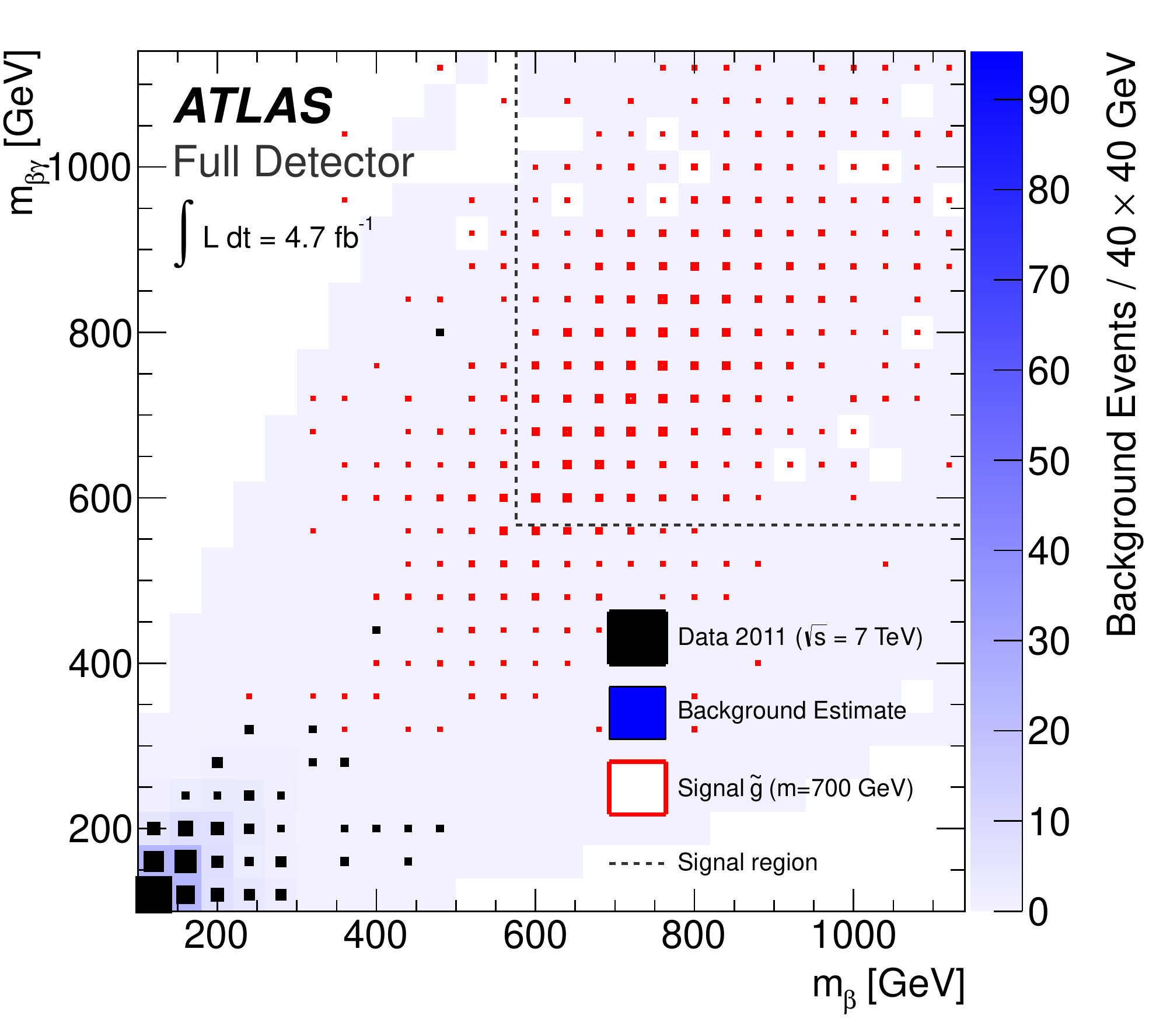}
  \includegraphics[width=0.49\linewidth]{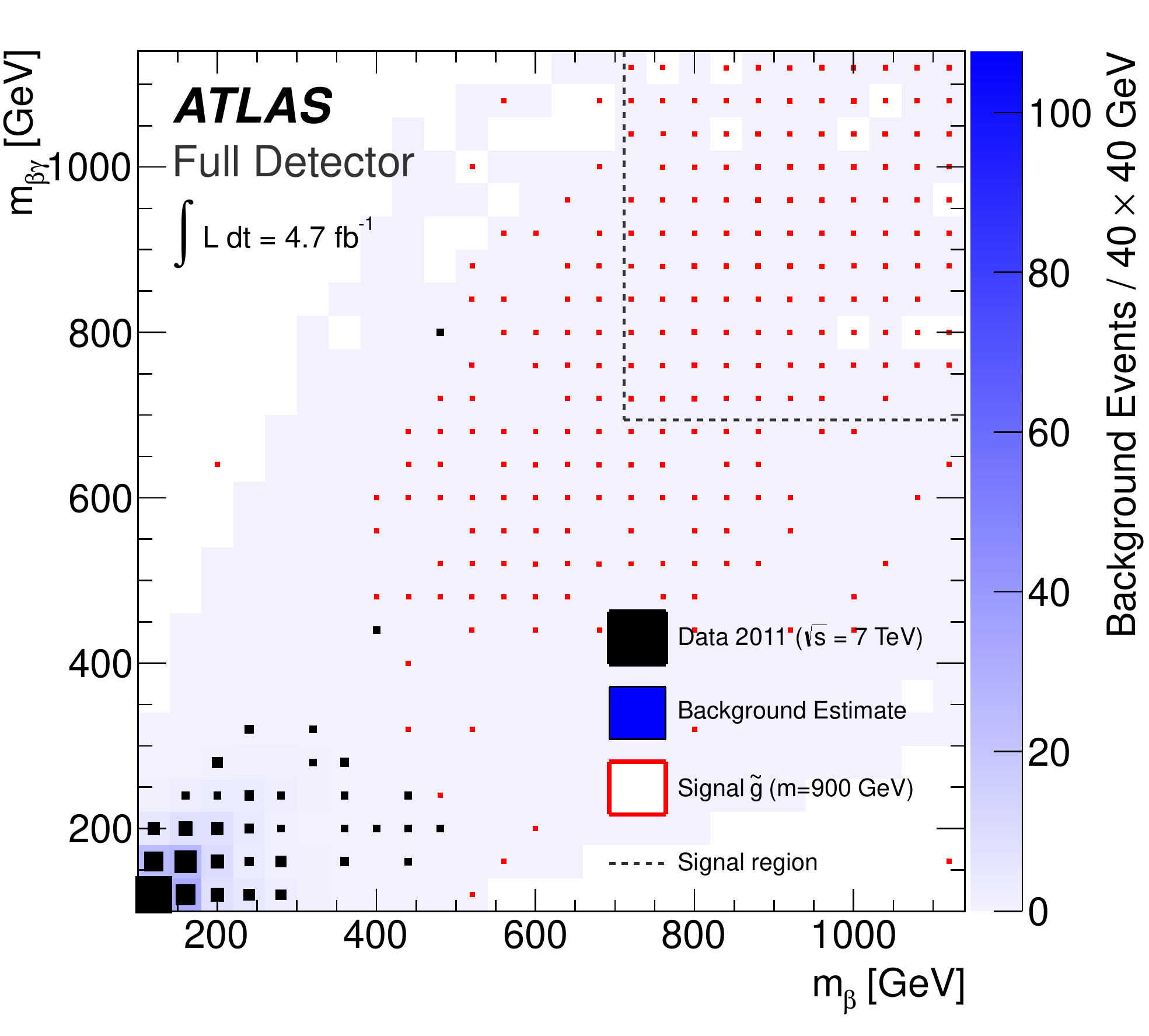}
  \includegraphics[width=0.49\linewidth]{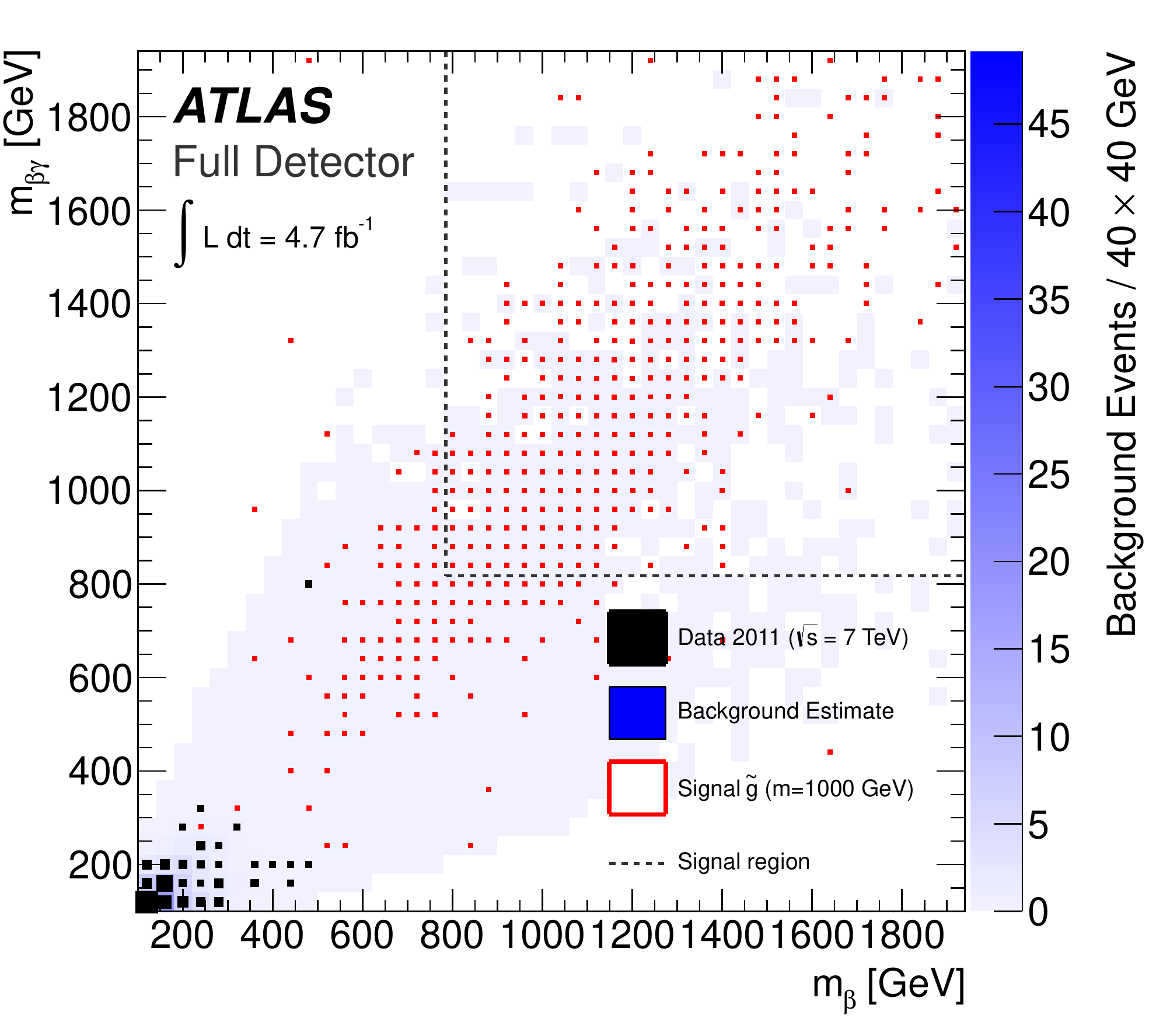}
  \caption{Examples of the $m_{\beta\gamma}$ vs. $m_{\beta}$ distributions for gluino \rhads with masses of 300, 700, 900 and 1000~\GeV, with the full-detector search. The figures show data, background and signal after $\beta$, $\beta\gamma$ and $p$ cuts. The dashed grey line highlights the counting window for each specific hypothesis, and is defined by requiring 90\% of the signal to be within the window.  There is one candidate in the $m_{\beta\gamma}$ overflow.}
  \label{fig:mass2d_full}
\end{figure*}

\begin{figure*}
  \centering
  \includegraphics[width=0.49\linewidth]{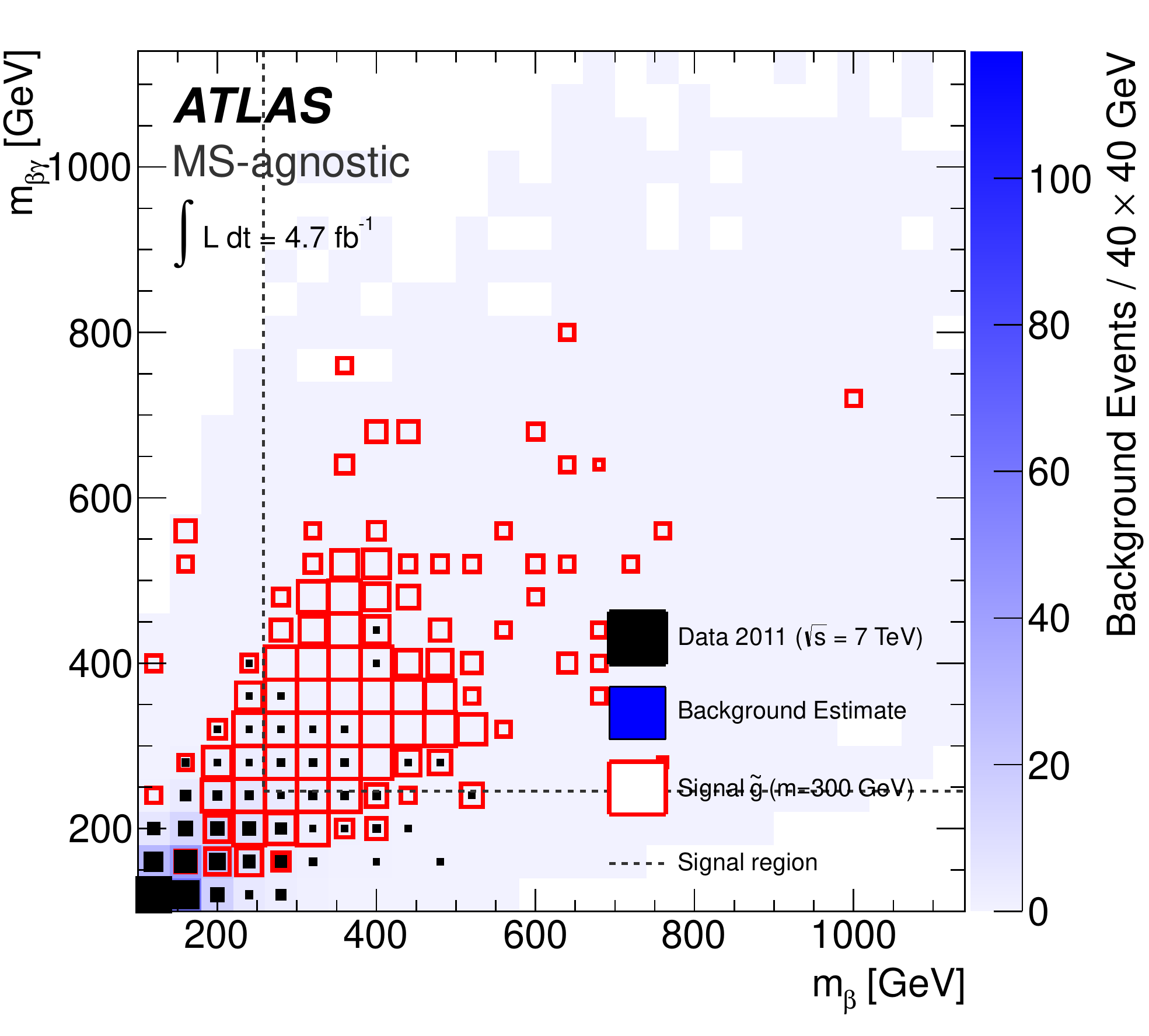}
  \includegraphics[width=0.49\linewidth]{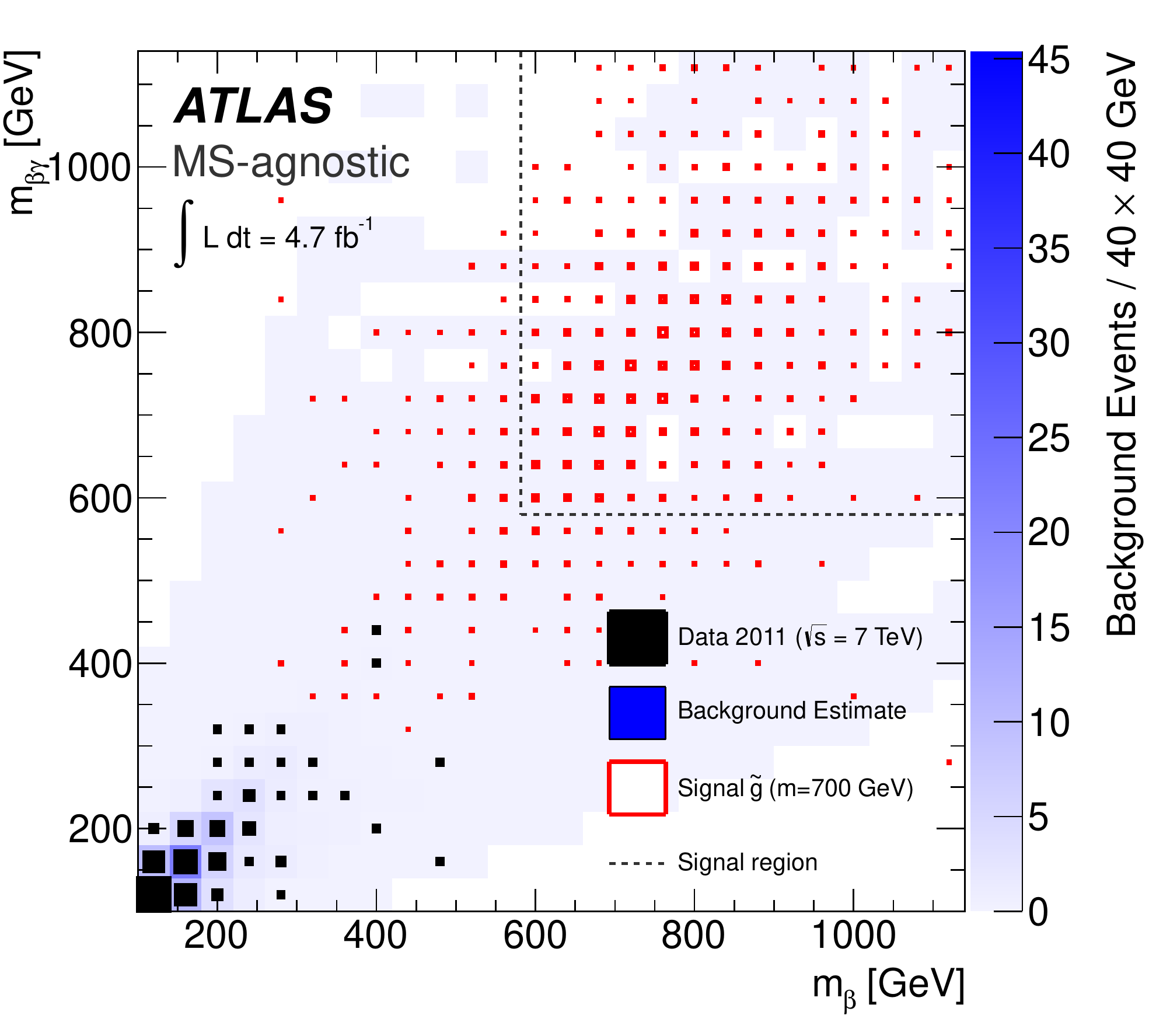}
  \includegraphics[width=0.49\linewidth]{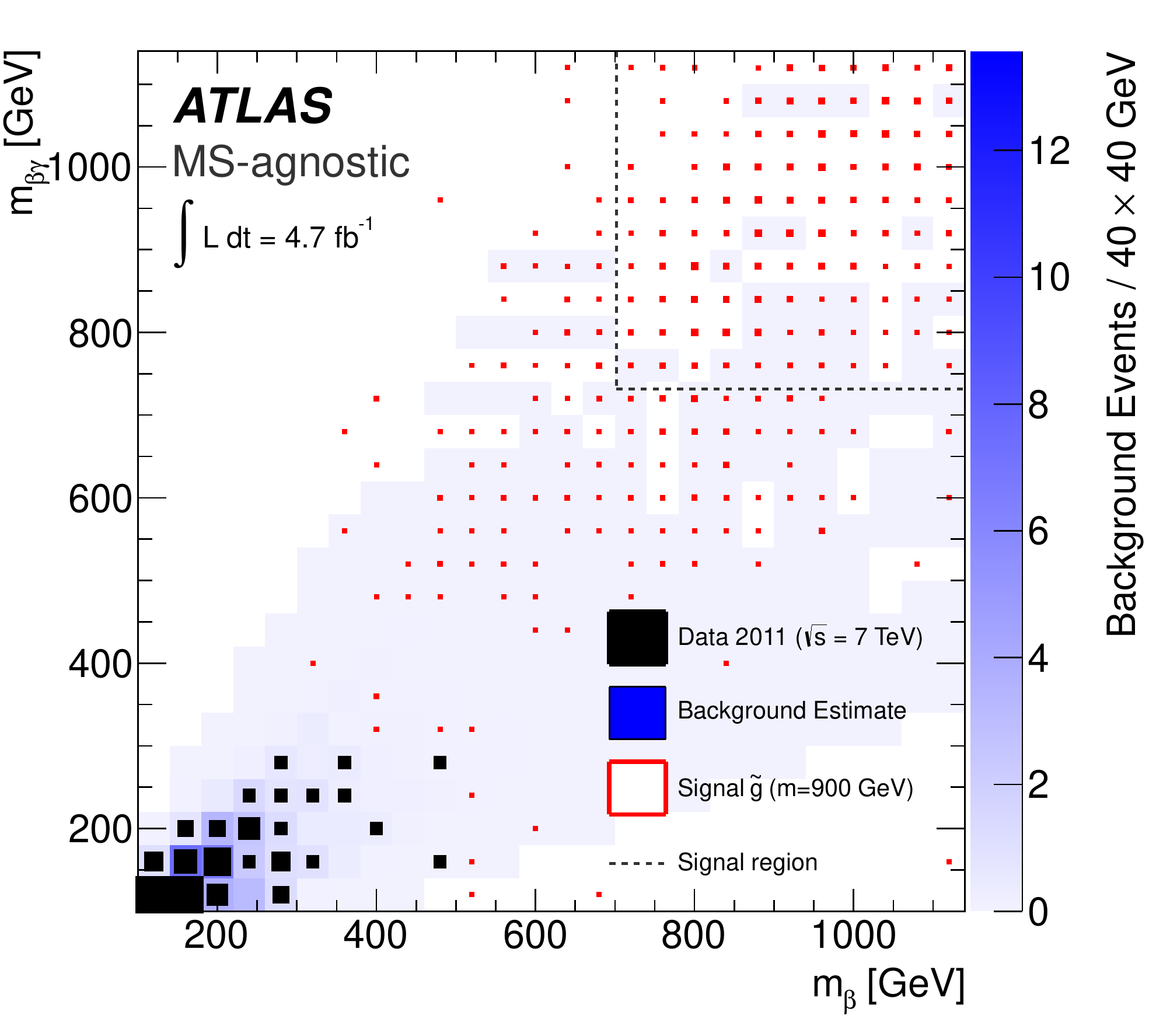}
  \includegraphics[width=0.49\linewidth]{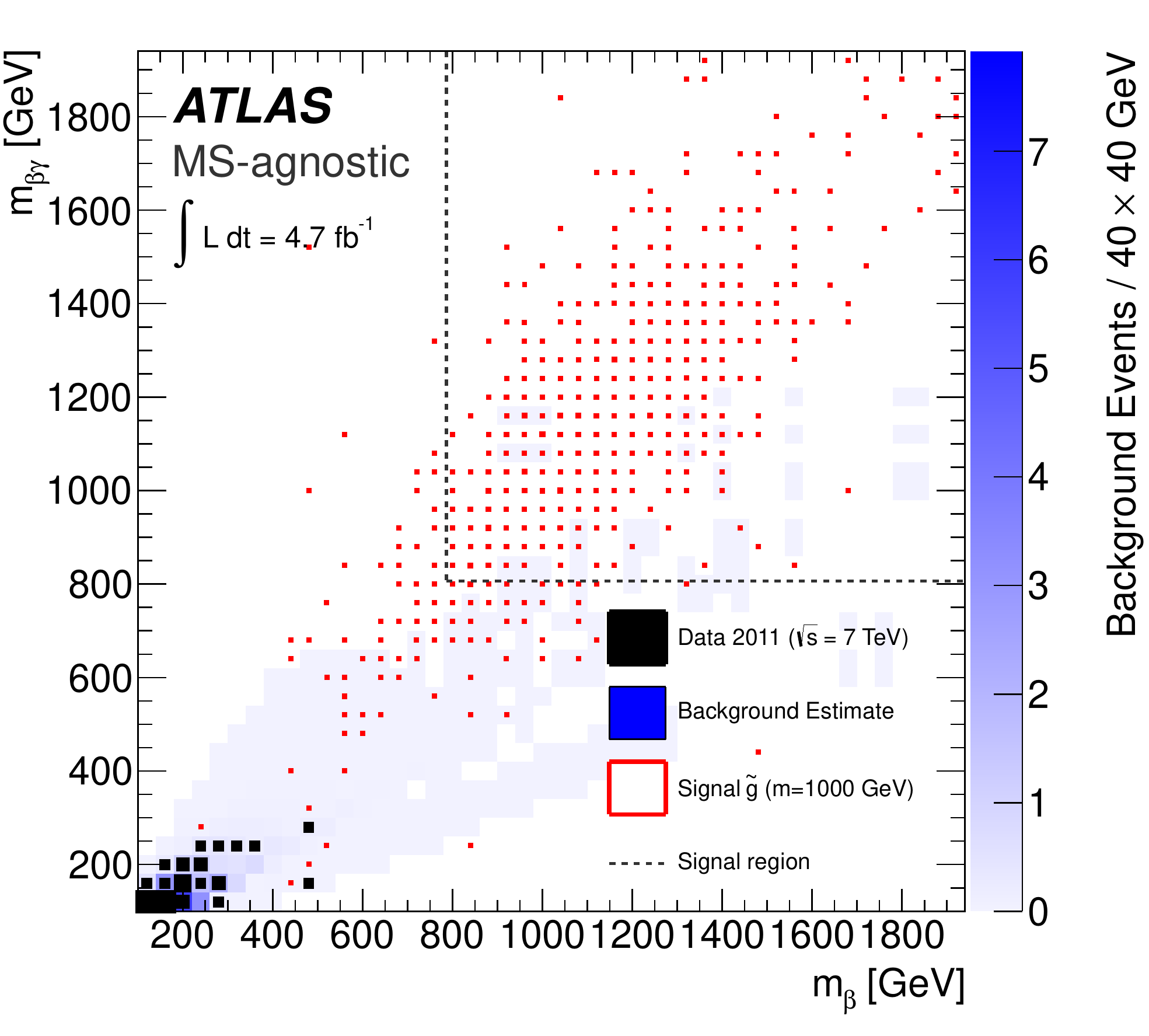}
  \caption{Examples of the $m_{\beta \gamma}$ vs. $m_{\beta}$ distributions for gluino \rhads with masses of 300, 700, 900 and 1000~\GeV, with the MS-agnostic search. The figures show data, background and signal after $\beta$, $\beta\gamma$ and $p$ cuts. The dashed grey line highlights the counting window for each specific hypothesis, and is defined by requiring 90\% of the signal to be within the window.}
  \label{fig:mass2d_agnostic}
\end{figure*}

\begin{figure*}
  \centering
  \includegraphics[width=0.49\linewidth]{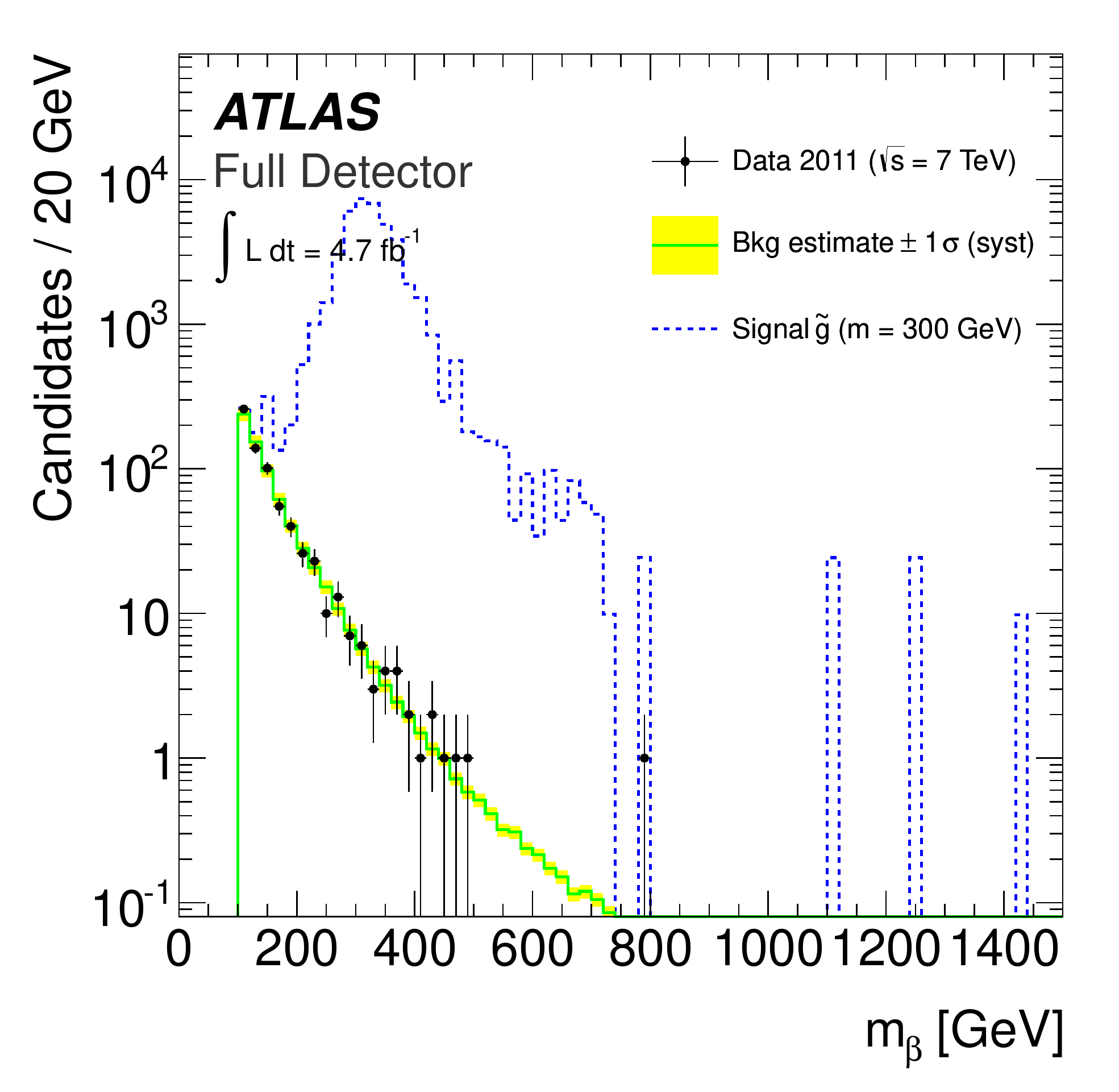}
  \includegraphics[width=0.49\linewidth]{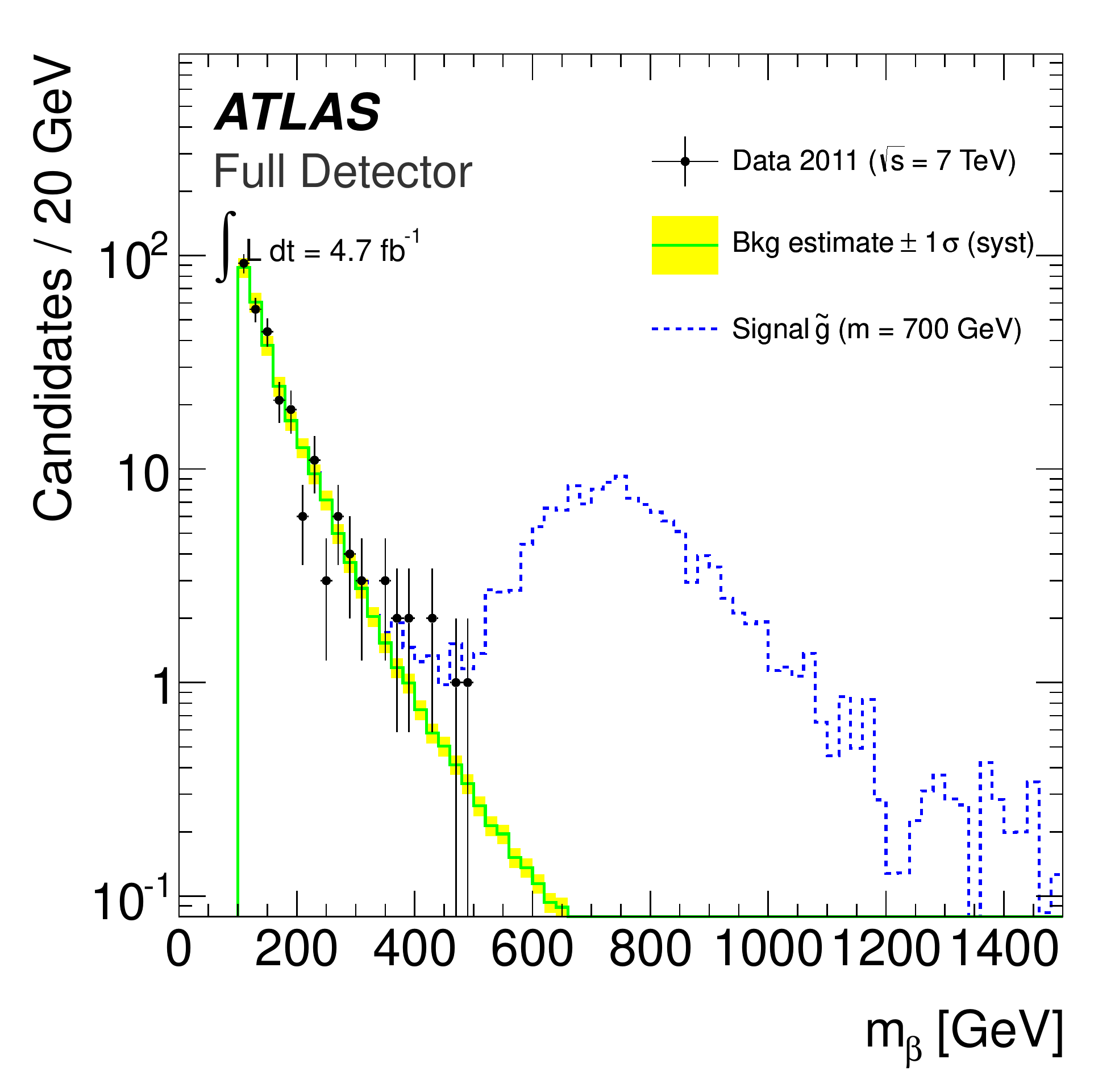}
  \includegraphics[width=0.49\linewidth]{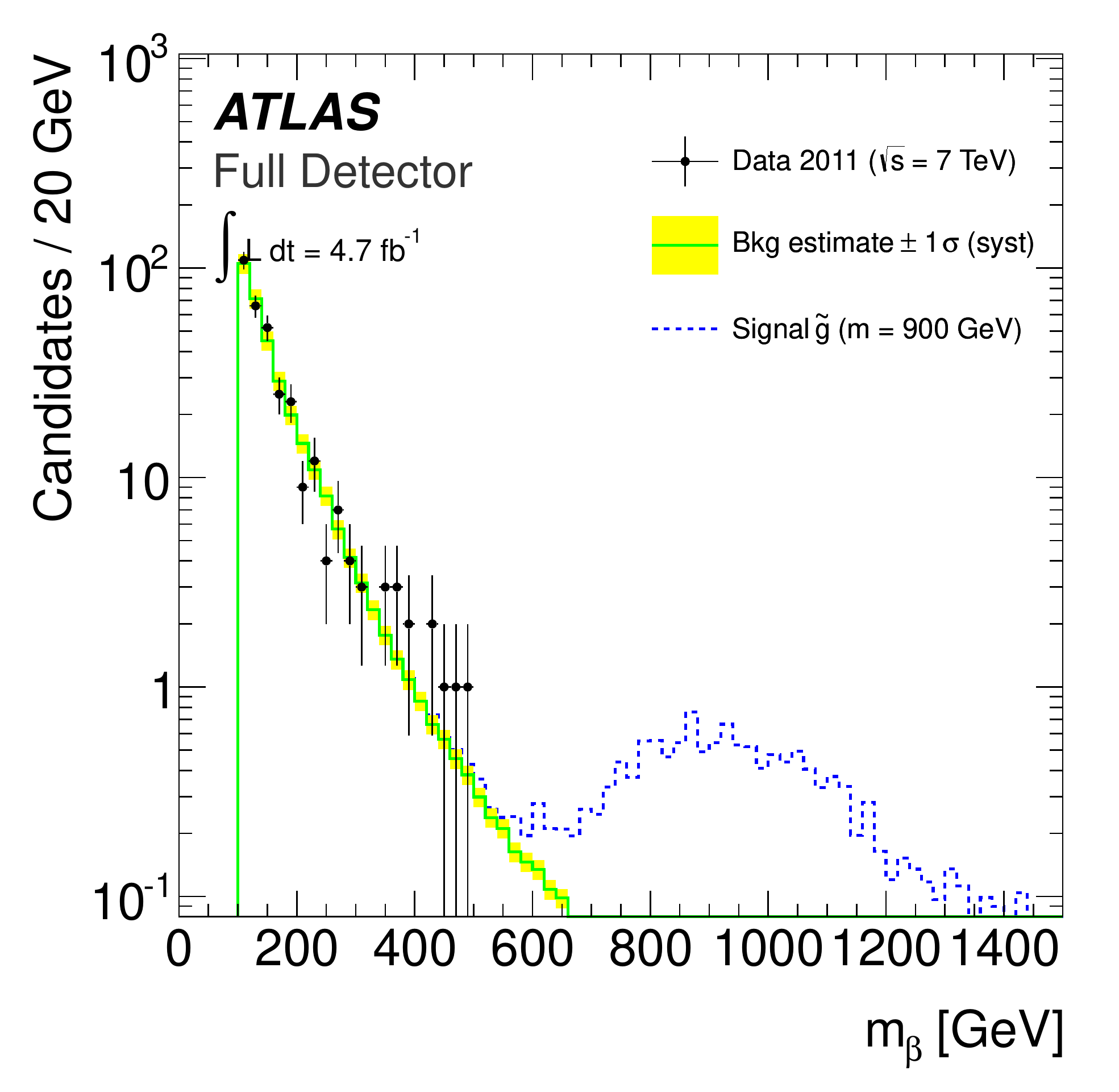}
  \includegraphics[width=0.49\linewidth]{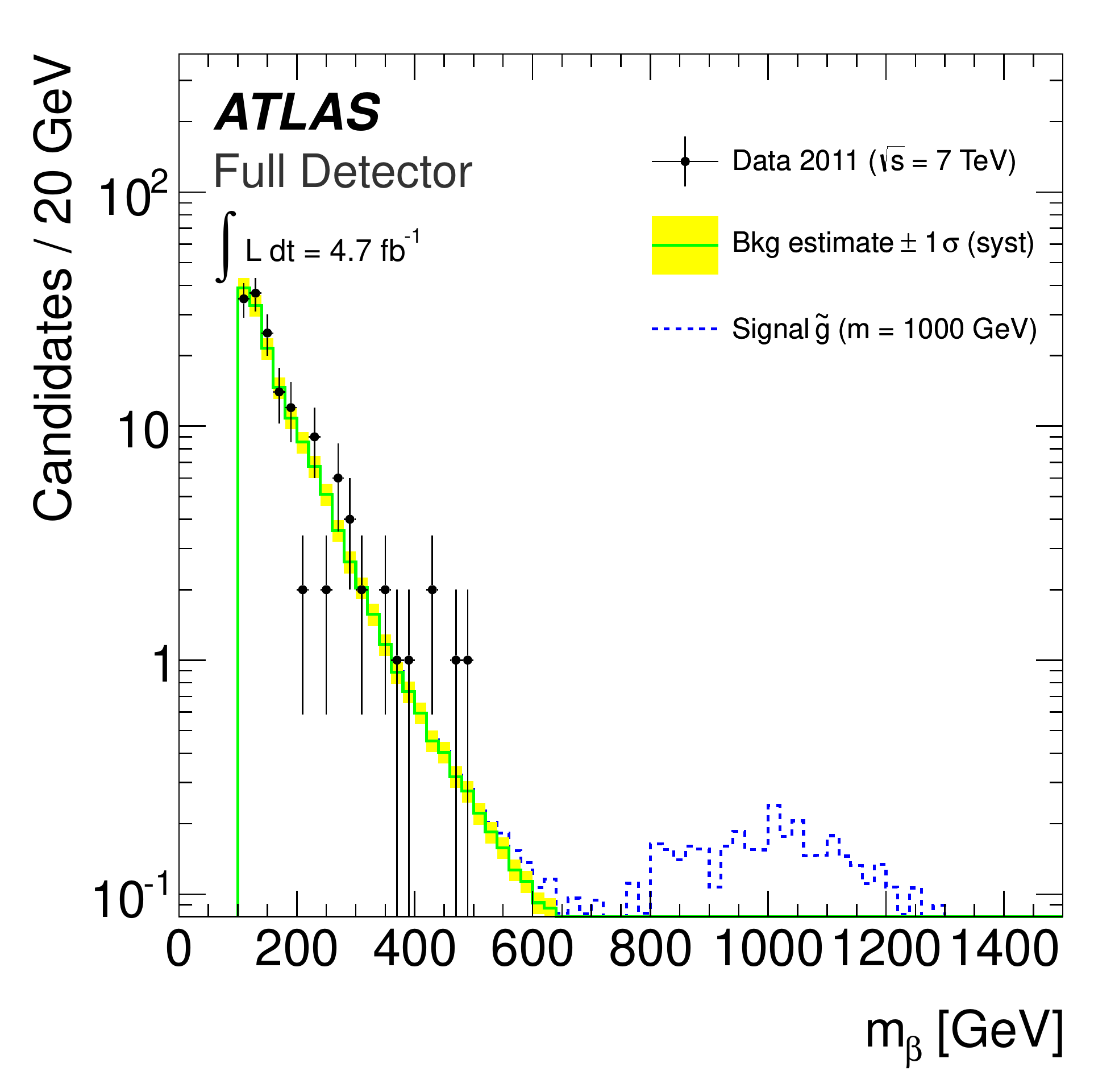}
  \caption{Observed data, background estimate and expected signal in the full-detector \rhad search, for various mass hypotheses (300, 700, 900 and 1000~\GeV). The 1D histograms are projections of the 2D signal range with unconstrained $m_{\beta\gamma}$.}
  \label{fig:mass1d_full}
\end{figure*}

\begin{figure*}
  \centering
  \includegraphics[width=0.49\linewidth]{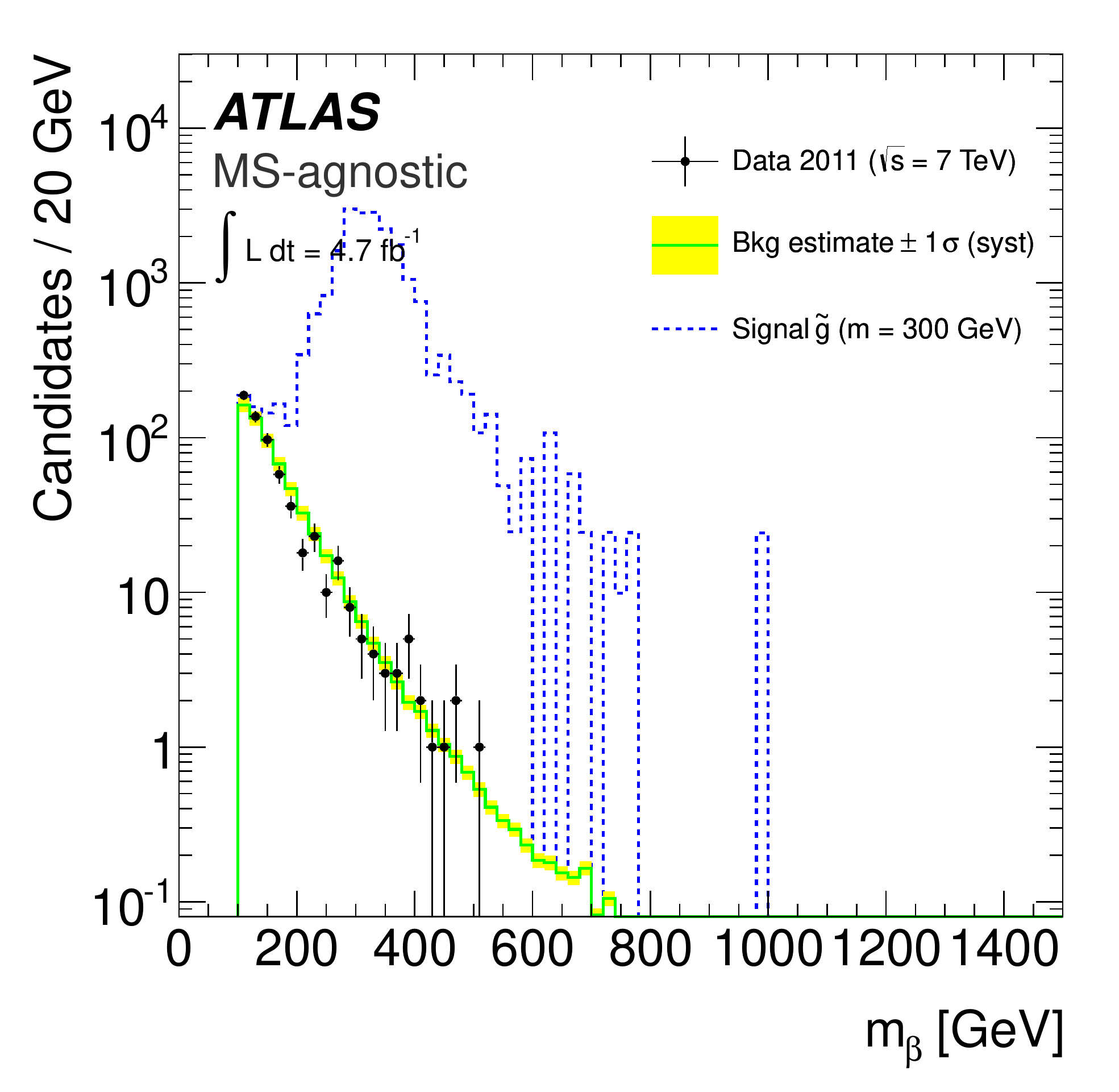}
  \includegraphics[width=0.49\linewidth]{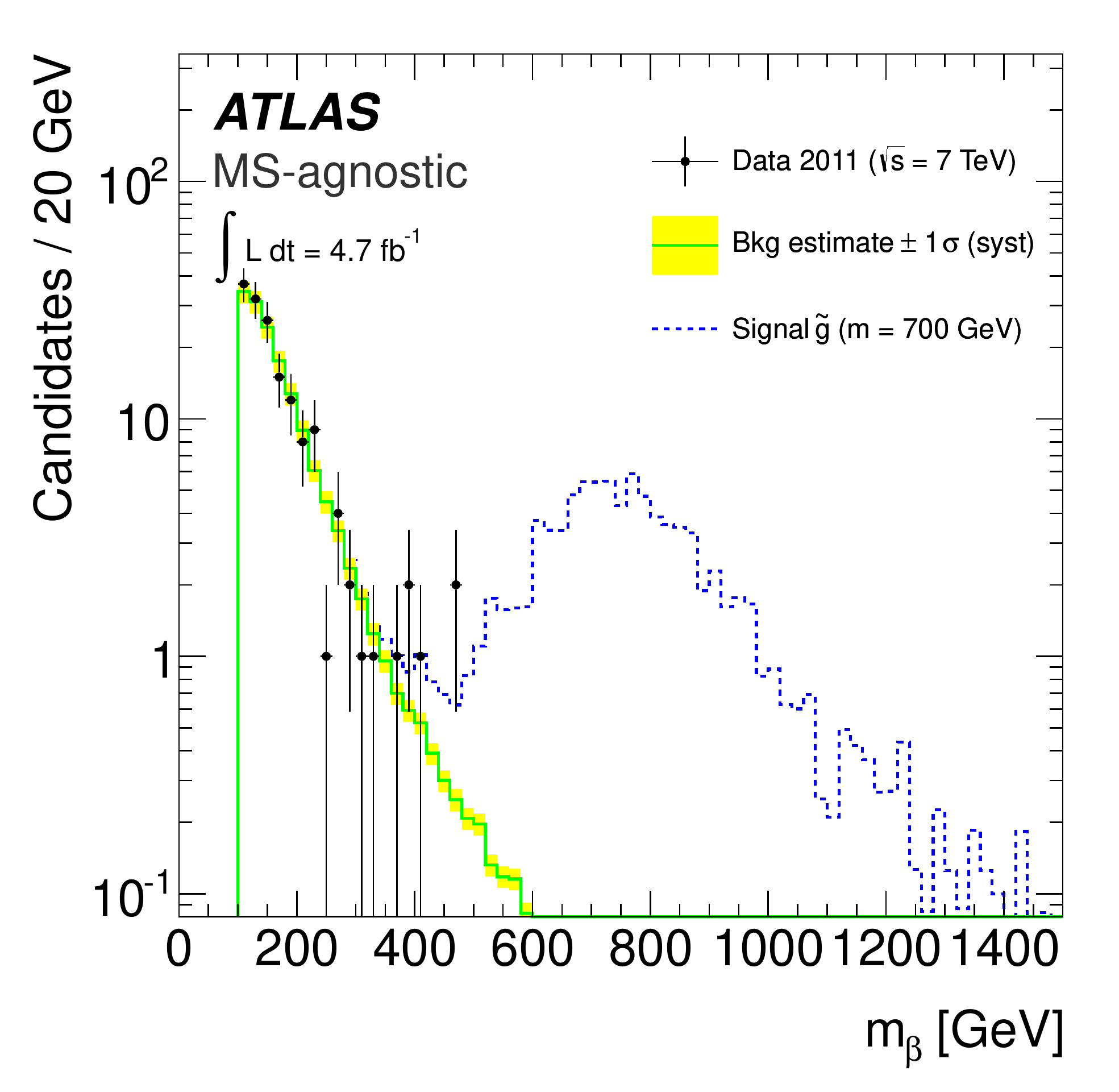}
  \includegraphics[width=0.49\linewidth]{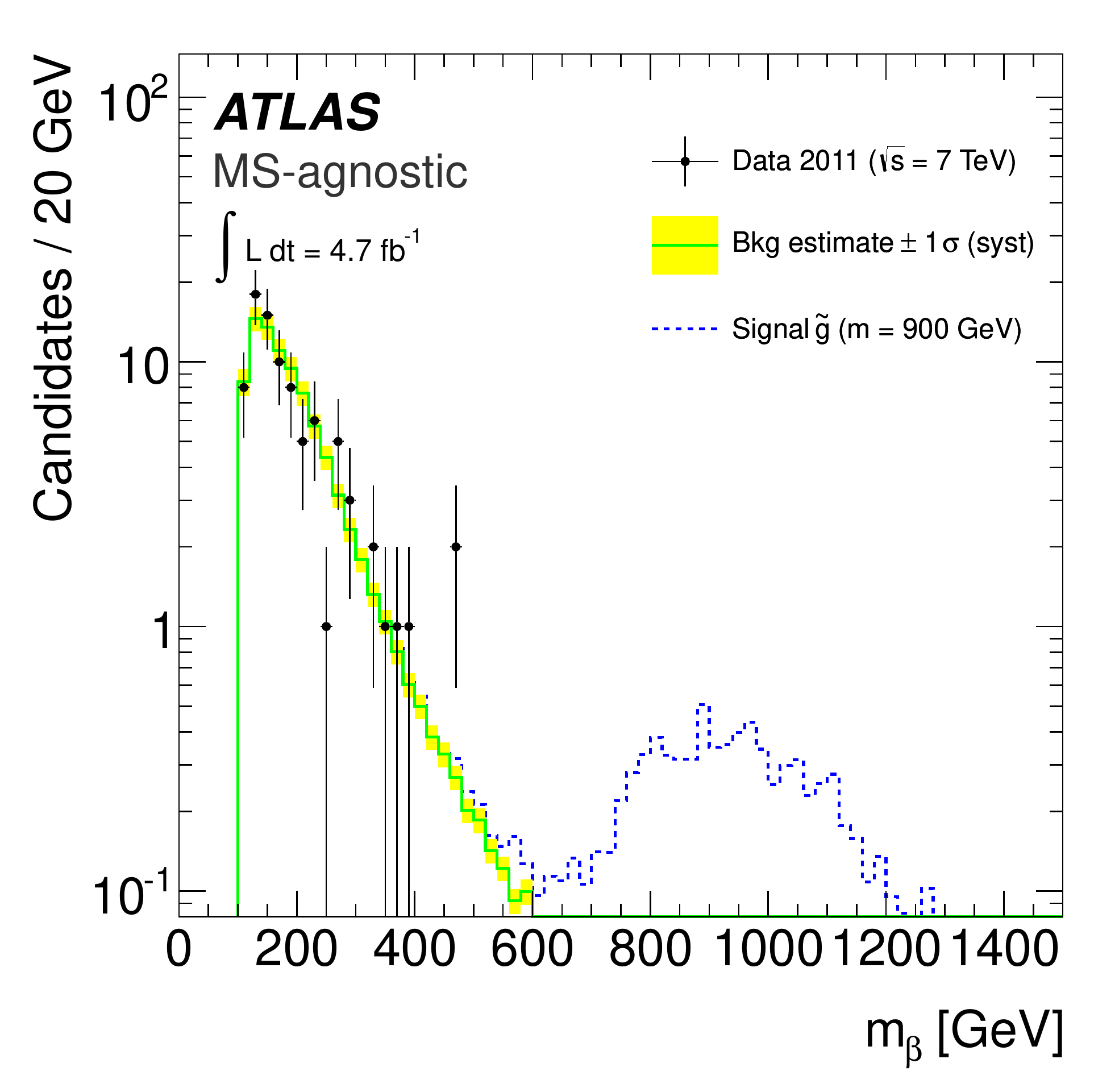}
  \includegraphics[width=0.49\linewidth]{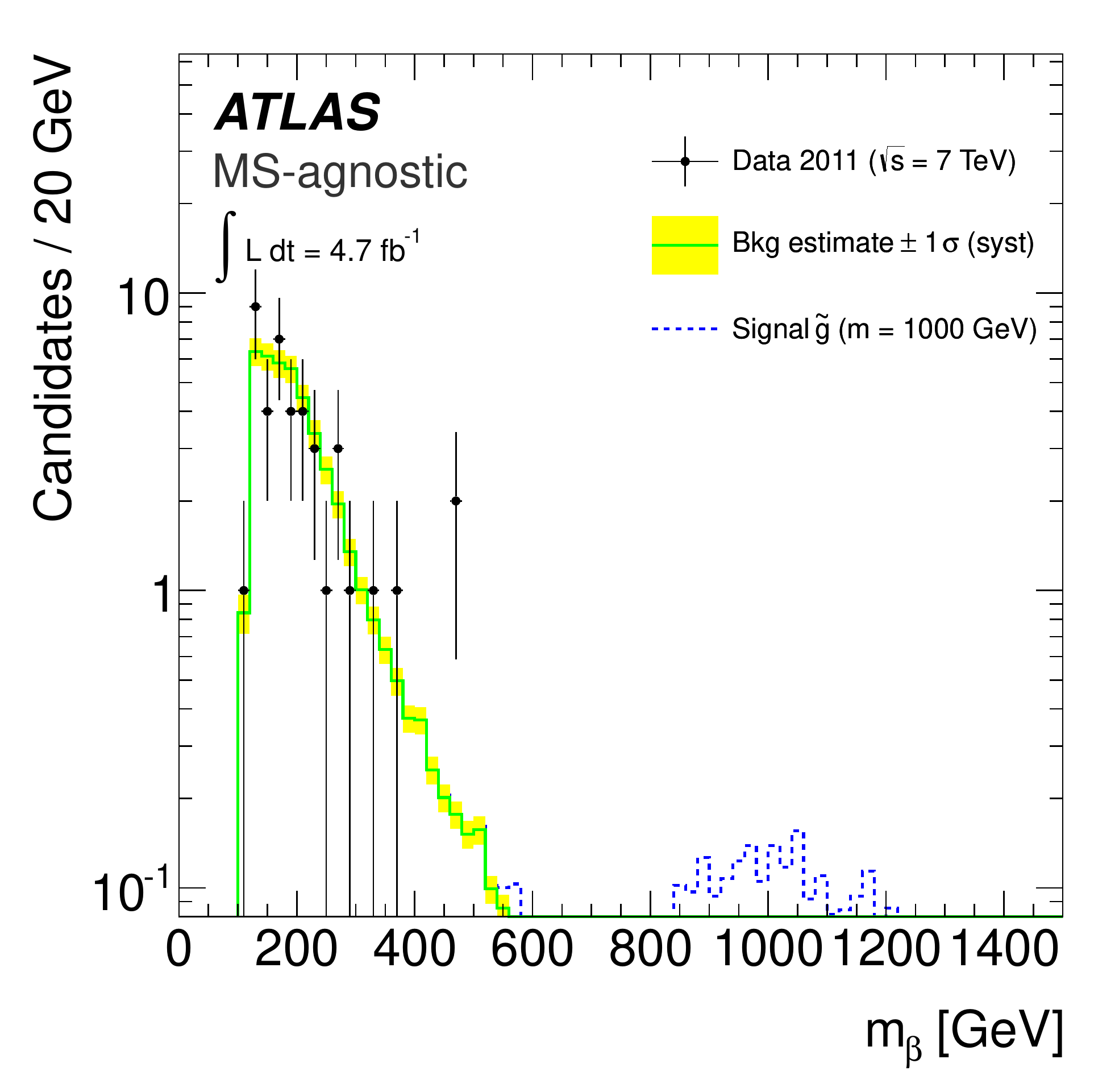}
  \caption{Observed data, background estimate and expected signal in the MS-agnostic \rhad search, for various mass hypotheses (300, 700, 900 and 1000~\GeV). The 1D histograms are projections of the 2D signal range with unconstrained $m_{\beta\gamma}$.}
  \label{fig:mass1d_agnostic}
\end{figure*}

\begin{figure*}
  \centering
  \includegraphics[width=0.49\linewidth]{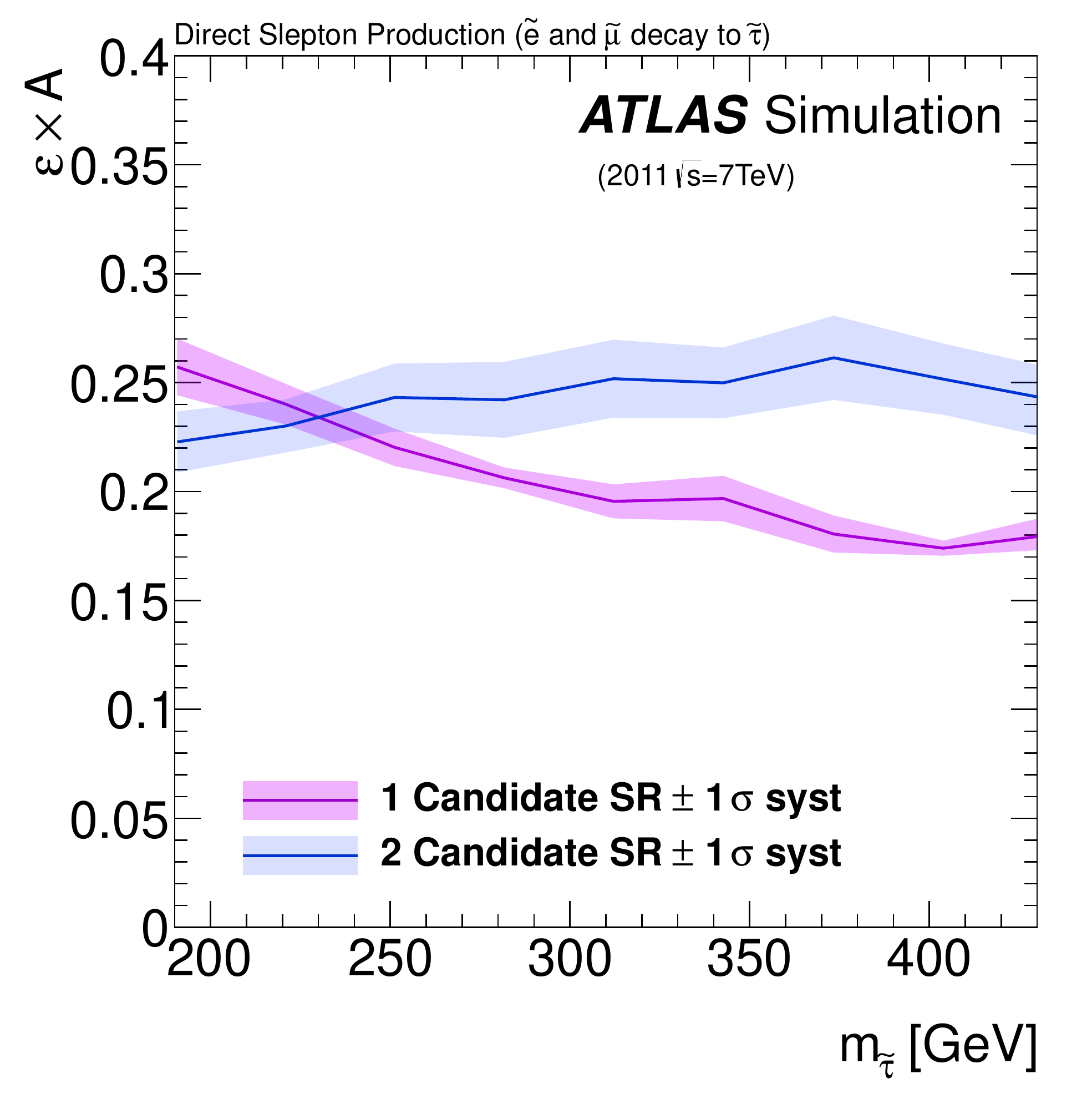}
  \caption{Efficiency times acceptance for directly produced slepton events in the one-candidate and two-candidate signal regions as a function of the stau mass.}
  \label{fig:efficiencies_slepton}
\end{figure*}

\begin{figure*}
  \centering
  \includegraphics[width=0.49\linewidth]{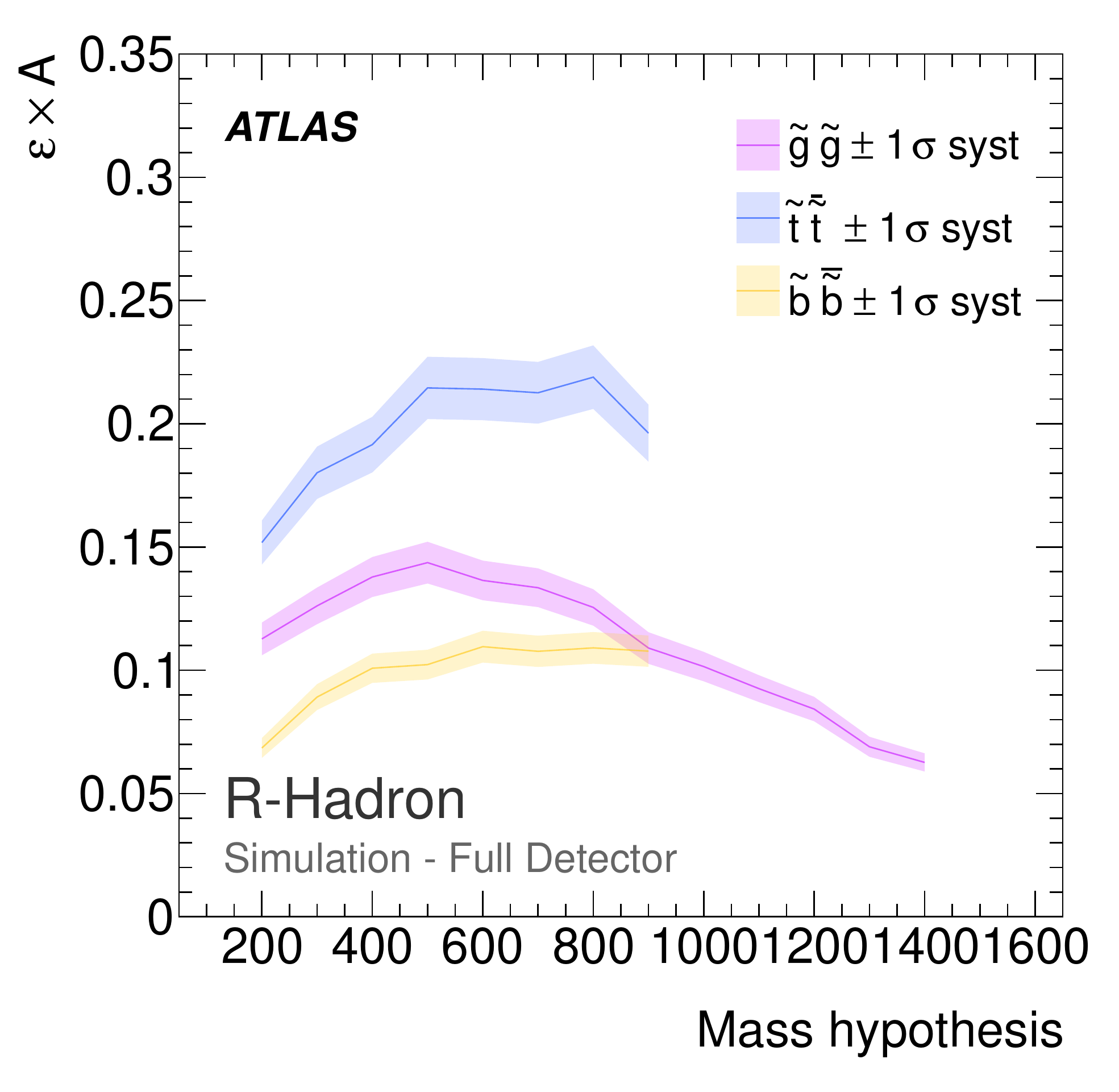}
  \includegraphics[width=0.49\linewidth]{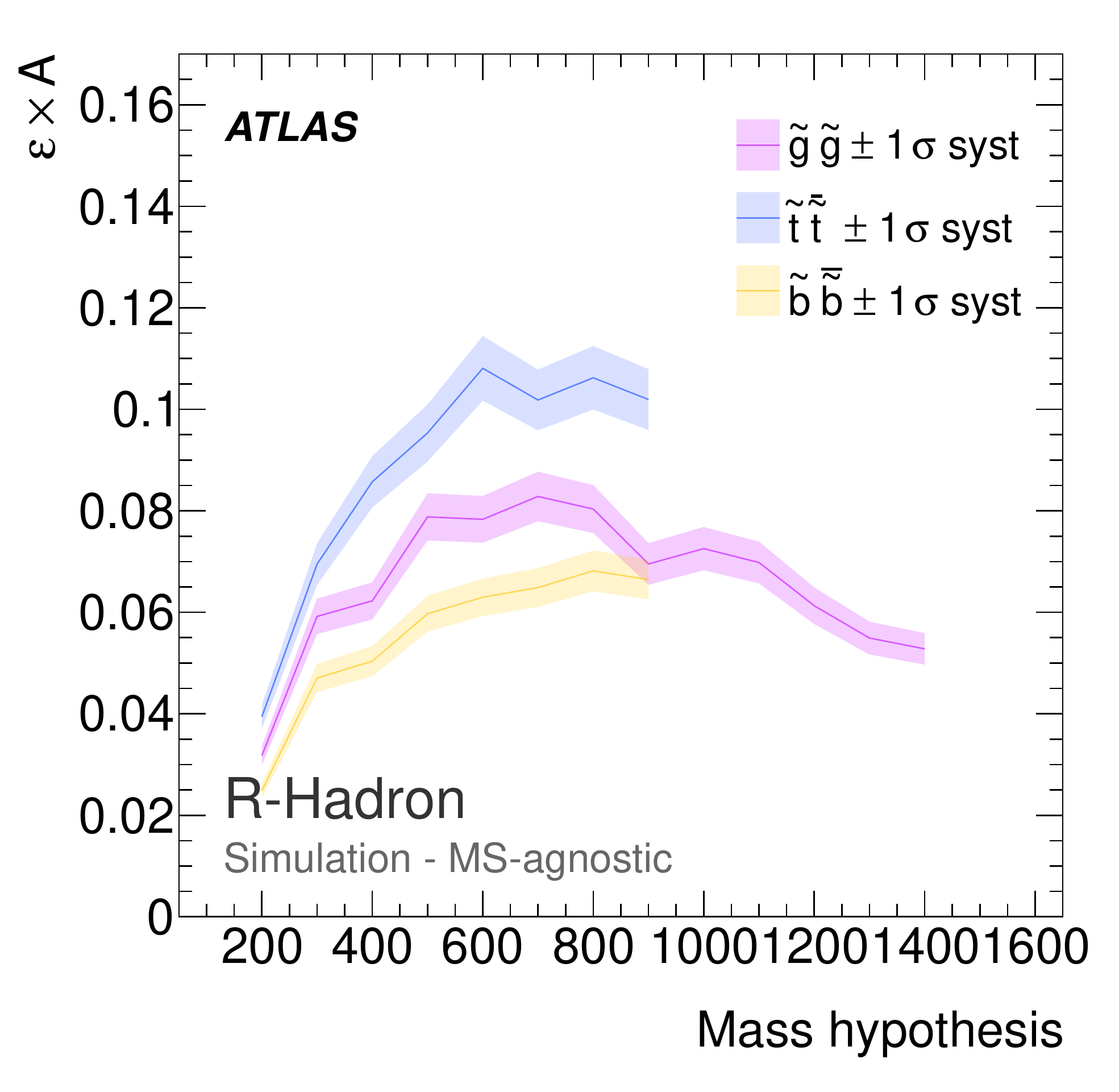}
  \caption{Efficiency times acceptance for the tested \rhad mass hypotheses. Each point on the x-axis represent a hypothesis mass, the efficiency for a given hypothesis is based on the number of signal events in the counting region after cuts, divided by the number of generated events. The band represents the 11.4\% systematic uncertainty on the efficiency.}
  \label{fig:efficiencies_rhad}
\end{figure*}

\begin{figure*}
  \centering
  \includegraphics[width=0.49\linewidth]{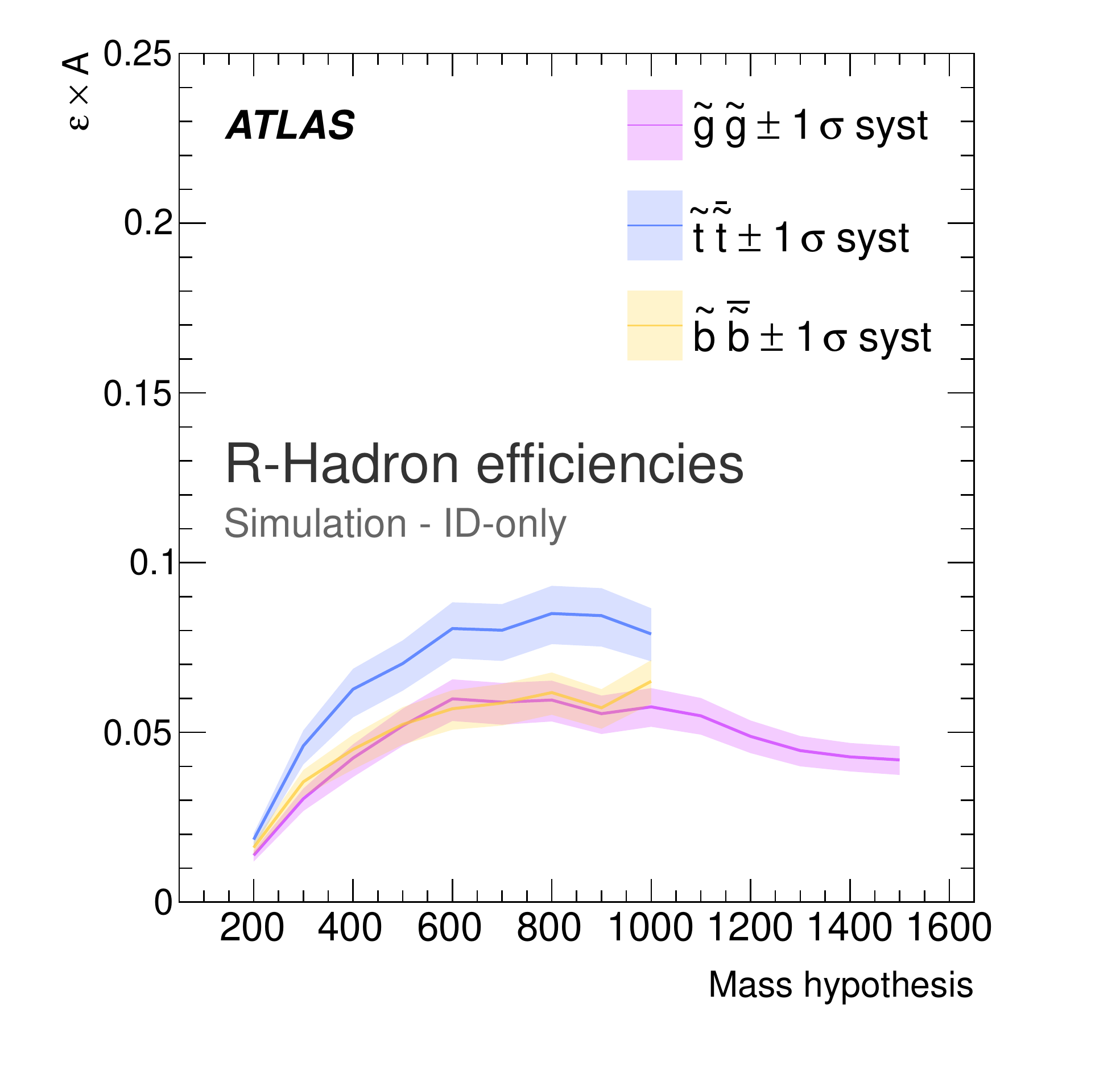}
  \caption{Efficiency times acceptance for the tested \rhad mass hypotheses. Each point on the x-axis represent a hypothesis mass, the efficiency for a given hypothesis is based on the number of signal events in the counting region after cuts, divided by the number of generated events. The band represents the systematic uncertainty on the efficiency, varying with mass as specified in Table~\ref{tab:systematics}.}
  \label{fig:efficiencies_rhadID}
\end{figure*}

\begin{figure*}
  \centering
  \includegraphics[width=0.49\linewidth]{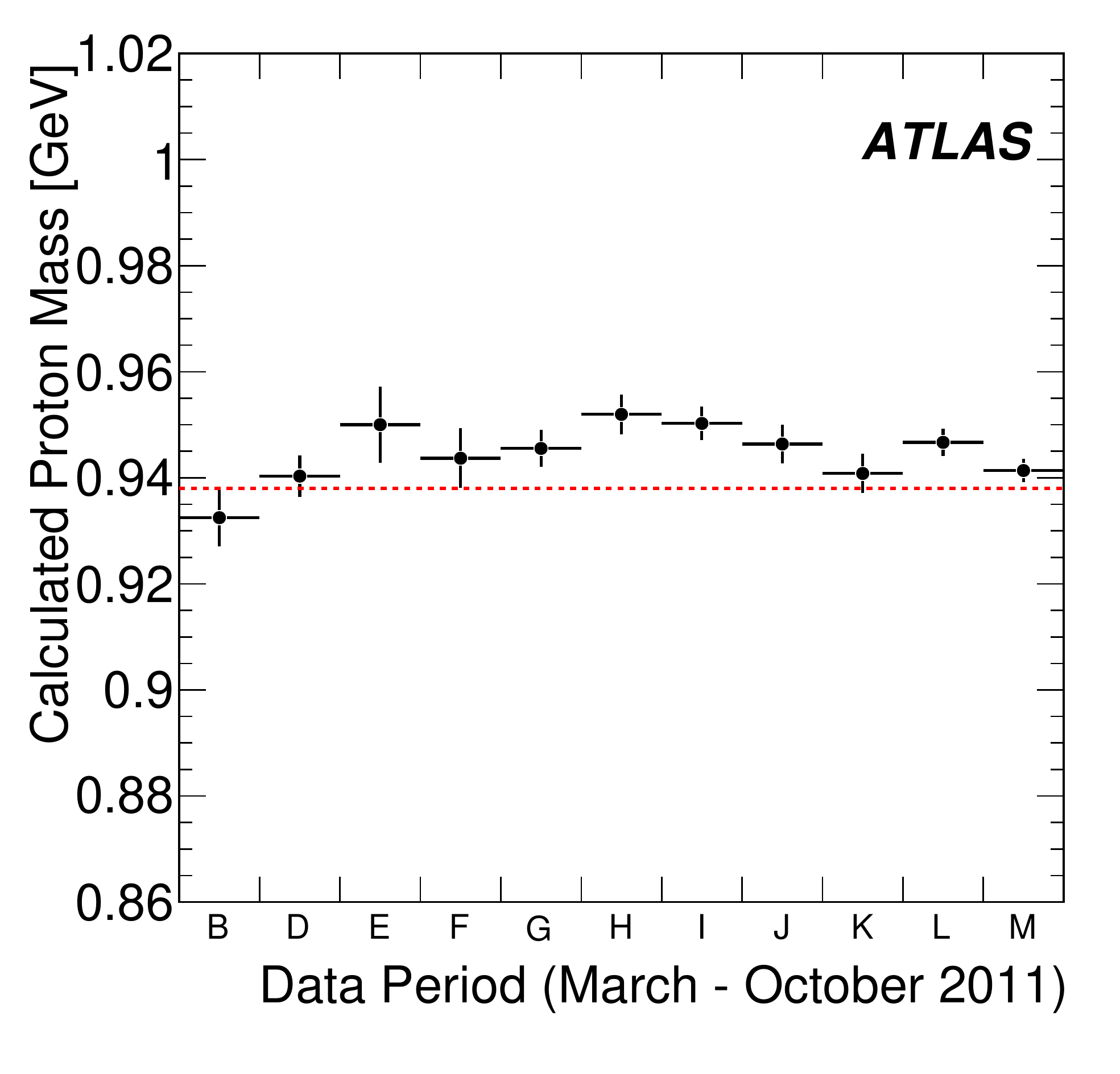}
  \caption{Proton mass calculated from the momentum measured in the inner detector and the specific energy loss measured in the pixel detector. The data, covering the 2011 statistics considered for this analysis, are subdivided in periods of similar data-taking conditions to illustrate the stability of the method. The red horizontal line represents the nominal proton mass value.}
  \label{fig:protonMassID}
\end{figure*}

\begin{figure*}
  \centering
  \includegraphics[width=0.49\linewidth]{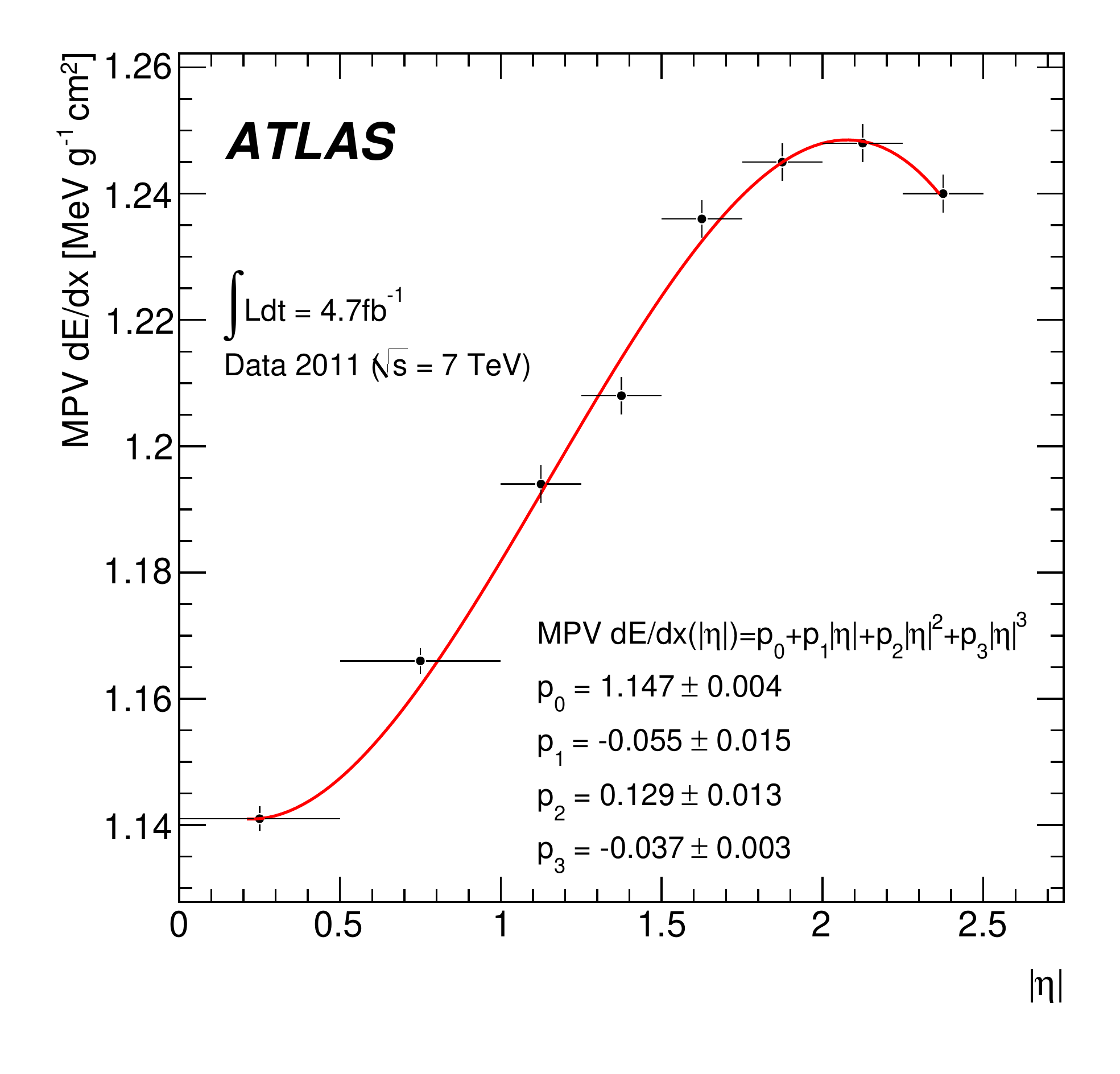}
  \caption{Plot of the \dedx most probable values (MPVs) obtained from data, selected as in the ``low-ionisation'' sample (see Section~\ref{subsec:RHbackground}), as a function of $\eta$ overlapped with a fit obtained using a third order polynomial.}
  \label{fig:MPVeta}
\end{figure*}

Appendix consists of Figs.~\ref{fig:dEdx_colour}--\ref{fig:MPVeta} and Tables~\ref{tab:aux2}--\ref{tab:aux9}.

\clearpage

\begin{table*}
  \centering
  \includegraphics[width=\linewidth]{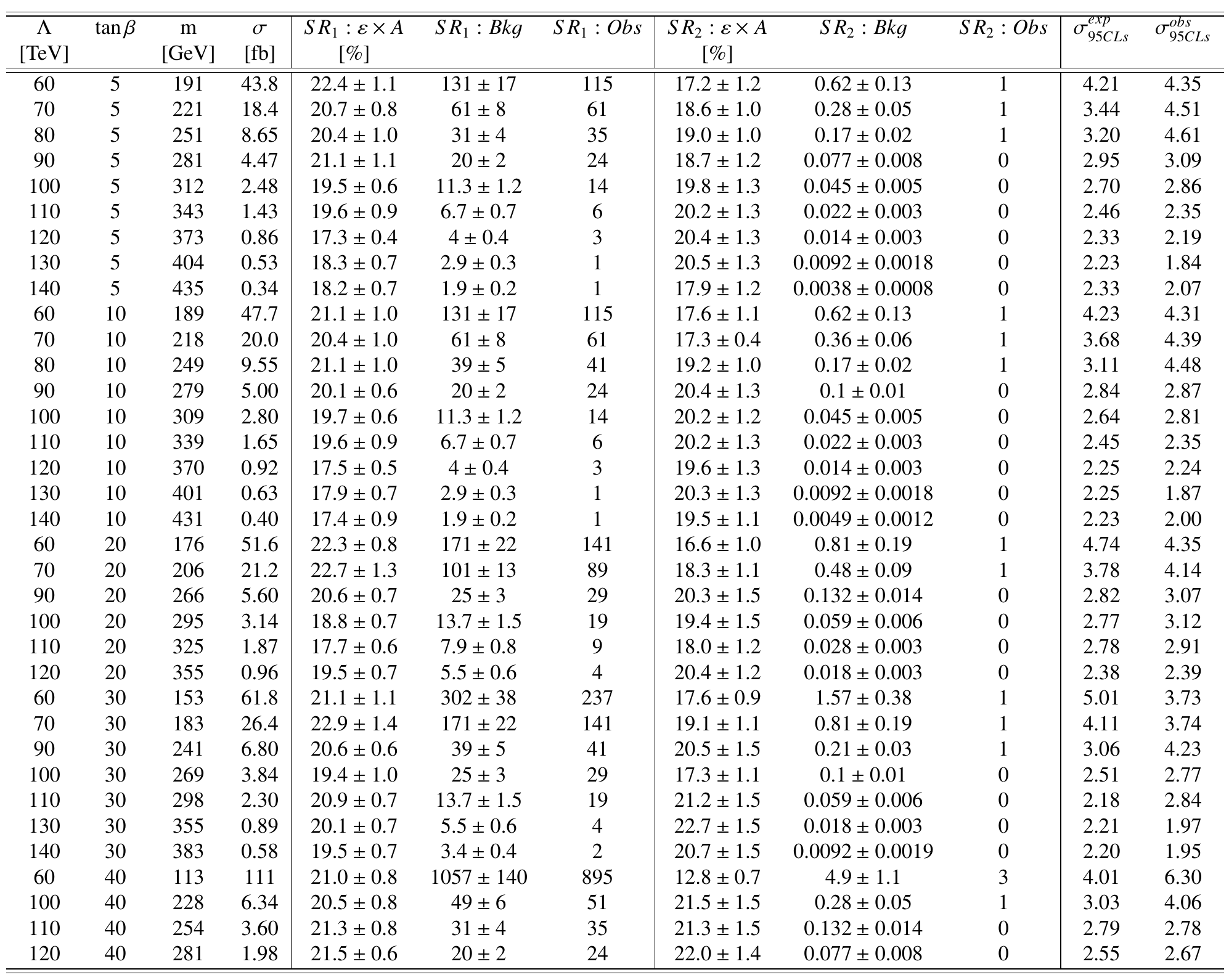}
  \caption{Signal efficiency, number of expected background events, events observed in data, expected and observed limit for the one candidate ($SR_1$) and two candidates ($SR_2$) signal regions in various GMSB scenarios.}
  \label{tab:aux2}
\end{table*}

\begin{table*}
  \centering
  \includegraphics[width=\linewidth]{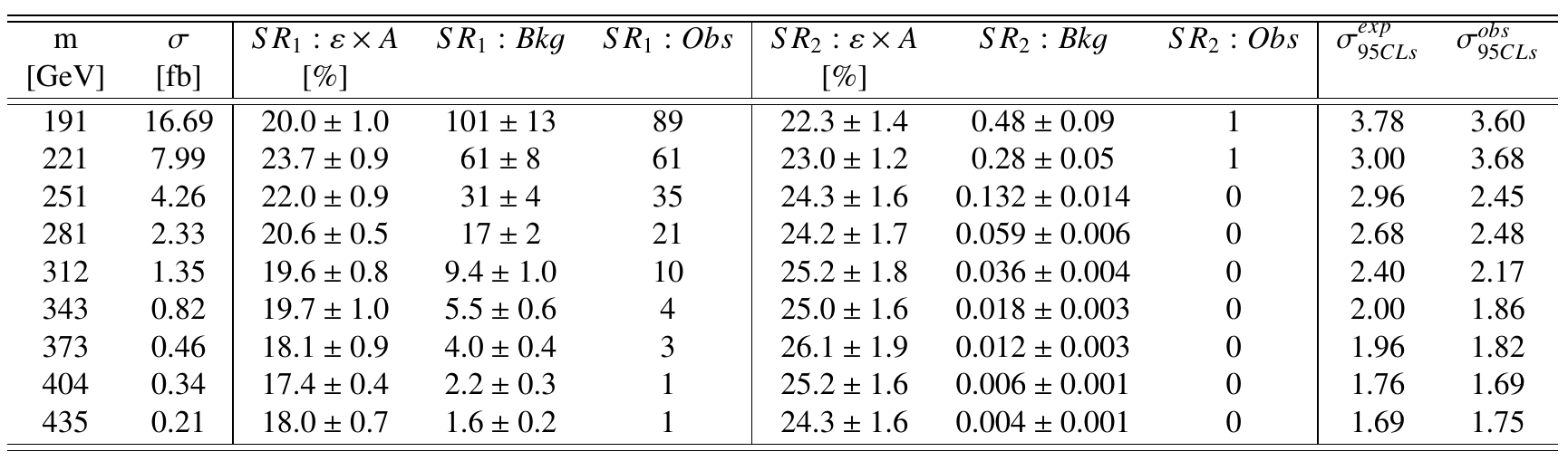}
  \caption{Signal efficiency, number of expected background events, events observed in data, expected and observed limit for the one candidate ($SR_1$) and two candidates ($SR_2$) signal regions for direct slepton production.}
\end{table*}

\begin{table*}
  \centering
  \input{tabaux_04.tex}
  \caption{Signal efficiency, number of expected background events, events observed in data, expected (with $1\sigma$ uncertainties) and observed limit for various gluino \rhad mass hypotheses in the full-detector search.}
\end{table*}

\begin{table*}
  \centering
  \input{tabaux_05.tex}
  \caption{Signal efficiency, number of expected background events, events observed in data, expected (with $1\sigma$ uncertainties) and observed limit for various sbottom \rhad mass hypotheses in the full-detector search.}
\end{table*}

\begin{table*}
  \centering
  \input{tabaux_06.tex}
  \caption{Signal efficiency, number of expected background events, events observed in data, expected (with $1\sigma$ uncertainties) and observed limit for various stop \rhad mass hypotheses in the full-detector search.}
\end{table*}

\begin{table*}
  \centering
  \input{tabaux_07.tex}
  \caption{Signal efficiency, number of expected background events, events observed in data, expected (with $1\sigma$ uncertainties) and observed limit for various gluino \rhad mass hypotheses in the MS-agnostic search.}
\end{table*}

\begin{table*}
  \centering
  \input{tabaux_08.tex}
  \caption{Signal efficiency, number of expected background events, events observed in data, expected (with $1\sigma$ uncertainties) and observed limit for various sbottom \rhad mass hypotheses in the MS-agnostic search.}
\end{table*}

\begin{table*}
  \centering
   \input{tabaux_09.tex}
  \caption{Signal efficiency, number of expected background events, events observed in data, expected (with $1\sigma$ uncertainties) and observed limit for various stop \rhad mass hypotheses in the MS-agnostic search.}
  \label{tab:aux9}
\end{table*}

\onecolumn
\clearpage
\input{atlas_authlist}
\end{document}

%% file: Acknowledgements.tex
\section{Acknowledgements}

We thank CERN for the very successful operation of the LHC, as well as the
support staff from our institutions without whom ATLAS could not be
operated efficiently.

We acknowledge the support of ANPCyT, Argentina; YerPhI, Armenia; ARC,
Australia; BMWF and FWF, Austria; ANAS, Azerbaijan; SSTC, Belarus; CNPq and FAPESP,
Brazil; NSERC, NRC and CFI, Canada; CERN; CONICYT, Chile; CAS, MOST and NSFC,
China; COLCIENCIAS, Colombia; MSMT CR, MPO CR and VSC CR, Czech Republic;
DNRF, DNSRC and Lundbeck Foundation, Denmark; EPLANET, ERC and NSRF, European Union;
IN2P3-CNRS, CEA-DSM/IRFU, France; GNSF, Georgia; BMBF, DFG, HGF, MPG and AvH
Foundation, Germany; GSRT and NSRF, Greece; ISF, MINERVA, GIF, DIP and Benoziyo Center,
Israel; INFN, Italy; MEXT and JSPS, Japan; CNRST, Morocco; FOM and NWO,
Netherlands; BRF and RCN, Norway; MNiSW, Poland; GRICES and FCT, Portugal; MERYS
(MECTS), Romania; MES of Russia and ROSATOM, Russian Federation; JINR; MSTD,
Serbia; MSSR, Slovakia; ARRS and MVZT, Slovenia; DST/NRF, South Africa;
MICINN, Spain; SRC and Wallenberg Foundation, Sweden; SER, SNSF and Cantons of
Bern and Geneva, Switzerland; NSC, Taiwan; TAEK, Turkey; STFC, the Royal
Society and Leverhulme Trust, United Kingdom; DOE and NSF, United States of
America.

The crucial computing support from all WLCG partners is acknowledged
gratefully, in particular from CERN and the ATLAS Tier-1 facilities at
TRIUMF (Canada), NDGF (Denmark, Norway, Sweden), CC-IN2P3 (France),
KIT/GridKA (Germany), INFN-CNAF (Italy), NL-T1 (Netherlands), PIC (Spain),
ASGC (Taiwan), RAL (UK) and BNL (USA) and in the Tier-2 facilities
worldwide.

%% file: tabaux_04.tex
\begin{tabular}{c|ccc|ccc|c}
\hline\noalign{\smallskip}
$m_{\tilde{g}}$	& Signal eff.	 & Exp. bkg	 & Obs. data	 & Exp. limit	 & $+ \sigma$	 & $- \sigma$	 & Obs. limit \\
$[\mathrm{GeV}]$	 & [\%]	 & [events]	 & [events]	 & [fb]	 & [fb]	 & [fb]	 & [fb] \\

\noalign{\smallskip}\hline\noalign{\smallskip}
$200$		 & $11.3$	 & $131.2$	 & $135$	 & $89$	 & $128$	 & $65$	 & $94$\\
$300$		 & $12.6$	 & $18.7$	 & $15$	 & $19.0$	 & $27.7$	 & $13.6$	 & $14.6$\\
$400$		 & $13.8$	 & $5.2$	 & $3$	 & $9.5$	 & $14.0$	 & $6.6$	 & $6.9$\\
$500$		 & $14.4$	 & $1.8$	 & $2$	 & $6.3$	 & $9.0$	 & $4.8$	 & $6.9$\\
$600$		 & $13.6$	 & $0.6$	 & $1$	 & $4.7$	 & $6.8$	 & $4.3$	 & $6.1$\\
$700$		 & $13.4$	 & $0.3$	 & $1$	 & $4.7$	 & $6.1$	 & $4.4$	 & $6.3$\\
$800$		 & $12.6$	 & $0.1$	 & $1$	 & $4.9$	 & $5.3$	 & $4.6$	 & $6.8$\\
$900$		 & $10.9$	 & $0.2$	 & $1$	 & $5.7$	 & $6.1$	 & $5.4$	 & $7.8$\\
$1000$		 & $10.2$	 & $0.1$	 & $1$	 & $6.1$	 & $6.4$	 & $5.7$	 & $8.5$\\
$1100$		 & $9.3$	 & $0.0$	 & $1$	 & $6.7$	 & $7.0$	 & $6.2$	 & $9.3$\\
$1200$		 & $8.4$	 & $0.0$	 & $1$	 & $7.3$	 & $7.7$	 & $6.9$	 & $10.2$\\
$1300$		 & $6.9$	 & $0.0$	 & $0$	 & $8.7$	 & $9.5$	 & $7.9$	 & $8.6$\\
$1400$		 & $6.3$	 & $0.0$	 & $0$	 & $9.1$	 & $9.7$	 & $8.5$	 & $9.1$\\

\noalign{\smallskip}\hline\noalign{\smallskip}
\end{tabular}

%% file: tabaux_05.tex
\begin{tabular}{c|ccc|ccc|c}
\hline\noalign{\smallskip}
$m_{\tilde{b}}$	& Signal eff.	 & Exp. bkg	 & Obs. data	 & Exp. limit	 & $+ \sigma$	 & $- \sigma$	 & Obs. limit \\
$[\mathrm{GeV}]$	 & [\%]	 & [events]	 & [events]	 & [fb]	 & [fb]	 & [fb]	 & [fb] \\
 
\noalign{\smallskip}\hline\noalign{\smallskip}
$200$		 & $6.8$	 & $97.1$	 & $94$	 & $114$	 & $164$	 & $80$	 & $107$\\
$300$		 & $8.9$	 & $22.0$	 & $18$	 & $29.9$	 & $43.2$	 & $20.8$	 & $23.0$\\
$400$		 & $10.1$	 & $6.3$	 & $5$	 & $14.2$	 & $20.5$	 & $9.9$	 & $12.0$\\
$500$		 & $10.2$	 & $1.4$	 & $2$	 & $8.2$	 & $11.9$	 & $6.3$	 & $9.8$\\
$600$		 & $11.0$	 & $0.9$	 & $2$	 & $6.7$	 & $9.5$	 & $5.5$	 & $9.8$\\
$700$		 & $10.8$	 & $0.3$	 & $1$	 & $5.9$	 & $7.8$	 & $5.3$	 & $7.8$\\
$800$		 & $10.9$	 & $0.3$	 & $1$	 & $5.8$	 & $7.3$	 & $5.4$	 & $7.7$\\
$900$		 & $10.8$	 & $0.2$	 & $1$	 & $5.7$	 & $7.2$	 & $5.5$	 & $7.9$\\

\noalign{\smallskip}\hline\noalign{\smallskip}
\end{tabular}

%% file: tabaux_06.tex
\begin{tabular}{c|ccc|ccc|c}
\hline\noalign{\smallskip}
$m_{\tilde{t}}$	& Signal eff.	 & Exp. bkg	 & Obs. data	 & Exp. limit	 & $+ \sigma$	 & $- \sigma$	 & Obs. limit \\
$[\mathrm{GeV}]$	 & [\%]	 & [events]	 & [events]	 & [fb]	 & [fb]	 & [fb]	 & [fb] \\
 
\noalign{\smallskip}\hline\noalign{\smallskip}
$200$		 & $15.2$	 & $128.4$	 & $122$	 & $64$	 & $94$	 & $46$	 & $58$\\
$300$		 & $18.0$	 & $21.9$	 & $18$	 & $14.7$	 & $21.1$	 & $10.3$	 & $11.5$\\
$400$		 & $19.2$	 & $4.6$	 & $4$	 & $6.4$	 & $9.5$	 & $4.8$	 & $6.1$\\
$500$		 & $21.5$	 & $2.0$	 & $2$	 & $4.4$	 & $6.4$	 & $3.4$	 & $4.5$\\
$600$		 & $21.4$	 & $0.6$	 & $2$	 & $3.0$	 & $4.4$	 & $2.9$	 & $5.2$\\
$700$		 & $21.3$	 & $0.4$	 & $1$	 & $3.0$	 & $4.2$	 & $2.7$	 & $3.9$\\
$800$		 & $21.9$	 & $0.3$	 & $1$	 & $2.8$	 & $3.6$	 & $2.6$	 & $3.8$\\
$900$		 & $19.6$	 & $0.2$	 & $1$	 & $3.2$	 & $3.8$	 & $3.0$	 & $4.3$\\

\noalign{\smallskip}\hline\noalign{\smallskip}
\end{tabular}

%% file: tabaux_07.tex
\begin{tabular}{c|ccc|ccc|c}
\hline\noalign{\smallskip}
$m_{\tilde{g}}$	& Signal eff.	 & Exp. bkg	 & Obs. data	 & Exp. limit	 & $+ \sigma$	 & $- \sigma$	 & Obs. limit \\
$[\mathrm{GeV}]$	 & [\%]	 & [events]	 & [events]	 & [fb]	 & [fb]	 & [fb]	 & [fb] \\
 
\noalign{\smallskip}\hline\noalign{\smallskip}
$200$		 & $3.2$	 & $45.2$	 & $47$	 & $138$	 & $199$	 & $100$	 & $146$\\
$300$		 & $5.9$	 & $22.2$	 & $17$	 & $44$	 & $59$	 & $32$	 & $33$\\
$400$		 & $6.2$	 & $2.1$	 & $0$	 & $11.7$	 & $14.4$	 & $9.5$	 & $9.3$\\
$500$		 & $7.9$	 & $0.7$	 & $0$	 & $7.9$	 & $9.8$	 & $7.0$	 & $7.1$\\
$600$		 & $7.8$	 & $0.7$	 & $0$	 & $8.0$	 & $9.3$	 & $7.0$	 & $7.4$\\
$700$		 & $8.3$	 & $1.1$	 & $0$	 & $7.6$	 & $9.7$	 & $6.5$	 & $6.6$\\
$800$		 & $8.0$	 & $0.8$	 & $0$	 & $9.5$	 & $10.9$	 & $7.4$	 & $11.2$\\
$900$		 & $7.0$	 & $0.0$	 & $0$	 & $8.2$	 & $8.7$	 & $7.8$	 & $8.2$\\
$1000$		 & $7.3$	 & $0.2$	 & $0$	 & $8.1$	 & $8.7$	 & $7.6$	 & $8.0$\\
$1100$		 & $7.0$	 & $0.4$	 & $0$	 & $8.3$	 & $9.9$	 & $7.8$	 & $8.0$\\
$1200$		 & $6.1$	 & $0.2$	 & $0$	 & $9.8$	 & $10.6$	 & $9.2$	 & $9.6$\\
$1300$		 & $5.5$	 & $0.2$	 & $0$	 & $10.6$	 & $11.4$	 & $9.7$	 & $10.5$\\
$1400$		 & $5.3$	 & $0.0$	 & $0$	 & $10.4$	 & $10.9$	 & $9.9$	 & $10.4$\\

\noalign{\smallskip}\hline\noalign{\smallskip}
\end{tabular}

%% file: tabaux_08.tex
\begin{tabular}{c|ccc|ccc|c}
\hline\noalign{\smallskip}
$m_{\tilde{b}}$	& Signal eff.	 & Exp. bkg	 & Obs. data	 & Exp. limit	 & $+ \sigma$	 & $- \sigma$	 & Obs. limit \\
$[\mathrm{GeV}]$	 & [\%]	 & [events]	 & [events]	 & [fb]	 & [fb]	 & [fb]	 & [fb] \\
 
\noalign{\smallskip}\hline\noalign{\smallskip}
$200$		 & $2.5$	 & $106.2$	 & $103$	 & $340$	 & $490$	 & $240$	 & $310$\\
$300$		 & $4.7$	 & $25.5$	 & $17$	 & $55$	 & $67$	 & $39$	 & $38$\\
$400$		 & $5.0$	 & $2.8$	 & $2$	 & $20.6$	 & $29.9$	 & $15.0$	 & $17.9$\\
$500$		 & $6.0$	 & $0.9$	 & $0$	 & $10.6$	 & $12.0$	 & $9.1$	 & $9.6$\\
$600$		 & $6.3$	 & $0.4$	 & $0$	 & $9.4$	 & $11.6$	 & $8.6$	 & $9.0$\\
$700$		 & $6.5$	 & $0.2$	 & $0$	 & $9.1$	 & $10.6$	 & $8.4$	 & $9.0$\\
$800$		 & $6.8$	 & $0.1$	 & $0$	 & $9.3$	 & $9.8$	 & $8.8$	 & $8.7$\\
$900$		 & $6.6$	 & $0.7$	 & $0$	 & $9.1$	 & $10.7$	 & $8.2$	 & $8.5$\\

\noalign{\smallskip}\hline\noalign{\smallskip}
\end{tabular}

%% file: tabaux_09.tex
\begin{tabular}{c|ccc|ccc|c}
\hline\noalign{\smallskip}
$m_{\tilde{t}}$	& Signal eff.	 & Exp. bkg	 & Obs. data	 & Exp. limit	 & $+ \sigma$	 & $- \sigma$	 & Obs. limit \\
$[\mathrm{GeV}]$	 & [\%]	 & [events]	 & [events]	 & [fb]	 & [fb]	 & [fb]	 & [fb] \\
 
\noalign{\smallskip}\hline\noalign{\smallskip}
$200$		 & $3.9$	 & $103.9$	 & $97$	 & $210$	 & $300$	 & $150$	 & $180$\\
$300$		 & $7.0$	 & $15.3$	 & $10$	 & $28.5$	 & $37.2$	 & $21.4$	 & $19.9$\\
$400$		 & $8.6$	 & $3.1$	 & $2$	 & $12.7$	 & $18.3$	 & $9.2$	 & $10.4$\\
$500$		 & $9.5$	 & $0.9$	 & $0$	 & $7.8$	 & $8.9$	 & $5.2$	 & $5.2$\\
$600$		 & $10.8$	 & $0.7$	 & $0$	 & $5.8$	 & $7.0$	 & $5.0$	 & $5.2$\\
$700$		 & $10.2$	 & $0.9$	 & $0$	 & $6.5$	 & $8.6$	 & $5.5$	 & $5.7$\\
$800$		 & $10.6$	 & $1.4$	 & $0$	 & $6.8$	 & $8.1$	 & $5.4$	 & $5.4$\\
$900$		 & $10.2$	 & $0.1$	 & $0$	 & $5.8$	 & $6.0$	 & $5.4$	 & $5.7$\\

\noalign{\smallskip}\hline\noalign{\smallskip}
\end{tabular}

%% file: atlas_authlist.tex
\begin{flushleft}
{\Large The ATLAS Collaboration}

\bigskip

G.~Aad$^{\rm 48}$,
T.~Abajyan$^{\rm 21}$,
B.~Abbott$^{\rm 111}$,
J.~Abdallah$^{\rm 12}$,
S.~Abdel~Khalek$^{\rm 115}$,
A.A.~Abdelalim$^{\rm 49}$,
O.~Abdinov$^{\rm 11}$,
R.~Aben$^{\rm 105}$,
B.~Abi$^{\rm 112}$,
M.~Abolins$^{\rm 88}$
% ...
\bigskip
\\
% ...
$^{11}$ Institute of Physics, Azerbaijan Academy of Sciences, Baku, Azerbaijan\\
$^{12}$ Institut de F{\'\i}sica d'Altes Energies and Departament de F{\'\i}sica de la Universitat Aut{\`o}noma de Barcelona and ICREA, Barcelona, Spain\\
% ...
$^{21}$ Physikalisches Institut, University of Bonn, Bonn, Germany\\
% ...
$^{48}$ Fakult{\"a}t f{\"u}r Mathematik und Physik, Albert-Ludwigs-Universit{\"a}t, Freiburg, Germany\\
$^{49}$ Section de Physique, Universit{\'e} de Gen{\`e}ve, Geneva, Switzerland\\
% ...
$^{88}$ Department of Physics and Astronomy, Michigan State University, East Lansing MI, United States of America\\
% ...
$^{105}$ Nikhef National Institute for Subatomic Physics and University of Amsterdam, Amsterdam, Netherlands\\
% ...
$^{111}$ Homer L. Dodge Department of Physics and Astronomy, University of Oklahoma, Norman OK, United States of America\\
$^{112}$ Department of Physics, Oklahoma State University, Stillwater OK, United States of America\\
% ...
$^{115}$ LAL, Universit{\'e} Paris-Sud and CNRS/IN2P3, Orsay, France\\
% ...
\end{flushleft}
